\documentclass[twocolumn]{revtex4}
\usepackage{mathrsfs}
\usepackage{hyperref}
\usepackage{graphicx,color}
\usepackage{amsmath,amssymb}

\begin{document}

\title{Diagonalizing the Hamiltonian of $\mathbf{\lambda\phi^4}$ Theory in 2 Space-Time Dimensions}

\author{Neil Christensen}
\affiliation{Illinois State University}

\begin{abstract}
We propose a new non-perturbative technique for calculating the scattering amplitudes of field-theory directly from the eigenstates of the Hamiltonian.  Our method involves a discretized momentum space and a momentum cutoff, thereby truncating the Hilbert space and making numerical diagonalization of the Hamiltonian achievable.  We show how to do this in the context of a simplified $\lambda\phi^4$ theory in two space-time dimensions.  We present the results of our diagonalization, its dependence on time, its dependence on the parameters of the theory and its renormalization.  
\end{abstract}

\maketitle

%%%%%%%%%%%%%%The S-Matrix
%\subsubsection*{The Union of Special Relativity and Quantum Mechanics}

Quantum field theory (QFT) allows us to write a manifestly local, Lorentz-invariant formulation of the fundamental physics governing relativistic quantum particles \cite{Weinberg:1995mt}.  The Standard Model (SM) of particle physics, a QFT, agrees with all experiments to date, including the high energy particle collisions at the Large Hadron Collider (LHC).  Its structure is sufficiently constraining that it forced theorists to predict the existence of the Higgs boson \cite{Englert:1964et,Higgs:1964pj,Higgs:1964ia,Guralnik:1964eu,Higgs:1966ev,Kibble:1967sv} five decades before its spectacular discovery in 2012 \cite{Aad:2012tfa,Chatrchyan:2012xdj}.  It is also sufficiently flexible to allow us to construct models for physics beyond the SM, including neutrino masses, supersymmetry, and dark matter to name a few.

However, it has become increasingly clear that its perturbation theory, traditionally encapsulated in Feynman diagrams, is extremely and unnecessarily inefficient.  
%
%However, there are two great challenges with this approach that suggest looking for new methods.  The first is that there are always processes of interest that are beyond the ability of these Feynman diagram calculators.  The second is that only the tree-level diagrams and, in some cases, one-loop diagrams have been automatized.
%
%The textbook technique of calculating scattering amplitudes is the Feynman-diagram perturbative method.  This method was extremely powerful and insightful when cutting-edge calculations involved only a few diagrams.  However, it is, by now, well known that there are major shortcoming with Feynman diagrams.  
In particular, the number of Feynman diagrams grows exponentially with the number of final state particles and with the number of loops.  For example, in a pure theory of gluons, it takes the following numbers of Feynman diagrams to calculate scattering rates at tree level: 4; 25; 220; 2,485; 34,300; 559,405; and 10,525,900 Feynman diagrams for 4, 5, 6, 7, 8, 9 and 10 external particles, respectively \cite{Feng:2011np}.
And, if we go to one loop, the scattering of just seven gluons involves 227,585 Feynman diagrams \cite{Feng:2011np}.  In the regimes of ultra-high numbers of Feynman diagrams, the individual diagrams lose their meaning and the calculation becomes challenging or impossible even for computers.

On the other hand, new methods to calculate the scattering amplitude that bypass Feynman diagrams altogether, and indeed sometimes QFT itself, are being discovered.
%In the last decades, phenomenal progress has been made with new research showing that Feynman diagrams are inherently and deeply inefficient.  
One highlight of this work is the discovery by Parke and Taylor \cite{Parke:1986gb} that the maximally-helicity-violating tree-level amplitude could be written as an expression with only one term no matter how many thousands of Feynman diagrams were involved.  Another important highlight is the discovery by Britto, Cachazo, Feng and Witten \cite{Britto:2005fq} that any tree-level helicity amplitude of gluons could be calculated using a simple recursion relation based on the asymptotic behavior of the complexified gluon momenta.  Since then, much exciting progress has been made and, indeed, continues \cite{Dixon:2013uaa,Benincasa:2013faa,ArkaniHamed:2008gz}.
There are even some indications that scattering amplitudes, although seemingly ordinary, may be leading us to a much deeper understanding of the combination of special relativity and quantum mechanics \cite{Arkani-Hamed:2013jha,Arkani-Hamed:2014dca}.

Notwithstanding the enormous progress made by the unitarity-based methods, it seems clear that new techniques for calculating scattering amplitudes that go beyond Feynman diagrams would be welcome.  In particular, an approach whose approximation scheme was non-perturbative would complement well the existing efforts.  In this paper, we begin an exploration of another potential method for calculating the S-matrix and thereby the scattering amplitudes.  We propose to diagonalize the Hamiltonian directly and use the eigenstates to calculate the scattering amplitude.  Of course, the Hamiltonian is infinite dimensional and so this is normally impossible.  However, very briefly, what we suggest is to discretize the Hamiltonian by discretizing momentum space (e.g. by reducing the volume of spacetime to a finite size with boundary conditions) and cutting the momentum off at an upper limit far above the scale of the calculation.  The resulting discrete Hamiltonian will be finite and diagonalizable using standard computational techniques.   
At best, this will, of course, be an approximation to the diagonalization of the full Hamiltonian.  However, it is a non-perturbative approximation and therefore, we believe, potentially complements standard methods.  
%This method of computationally diagonalizing a discretized Hamiltonian has been followed in atomic physics, however it has not previously been used to calculate scattering amplitudes in HEP. 

In slightly greater detail, our plan is to: a) Legendre transform the Lagrangian to obtain the Hamiltonian; b) Use direct products of free-particle states as a complete orthonormal basis; c) Insert the definitions of the fields in terms of the creation and annihilation operators of free-particle states; d) Discretize momentum space and therefore the Hilbert space; e) Truncate the basis states by cutting off the energy of the states and thereby make the Hilbert space finite; f) Calculate the numerical Hamiltonian matrix elements in the resulting finite Hilbert space; g) Numerically diagonalize the resulting Hamiltonian matrix; and h) Use the resulting eigenstates to calculate the S matrix.  These steps will be further explained in the context of a simplified $\lambda\phi^4$ theory in Sec.~\ref{sec:theory} and the results will be presented in the following sections.

Before getting into the details, we should note that there is a Ph.D. thesis from 1969 by A.~Suri \cite{Suri:1969zz}, where Wilson states \cite{Rebbi:1984tx} that he and his student diagonalized the Hamiltonian in a latticized space-time and calculated the scattering amplitudes.  Their work was seminal in the foundation of lattice gauge theory \cite{Peskin:2014lsa}.  Although the diagonalization of the Hamiltonian is a common goal, our method uses different techniques.  We do not latticize space-time.  Instead, we use of a complete set of free-particle product states and the insertion of creation and annihilation operators as a means to calculate the Hamiltonian that is then diagonalized.  Furthermore, we take advantage of the many orders-of-magnitude improvement in the speed and memory of modern computers since that thesis.  To the best of our knowledge, our suggested approach is unique.

We also note that we are not the first to diagonalize the Hamiltonian of field theory.  This has been a recently expanding area of research \cite{Salwen:1999pw,Lee:2000ac,Lee:2000xna,Borasoy:2001pb,Wagner:1,Wagner:2,Rychkov:2014eea,Rychkov:2015vap,Elias-Miro:2015bqk,Bajnok:2015bgw,Katz:2016hxp}.  Since the diagonalization of the scalar $\phi^4$ theory is the main result of our paper, we should state what we do that is new.  We plot the wavefunction in a novel way using the free-particle energy for the horizontal axis, giving a more clear understanding of its properties.  We find the dependence of the wavefunction phase on time.  We analyze in detail how the solutions depend on the parameters, both physical and unphysical and we devise a new approach to renormalization when diagonalizing the Hamiltonian.  Moreover, we describe the results of this renormalization on the eigenvalues and eigenstates.

%Before continuing the technical details, we note that the numerical diagonalization of scalar $\phi^4$ theory has been carried out in \cite{Wagner:1,Wagner:2,Rychkov:2014eea}.  However, our work is novel in several respects.  Among them are that \cite{Wagner:1,Wagner:2,Rychkov:2014eea} focused mainly on the properties of the vacuum whereas we consider it as just one (important) state among the many other eigenstates of the Hamiltonian.  Also, our Hamiltonian is Lorentz covariant, our approach to renormalization is different, and \cite{Wagner:1,Wagner:2,Rychkov:2014eea} were not interested in the S-matrix.  Also, \cite{Rychkov:2014eea} is more interested in strong coupling and the critical coupling.  On the other hand, we find the approach to renormalization by \cite{Rychkov:2014eea} very interesting.  It involves an improvement to convergence by approximating the contribution of the high energy states that are removed from the Hilbert space by the energy cutoff.  Finally, we note that \cite{Glimm:1968kh,Jaffe:2,Jaffe:3} studied various theoretical properties of this 2-dimensional theory. 

We organize this paper as follows.  In Sec.~\ref{sec:theory}, we describe in detail our Hamiltonian, its discretization and truncation.  In Sec.~\ref{sec:results}, we describe the results of our numerical diagonalization of the Hamiltonian with an illustrative set of parameters.  In Sec.~\ref{sec:dependence on parameters}, we show how our results depend on the parameters of the theory.  In Sec.~\ref{sec:renormalization}, we discuss the renormalization of the parameters and its effect on the eigenstates and their energies.  In Sec.~\ref{sec:conclusions}, we discuss our results and conclude.  
%Appendix~\ref{app:a and adagger} contains a review of some details of creation and annihilation operators in the context of our calculation, while in App.~\ref{Evac 1/Delta p}, we work out the dependence of the vacuum energy on $\lambda$ and the discrete momentum spacing that is used in the main text.

\section{\label{sec:theory}The Theory and Method}
For the present paper, we begin to develop our techniques in the simplest possible context.  We begin with a pure scalar field theory in 2 dimensions (1 spatial dimension) with a quartic coupling and a parity symmetry.  Our Lagrangian is
\begin{equation}
\mathcal{L} = \frac{1}{2}\partial_\mu\phi\partial^\mu\phi - \frac{1}{2}m^2\phi^2 - \frac{\lambda}{4!}\phi^4\ .
\label{eq:Lagrangian}
\end{equation}
Because of the parity symmetry, even and odd numbers of asymptotically-free particles decouple and we focus on basis states with an even number of free particles in the present work.  
The action resulting from our Lagrangian is
\begin{equation}
\mathcal{S} = \int d^2x \mathcal{L}\ .
\end{equation}
Since the action is dimensionless, $dx$ has dimension $1/E$ and
$\partial_\mu$ has dimension $E$, we find that $\phi$ is
dimensionless.  This implies that $\lambda$ has dimension $E^2$.

In order to diagonalize the Hamiltonian, we need a complete basis set for which we can calculate the Hamiltonian.  We use direct products of free particles for our basis.  We label them as $|\rangle, |(p,n)\rangle,|(p_1,n_1),(p_2,n_2)\rangle,...$ where $p, p_1$ and $p_2$ are the momenta and $n, n_1$ and $n_2$ are the multiplicities for those momenta.  As previously described, because our theory has a parity symmetry, even and odd numbers of free particles will decouple.  For computational efficiency, therefore, we diagonalize the Hamiltonian separately for basis state with even and odd numbers of particles.  However, we will consider both even and odd numbers of free particles together in the present discussion for convenience.  The free energies for our basis states are defined by $\omega=\sum_in_i\sqrt{p_i^2+m^2}$, which is simply the sum of the free energies for each free particle in the basis state.  

We next define a creation operator to have the property that it adds a free-particle to the direct product of free-particle states already present
\begin{equation}
a^\dagger(p_i)|\cdots(p_i,n_i)\cdots\rangle = \sqrt{n_i+1} |\cdots(p_i,n_i+1)\cdots\rangle\ .
\end{equation}
Some details of the creation operator can be found in Appendix~\ref{app:a and adagger} where we also show that this results in the hermitian conjugate annihilating a free particle as in
\begin{equation}
a(p_i)|\cdots(p_i,n_i)\cdots\rangle = \sqrt{n_i}|\cdots(p_i,n_i-1)\cdots\rangle\ .
\end{equation}
Consequently, these operators satisfy the commutation property
\begin{equation}
\left[a(p),a^\dagger(p')\right] = 2\pi\delta(p-p')\ .
\end{equation}
With these operators, a field theory can be constructed in the usual way (see Weinberg \cite{Weinberg:1995mt}) by writing the scalar field as a linear combination of these two operators
\begin{equation}
\phi(x) = \int\frac{dp}{2\pi}\frac{1}{\sqrt{2\omega}}\left[a(p)e^{i \left(\omega t-p x\right)}+a^\dagger(p)e^{-i \left(\omega t-p x\right)}\right]\ .
\label{eq:phi(x) def}
\end{equation}

We next Legendre transform the Lagrangian to obtain the Hamiltonian.  We first find the conjugate momentum
\begin{equation}
\pi = \frac{\partial \mathcal{L}}{\partial\dot{\phi}} = \dot{\phi}\ ,
\end{equation}
and use it to obtain
\begin{eqnarray}
H &=& \int dx\left(\pi\dot{\phi}-\mathcal{L}\right)\\
&=& \int dx\left[\frac{1}{2}\left(\frac{\partial\phi}{\partial t}\right)^2+\frac{1}{2}\left(\frac{\partial\phi}{\partial x}\right)^2+\frac{1}{2}m^2\phi^2+\frac{\lambda}{24}\phi^4\right]\ .\nonumber
\end{eqnarray}
We then insert our definition of the scalar field in terms of the creation and annihilation operators to obtain our Hamiltonian operator.  

Before writing our resulting Hamiltonian operator, we switch to discrete momentum space.  We take $\int dp/(2\pi)\to \sum_p \Delta p$ which implies $2\pi\delta(p)\to\delta_p/\Delta p$.  Then, the commutator of creation and annihilation operators become
\begin{equation}
\left[ a_p , a^\dagger_{p'}\right] = \delta_{p,p'}\ ,
\end{equation}
which implies $a(p)\to a_p/\sqrt{\Delta p}$ and $a^\dagger(p)\to a^\dagger_p/\sqrt{\Delta p}$.
After normal ordering, a straightforward calculation gives us the Hamiltonian
\begin{eqnarray}
H &=&
\sum_p \omega a^\dagger_p a_p 
+\frac{\lambda\Delta p}{16}\sum_{p'}\frac{1}{\omega'}\sum_p\frac{1}{\omega}\Big[\nonumber\\
&&a_pa_{-p}e^{2i \omega t}+2a^\dagger_pa_p+a^\dagger_pa^\dagger_{-p}e^{-2i \omega t}\Big]\nonumber\\
&&+\frac{\lambda\Delta p}{96}\sum_{p_1p_2p_3p_4}\frac{1}{\sqrt{\omega_1\omega_2\omega_3\omega_4}}\Big[\nonumber\\
&&a_{p_1}a_{p_2}a_{p_3}a_{p_4} e^{i\left(\omega_1+\omega_2+\omega_3+\omega_4\right)t}\delta_{p_1+p_2+p_3+p_4}\nonumber\\
&&+4a^\dagger_{p_1}a_{p_2}a_{p_3}a_{p_4} e^{i\left(-\omega_1+\omega_2+\omega_3+\omega_4\right)t}\delta_{-p_1+p_2+p_3+p_4}\nonumber\\
&&+6a^\dagger_{p_1}a^\dagger_{p_2}a_{p_3}a_{p_4} e^{i\left(-\omega_1-\omega_2+\omega_3+\omega_4\right)t}\delta_{-p_1-p_2+p_3+p_4}\nonumber\\
&&+4a^\dagger_{p_1}a^\dagger_{p_2}a^\dagger_{p_3}a_{p_4} e^{i\left(-\omega_1-\omega_2-\omega_3+\omega_4\right)t}\delta_{-p_1-p_2-p_3+p_4}\nonumber\\
&&+a^\dagger_{p_1}a^\dagger_{p_2}a^\dagger_{p_3}a^\dagger_{p_4} e^{i\left(-\omega_1-\omega_2-\omega_3-\omega_4\right)t}\delta_{-p_1-p_2-p_3-p_4}\Big]\ ,\nonumber\\
\label{eq:Discrete Hamiltonian}
\end{eqnarray}
where we have dropped an overall constant term that does not affect the dynamics.  We can see that the effectively free part of this Hamiltonian (the part that is quadratic in the creation and annihilation operators) not only contains the free-particle energy $\omega$ but also the second term proportional to $\lambda$.  This comes from the normal ordering removing creation and annihilation operators from the original Hamiltonian.  Eq.~(\ref{eq:Discrete Hamiltonian}) is the Hamiltonian that we will diagonalize in this work.

Since our Hamiltonian conserves momentum (as it must), we need only consider states with total momentum equal to zero.  The Hamiltonian matrix elements for any other center of momentum can be obtained from these by an appropriate Lorentz transformation.

In order to diagonalize this Hamiltonian, we find its matrix elements in the free-particle product basis by sandwiching this operator between the basis states.  For example, the free-vacuum matrix element is given by
\begin{equation}
\langle|H|\rangle = 0\ ,
\end{equation}
since every operator of the Hamiltonian annihilates either the free vacuum on the right or the left.  Other elements are, of course, non zero.  As we mentioned, the basis states with an odd number of free particles decouple.  We can now see that this is because every term of the Hamiltonian contains an even number of creation and annihilation operators.  Since we only consider basis states with zero total momentum, there is only one basis state with one free particle to consider.  It is $|(0$GeV$,1)\rangle$ and the only nonzero Hamiltonian matrix element involving it is the diagonal term
\begin{equation}
\langle(0\mbox{GeV},1)|H|(0\mbox{GeV},1)\rangle = m + \frac{\lambda\Delta p}{8 m}\sum_{p'}\frac{1}{\omega'}\ .
\label{eq:1 part energy}
\end{equation}
The matrix element between the free vacuum $|\rangle$ and $|(0$GeV$,2)\rangle$ is
\begin{equation}
\langle|H|(0\mbox{GeV},2)\rangle = \sqrt{2}\frac{\lambda\Delta p}{16m}\sum_{p'}\frac{1}{\omega'}e^{2 i m t}\ ,
\end{equation}
while
\begin{equation}
\langle(0\mbox{GeV},2)|H|\rangle = \sqrt{2}\frac{\lambda\Delta p}{16m}\sum_{p'}\frac{1}{\omega'}e^{-2 i m t}
\end{equation}
is its hermitian conjugate.
  The matrix element between the free vacuum $|\rangle$ and one of the basis states with a product of two distinct free particles $|(-p,1),(p,1)\rangle$ is
\begin{equation}
\langle|H|(-p,1),(p,1)\rangle = \frac{\lambda\Delta p}{8\omega}\sum_{p'}\frac{1}{\omega'}e^{2 i \omega t}\ ,
\end{equation}
where $\omega=\sqrt{p^2+m^2}$ and its hermitian conjugate is the complex conjugate.  This accounts for the top row and first column of the matrix.  
The third row contains $\langle(0$GeV$,2)|$ on the left.  The diagonal entry of this row is given by
\begin{equation}
\langle(0\mbox{GeV},2)|H|(0\mbox{GeV},2)\rangle = 2m 
+ \frac{\lambda\Delta p}{4m}\sum_{p'}\frac{1}{\omega'}
+ \frac{\lambda\Delta p}{8m^2}\ ,
\label{eq:<0,2|H|0,2>}
\end{equation}
slightly greater than double Eq.~(\ref{eq:1 part energy}), while the other entries are 
\begin{equation}
\langle(0\mbox{GeV},2)|H|(-p,1),(p,1)\rangle = 
\sqrt{2}\frac{\lambda\Delta p}{8m\omega}e^{-2i(m-\omega)t}\ .
\end{equation}

The interior of the matrix is given by both the bra and ket containing a product of two distinct free particles.  
\begin{displaymath}
\langle(-p,1),(p,1)|H|(-p',1),(p',1)\rangle = \hspace{2in}
\end{displaymath}
\begin{equation}
2\left(\omega+\frac{\lambda\Delta p}{8\omega}\sum_{p''}\frac{1}{\omega''}\right)\delta_{p-p'}
+\frac{\lambda \Delta p}{4\omega\omega'}e^{-2i(\omega-\omega')t}\ .
\label{eq:<-p,p|H|-q,q>}
\end{equation}

Other Hamiltonian matrix elements are similarly obtained with greater numbers of free particles in the basis states.

\section{\label{sec:results}The Results}

We implemented the Hamiltonian described in Eq.~(\ref{eq:Discrete Hamiltonian}) and diagonalized it using the numerical techniques described in App.~\ref{app:numerical techniques}.  For this section, we chose our illustrative constants to be 
\begin{equation}
\Delta p=0.05\mbox{GeV},\quad \lambda=0.1\mbox{GeV}^2 \quad \mbox{and}\quad m=1\mbox{GeV}\ ,
\label{eq:parameters}
\end{equation}
 and cut off our single particle momenta at $\pm10$GeV.  By design, we took $\Delta p$ to be small compared to $m$ and $\lambda$ while the cutoff momentum was taken to be large compared to the same.  We also took $\lambda$ to be small compared to $m$.  Altogether, we took $\Delta p\ll\sqrt{\lambda}\ll m\ll p_{cut}$.  The absolute scale of energy is arbitrary; we have chosen GeV.

In order to build up our understanding of this field-theory-Hamiltonian diagonalization technique and its consequences, in the present work, we have restricted our basis states to only include two or fewer free particles.   This will allow us to develop our results in as simple a system as possible and enable us to draw conclusions about the results more easily.  However, our intention in future works is to go beyond this limitation and consider more complex systems with more complete Hilbert spaces.
Therefore, in this work, our basis states consist of the free vacuum $|\rangle$, the one-free-particle state $|(0$GeV$,1)\rangle$, and two-free-particle states with each particle's momentum below $\pm10$GeV as described above.  There are 203 of these basis states ordered in terms of their free energy.  We give a list of the basis states used in our numerical code in Table~\ref{tab:basis states}.
\begin{table}
\begin{center}
\begin{tabular}{|lll|}
\multicolumn{3}{c}{Free-Particle Basis States}\\
\hline
&Basis State&Free Energy\\
\hline
0&$|\rangle$&0\\
s&$|(0$GeV$,1)\rangle$&1GeV\\
1&$|(0$GeV$,2)\rangle$&2GeV\\
2&$|(-0.05$GeV$,1),(0.05$GeV$,1)\rangle$\hspace{0.1in} &2.003GeV\\
3&$|(-0.1$GeV$,1),(0.1$GeV$,1)\rangle$&2.010GeV\\
4&$|(-0.15$GeV$,1),(0.15$GeV$,1)\rangle$&2.022GeV\\
&\vdots&\vdots\\
50&$|(-2.45$GeV$,1),(2.45$GeV$,1)\rangle$&5.292GeV\\
&\vdots&\vdots\\
201 & $|(-10$GeV$,1),(10$GeV$,1)\rangle$&20.10GeV\\
\hline
\end{tabular}
\end{center}
\caption{\label{tab:basis states}A list of the basis states and their associated free-particle energies used in this analysis.  The first number in parentheses is the momentum of a free-particle state while the second number is the multiplicity of that momentum.  These states are formed as a direct product of free-particle states.}
\end{table}

We will now describe the results of our diagonalization.  We will begin in Subsection~\ref{sec:results:energy} with a description of the energy spectrum.  In Subsections~\ref{sec:Results:absolute wave function} and \ref{sec:Results:wave function phases}, we will describe the wavefunctions of the eigenstates.  In Subsection~\ref{sec:Results:absolute wave function}, we will describe the absolute value of the coefficients of the basis states for the wavefunctions while in Subsection~\ref{sec:Results:wave function phases}, we will describe their phases.   In Subsection~\ref{sec:scattering}, we will briefly describe the S matrix and scattering in the context of the current truncated Hilbert space.

\subsection{\label{sec:results:energy}Energy Spectrum}
\begin{figure}
\begin{center}
\hspace{0.12in} \includegraphics[scale=0.8]{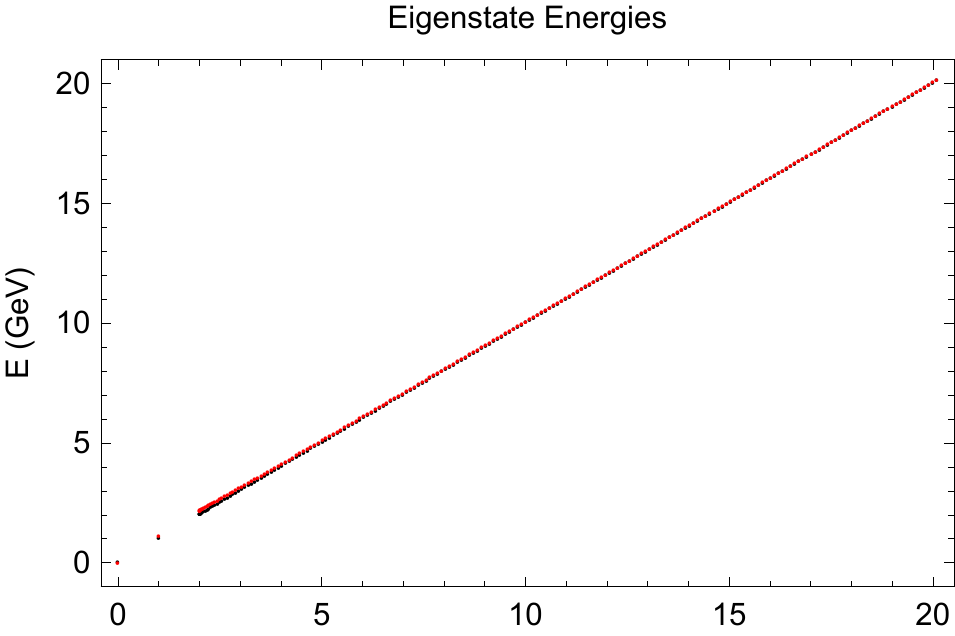}\\
\includegraphics[scale=0.85]{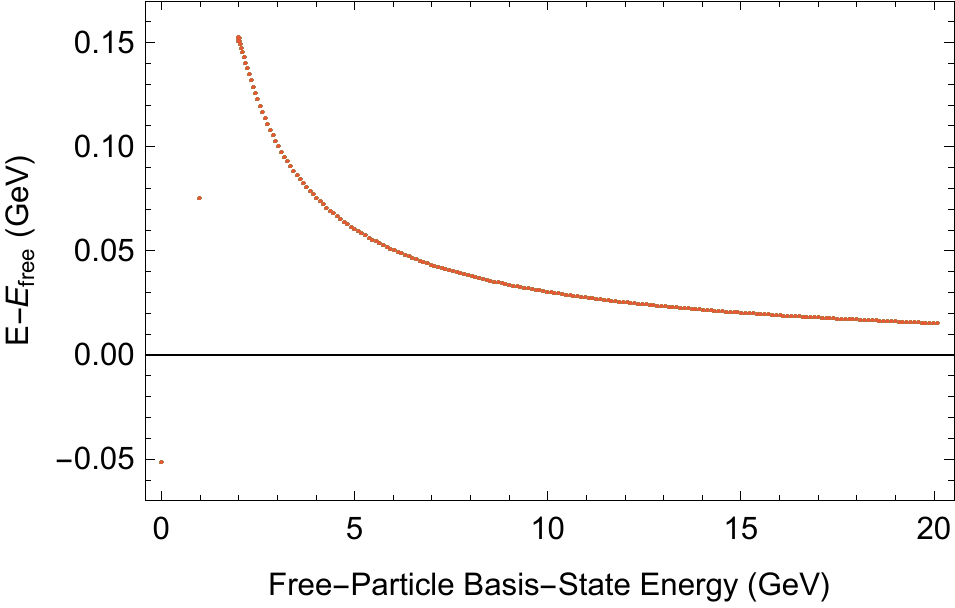}
\end{center}
\caption{\label{fig:EnSpectrum2}A plot of the energy eigenvalues of the Hamiltonian (red dots) which were obtained by numerical diagonalization of the discretized form of the Hamiltonian in Eq.~(\ref{eq:Discrete Hamiltonian}) with the parameters given in Eq.~(\ref{eq:parameters}).   The black dots are the free energies of the basis states.  The bottom plot displays the difference between the energy eigenvalues and the free energies of the basis states.}
\end{figure}

After diagonalizing, the eigenstates have definite energy corresponding with the eigenvalues.  We show a plot of the energies obtained in the top plot of Fig.~\ref{fig:EnSpectrum2}.
We have plotted the eigenvalues in red while in black we have plotted the free-particle energies of the basis states for comparison.  Along the horizontal axis, we have used the free-particle basis-state energies for reference.  Because of this, the free-particle basis-state energies in black form a diagonal line with a slope of $1$ and a gap between the free vacuum, the free 1-particle state and the free 2-particle states.  For the coordinates of each eigenstate, on the other hand, we have given it an abscissa equal to the energy of the free-particle basis state that dominates its eigenstate (see the next subsections), while the ordinate is the energy eigenvalue.  We have found that with the parameters listed in Eq.~(\ref{eq:parameters}), the lowest energy eigenvalue (associated with the vacuum) has a value of -0.0517GeV, the single-particle state (associated with $|(0$GeV$,1)\rangle$) has an energy of 1.075GeV, and the lowest-energy two-particle eigenstate (associated most closely with $|(0$GeV$,2)\rangle$) has an energy of 2.151GeV, double that of the single-particle state.  This is because the lowest-energy two-particle state is dominated by two free particles at rest and the diagonal entry for two particles at rest in Eq.~(\ref{eq:<0,2|H|0,2>}) is essentially double that of the diagonal entry of a single particle at rest in Eq.~(\ref{eq:1 part energy}).  The energies of the other two-particle states grow nearly linearly from there.  

We note that the eigenvalue energies lie very close to the free energies of their dominant basis state.  However, they are not exactly equal.  To further clarify the relationship with the free energies of the basis states, we have plotted the difference between the eigenvalues and the basis-state energies in the bottom plot of Fig.~\ref{fig:EnSpectrum2}.  We see that the vacuum is slightly below the free energy of the free vacuum $|\rangle$ as mentioned previously.  The fact that its energy is very close to $-\Delta p$ is a coincidence as we will see in Sec.~\ref{sec:Delta p}.  The single-particle state energy is slightly above the free energy of a single particle, the lowest two-particle eigenstate doubles the difference with the free energy of $|(0$GeV$,2)\rangle$, and the higher energy two-particle states have energies that asymptotically approach the free energies of their associated basis states.  The reason the energies are greater than the free energies of the associated basis states is due to the interactions and the nonperturbative nature of our calculation as can be seen in Eq.~(\ref{eq:1 part energy}) for the single-particle state.  In Eq.~(\ref{eq:<0,2|H|0,2>}), on the other hand, we see that the first two terms are exactly double those in Eq.~(\ref{eq:1 part energy}), leading to the very-nearly double eigenvalue.

We also note that there is a gap between the vacuum and the one-particle eigenstate and then another between the one-particle state and the continuum of two-particle states.  As expected, this is due to the mass of the particle.  In fact, the mass gap is the difference between the single-particle state energy and that of the vacuum.  Therefore, we take the physical value of the mass in this calculation to be 1.075GeV+0.052GeV=1.127GeV.  And, similarly for the other eigenstates.  The physical energy is the difference between the energy eigenvalue and the vacuum energy.  So, the physical energy of the first two-particle state is 2.151GeV+0.052GeV=2.203GeV, a little less than double the physical energy of the single-particle state.  This may be a sign of binding energy.  Further investigation is required.

Additionally, we have diagonalized the Hamiltonian for hundreds of unique nonzero times and find that the eigenvalues exactly agree for all time.  We find, therefore, that the energy eigenvalues are independent of time, as we would expect from energy conservation.

\subsection{\label{sec:Results:absolute wave function}Absolute Wave Functions}
\begin{figure}[!]
\begin{center}
\includegraphics[scale=0.85]{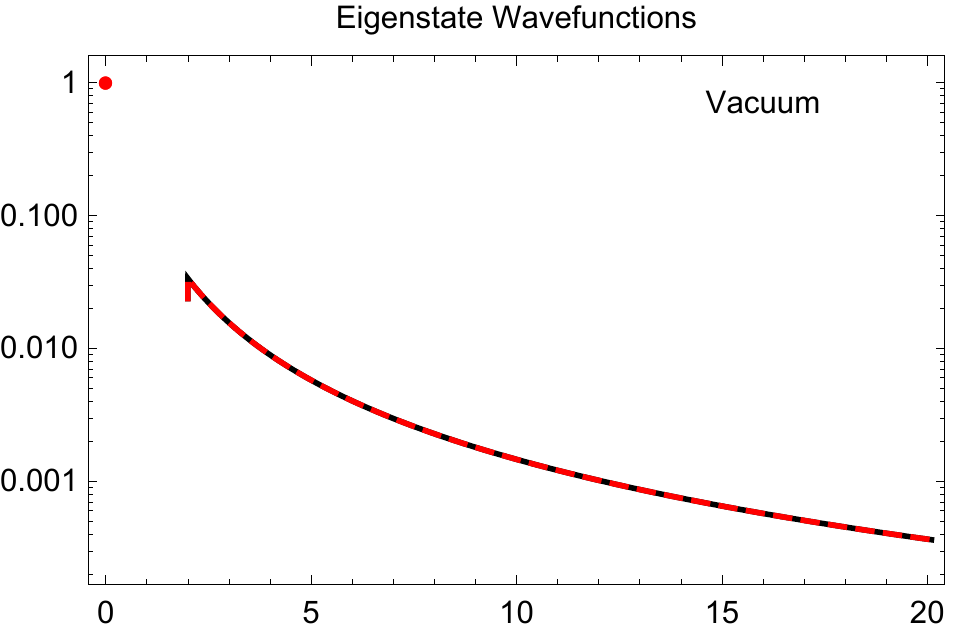}\\
\includegraphics[scale=0.85]{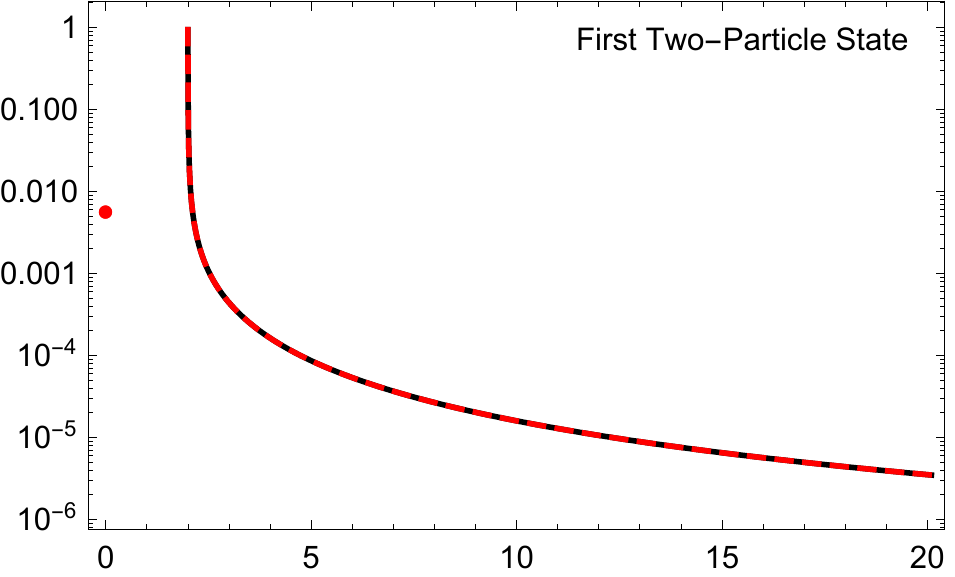}\\
\includegraphics[scale=0.85]{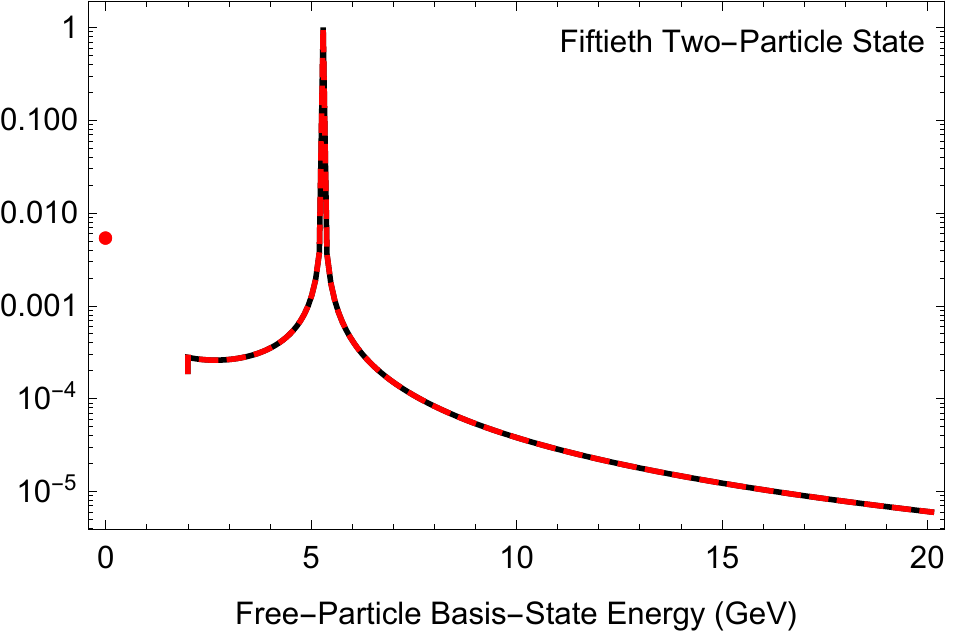}
\end{center}
\caption{\label{fig:VacuumWaveFunction2}A plot of the absolute values of the coefficients of the wavefunction for a few eigenstates.  The horizontal axis gives the basis-state free-particle energies as specified in the right column of Table~\ref{tab:basis states}.  These wavefunctions were obtained by a numerical diagonalization of the Hamiltonian in Eq.~(\ref{eq:Discrete Hamiltonian}) with the parameters given in Eq.~(\ref{eq:parameters}).  The solid black line is for $t=0$ while the dashed red line (directly on top of the solid black line) is for $t=1$GeV$^{-1}$.  The top plot is for the vacuum state, the middle plot is for the first excited two-particle state and the bottom plot is for the fiftieth excited two-particle state.}
\end{figure}

Since the coefficients of the basis states are, in general, complex, we begin by describing the modulus of the coefficients.  We present the absolute values of the coefficients of the wave function for a few eigenstates in Fig.~\ref{fig:VacuumWaveFunction2}.
We found that the moduli of the wavefunctions are independent of time.  This can be seen in Fig.~\ref{fig:VacuumWaveFunction2} where the wavefunction for two different times were plotted and are exactly on top of each other.  In fact, we numerically diagonalized the Hamiltonian for a few hundred different times and found exact agreement of the modulus of the coefficients for all time.  This is expected since the Hamiltonian is the generator of time evolution.  Since these solutions are eigenstates of the Hamiltonian, we expect that, at most, their phases change as a function of time.  We will discuss the phases in Subsection~\ref{sec:Results:wave function phases}.

There are several general features of these wavefunctions.  We see that there is no contribution from the free single-particle basis state $|(0$GeV$,1)\rangle$.  This is because the basis states with odd numbers of free particles decouple from those with even numbers of free particles.  We also see that there is a gap between the contribution of the free vacuum and the free two-particle states.  This is because we have plotted these eigenstates as functions of the free-particle energies of the basis states and there is a gap in the free-particle energies between the zero-particle and two-particle basis states.  However, even if we plotted the eigenstate as a function of the ordinal number of the basis state given in the left column of Table~\ref{tab:basis states} (but leaving out the single-particle state), there would be no gap and the line would be continuous, but there would be a kink in the wavefunction at the free-particle vacuum.  This is because the free vacuum is fundamentally different than the continuum of two-particle basis states.  So, in general, we do not expect its contribution to any particle eigenstate to be infinitesimally close to the two-particle basis states adjacent to it.  

In general, we also find that there is a kink between the contribution from the basis-state $|(0\mbox{GeV},2)\rangle$ and $|(-0.05\mbox{GeV},1),(0.05\mbox{GeV},1)\rangle$, which are the first two basis states in the continuum.  Presumably, this is because the basis state $|(0\mbox{GeV},2)\rangle$ is a direct product of two particles at rest, whereas the other two-particle basis states have particles with equal but opposite non-zero momenta.  

We also see that, for each eigenstate, there is one basis state that dominates the wavefunction while all other basis states contribute at approximately the percent or less level.  This can be understood in terms of the coupling.  If we turned the coupling off completely, the theory would be absolutely free and the eigenstates would be delta functions centered at each of the free-particle basis states.  When we turn the coupling on to a very small value, the states adjacent to this basis state suddenly turn on their contribution, continuously with the coupling.  Because of this, the contribution of the adjacent basis states is small in proportion to the smallness of the coupling constant.  We, furthermore, see that this creates a kink in the eigenstate at the basis states directly to the sides of the former delta function.  This is because as we turn the coupling on from zero, the contribution from these states immediately turns on, but continuously in the continuum of states to the sides of the former delta function.  That is, they all come up together, forming a kink.  This kink tends to be mild, however.

The top plot of Fig.~\ref{fig:VacuumWaveFunction2} gives the wavefunction of the vacuum, the eigenstate of lowest energy.  As expected, we can see that it is strongly peaked at the free vacuum $|\rangle$.  Its coefficient is 0.991.  The next most important contribution to the vacuum is from the basis state $|(-0.05\mbox{GeV},1),(0.05\mbox{GeV},1)\rangle$, containing two free particles, each with $\pm0.05$GeV of momentum.  Its contribution to the vacuum is a few percent.  Beyond this, the contribution falls off towards zero.  We note that although the fall off appears to be gradual, this is on a log plot.  By the time the two-hundred-third basis state is reached, the contribution to the wavefunction has fallen to a few hundredths of a percent.  

The middle plot of Fig.~\ref{fig:VacuumWaveFunction2} gives the wavefunction of the first two-particle state, the state whose energy is directly above that of the vacuum and single-particle state, and which has slightly more energy than the free-particle energy of its dominant basis state.  We can see that it is strongly peaked at the lowest free-particle basis state with two particles, namely $|(0\mbox{GeV},2)\rangle$, where each free particle is at rest.  Its coefficient is 0.937.    The next most important contribution to this eigenstate is from the basis state $|(-0.05\mbox{GeV},1),(0.05\mbox{GeV},1)\rangle$ which contributes at the few percent level.  Beyond this, the contribution to this eigenstate falls off towards zero.

The bottom plot of Fig.~\ref{fig:VacuumWaveFunction2} gives the wavefunction of the fiftieth two-particle state, the state with the fiftieth energy level above that of the vacuum and single-particle state.  We see that it is strongly peaked at the fiftieth free-particle product state with two particles, namely $|(-2.45\mbox{GeV},1),(2.45\mbox{GeV},1)\rangle$, whose free-particle energy is $5.3$GeV.  Its coefficient is 0.99996.  All other basis states contribute at the less than one percent level.  The highest of these is the free vacuum at the half percent level.  This drops to a hundredth of a percent for $|(0\mbox{GeV},2)\rangle$, followed by a small discrete jump and then a gradual increase towards the peak reaching a pre-peak maximum of a few tenths of a percent.  At this point, and on the other side of the peak, is a mild kink due to the nonzero coupling bringing up the contribution of the adjacent states together.  As in the previous two cases, we see that the contribution to this eigenstate continues to fall off at higher basis states, eventually tending towards zero.

We also note that due to the decoupling nature of the basis states with an odd number of free particles and the fact that we only included basis states with two or fewer free particles, the single particle state $|(0$GeV$,1)\rangle$ is the only basis state to be diagonal in its block of the Hamiltonian.  Therefore, it is already diagonalized and its wavefunction is a delta function at this basis state.

\subsection{\label{sec:Results:wave function phases}Wave Function Phases}
\begin{figure}[h!]
\begin{center}
\includegraphics[scale=0.85]{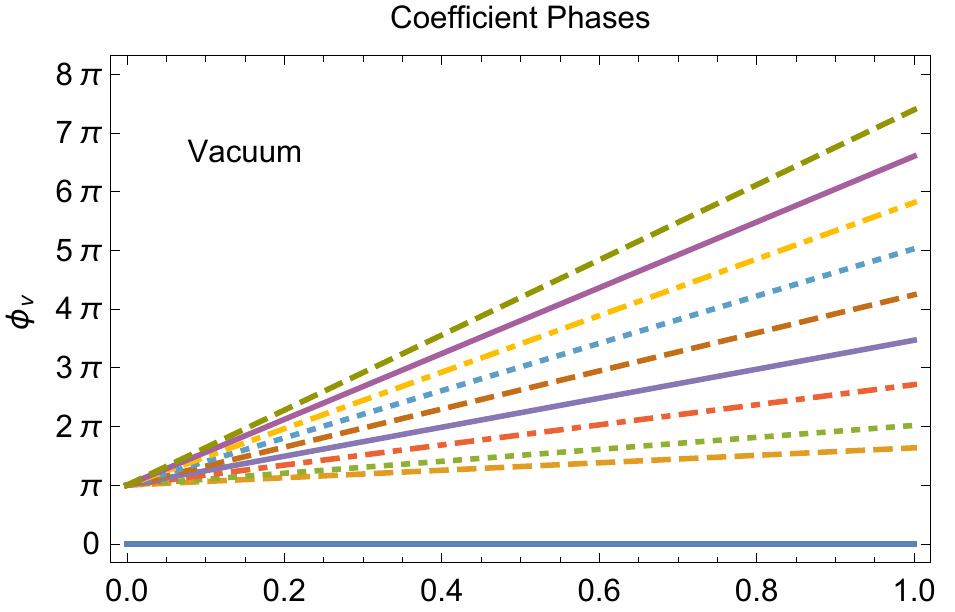}\\
\includegraphics[scale=0.85]{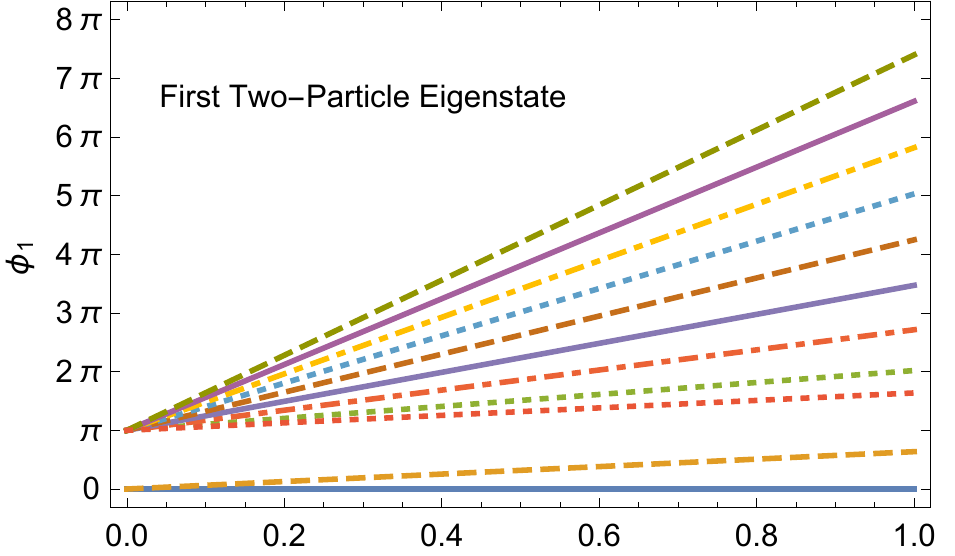}\\
\includegraphics[scale=0.85]{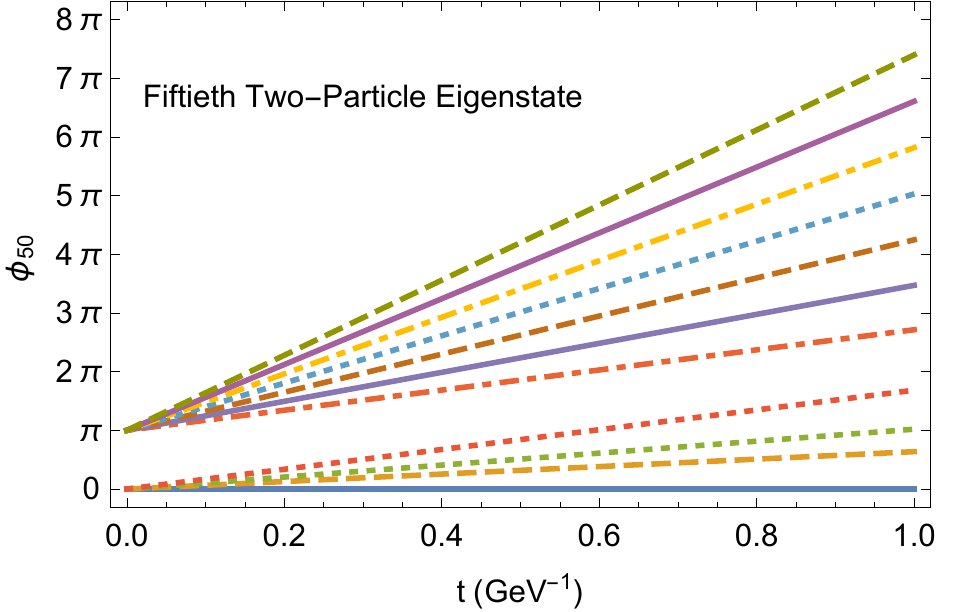}
\end{center}
\caption{\label{fig:wave function phases}A plot of the wavefunction phases for a few eigenstates as a function of time.  These wavefunctions were obtained by a numerical diagonalization of the Hamiltonian in Eq.~(\ref{eq:Discrete Hamiltonian}) with the parameters given in Eq.~(\ref{eq:parameters}).  The top plot is for the vacuum state, the middle plot is for the first two-particle state and the bottom plot is for the fiftieth two-particle state.  An explanation of the lines is given in Subsection~\ref{sec:Results:wave function phases}.}
\end{figure}

Now that we have an understanding of the absolute values of the coefficients of the wavefunctions, we turn our attention to the phases of these same coefficients.  Since the overall phase of any eigenstate is arbitrary, we have normalized it by taking the phase of the coefficient of the free-vacuum basis state $|\rangle$ to be $0$. So, all other phases are relative to the coefficient of the free-vacuum.  We have plotted a selection of these phases for three eigenstates in Fig.~\ref{fig:wave function phases}.
For each plot, time is plotted along the horizontal axis while the phases are along the vertical axis.  We have plotted the phase for just ten or eleven coefficients (out of the 203 coefficients of the basis states listed in Table~\ref{tab:basis states}) to make the plot less busy, but the other coefficients follow the same pattern, filling in the gaps with one exception.  The gap between $0$ and $\pi$ on the left of the plot is not filled in by the other coefficients.  

The organization of the lines is as follows:  The bottom, solid blue line is the phase of the coefficient of the free-vacuum basis state $|\rangle$ for each eigenstate.  By our normalization convention, it is always $0$.  Each successively higher line is for a correspondingly higher basis state.  As a result, we find that each successively higher basis state has a phase that increases with time at a correspondingly higher constant rate.  Directly above the free-vacuum line at $0$, the dashed gold line is the phase of the coefficient of the basis state $|(0$GeV$,2)\rangle$.  We note that in the top plot, this dashed gold line begins at $\pi$ while in the bottom two plots, this dashed gold line begins at $0$.  We find that, as a general rule, the phase of all coefficients for basis states below and including the dominant basis state of the eigenstate, begin at $0$ while the phases of the coefficients of the basis states above the dominant basis state begin at $\pi$.  Therefore, we find that for the vacuum eigenstate, shown in the top plot of Fig.~\ref{fig:wave function phases}, only the free-vacuum basis-state phase begins at $0$.  All other phases begin at $\pi$ since the free-vacuum basis state dominates the vacuum.  On the other hand, in the middle plot of Fig.~\ref{fig:wave function phases}, both the coefficients of the free vacuum $|\rangle$ and the dominant basis state $|(0$GeV$,2)\rangle$ begin at $0$ while all others begin at $\pi$.  The dotted red line of the middle plot is the phase of the basis state $|(-0.05$GeV$,1),(0.05$GeV$,1)\rangle$, which is the next higher basis state.  Finally, the bottom plot, for the fiftieth excited eigenstate, has all phases for basis states below and including the dominant basis state $|(-2.45\mbox{GeV},1),(2.45\mbox{GeV},1)\rangle$ beginning at $0$.  The phases for all basis states above this, on the other hand, begin at $\pi$.

The rest of the lines are as follows:  the dotted green line above the dashed gold line is for the basis state $|(-1.25$GeV$,1),(1.25$GeV$,1)\rangle$, the dot-dashed red line above that is for $|(-2.5\mbox{GeV},1),(2.5\mbox{GeV},1)\rangle$, the solid blue line above that is for $|(-3.75$GeV$,1),(3.75$GeV$,1)\rangle$, and so on.  Each higher line is for a basis state with the momenta increased by $1.25$GeV, with the final dashed green line at the top being for the basis state $|(-10$GeV$,1),(10$GeV$,1)\rangle$.  In addition to the lines separated by $1.25$GeV, we have added a dotted red curve to the middle and bottom plots.  For the middle plot, this dotted red line is the third line from the bottom and corresponds with the basis state $|(-0.05$GeV$,1),(0.05$GeV$,1)\rangle$.  Its purpose is to include the basis state directly above the dominant basis state.  In the bottom plot, the dotted red line is for the dominant basis state $|(-2.45\mbox{GeV},1),(2.45\mbox{GeV},1)\rangle$.

\begin{figure}
\begin{center}
\includegraphics[scale=0.85]{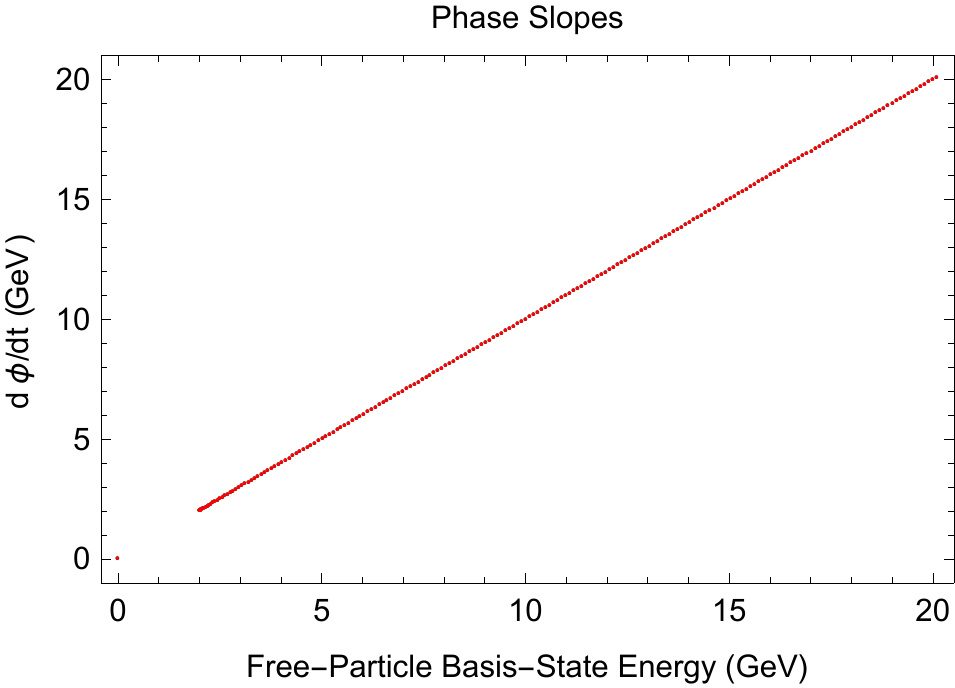}
\end{center}
\caption{\label{fig:phase plot}A plot of the rate of change of the phases of the coefficients of each basis state.   The horizontal axis is the free energy of the basis states given in Table~\ref{tab:basis states}.  The eigenstates giving these phases were obtained by a numerical diagonalization of the Hamiltonian in Eq.~(\ref{eq:Discrete Hamiltonian}) with the parameters given in Eq.~(\ref{eq:parameters}).  The red dots are the phase rates of change while the black dots (that are exactly covered) are the free energies of the basis states.}
\end{figure}

We further find that the slopes of each line corresponding to the same basis state are the same for all eigenstates.  That is to say, for example, the gold dashed line corresponding to the phase of the basis state $|(0$GeV$,2)\rangle$ has the same slope in all three plots of Fig.~\ref{fig:wave function phases}, and indeed for all other eigenstates.  This is true whether the phase begins at $0$ or $\pi$.  As another example, the phase of the basis state $|(-1.25$GeV$,1),(1.25$GeV$,1)\rangle$ shown in dotted green in all three plots of Fig.~\ref{fig:wave function phases} has a slope of approximately $\pi$ in all three plots whether it begins at $\pi$ as in the top and middle plot or begins at $0$ as it does in the bottom plot.  In fact, we plot the slopes of these phase rates as red dots in Fig.~\ref{fig:phase plot}.
Along with these phase slopes, we plot the free energies of the basis states in black.  The black points are exactly covered by the red dots.  As a result, we see that the phases of each of the coefficients of the basis states change in time with a rate equal to the free energy of that basis state.

It might be expected that the eigenstates would be constant in time up to an overall phase that changes according to the eigenenergy as in $\mbox{exp}(-i H t)\psi_E = \mbox{exp}(-i E t)\psi_E$.  This would be true if the Hamiltonian were independent of time.  However, our Hamiltonian, as seen in Eq.~(\ref{eq:Discrete Hamiltonian}), is not independent of time.  Its dependence on time is determined by Lorentz covariance.    Nevertheless, we find that the dependence of the eigenstates on time is not complicated.  There is a universal unitary operator that advances the eignestates from one time to another as in $\psi_E(t+\Delta t)=U(\Delta t)\psi_E(t)$.  It is given by
\begin{equation}
U(\Delta t) = \left(\begin{array}{cccc}
e^{i\omega_v\Delta t} & 0 & 0 & \cdots\\
0 & e^{i\omega_1\Delta t} & 0 & \cdots\\
0 & 0 & e^{i\omega_2\Delta t} & \cdots\\
\vdots & \vdots & \vdots & \ddots
\end{array}\right)\ ,
\end{equation}
where $\omega_i$ is the free-particle energy of the basis state.  Consequently, we find, in this basis, that the Hamiltonian at time $t$ is related to the Hamiltonian at time $t=0$ by
\begin{equation}
H(t) = U(t)H(0)U^{-1}(t)\ .
\end{equation}

\subsection{\label{sec:scattering}The S Matrix and Scattering}
One of our long-term goals is to determine whether the S matrix and the associated scattering amplitudes can be calculated from the results of numerical diagonalization of the Hamiltonian.  In the present case with only up to two free particles in the basis states, we find that there are no degeneracies.  Every eigenvalue occurs once and only once.  Therefore, as is well known, the eigenstates are orthogonal.  In particular, the S matrix is defined to be transitions of in states of definite energy to out states of definite energy (see, for example, \cite{Weinberg:1995mt}).  However, since our eigenvalues are all unique, we find that the S matrix is just a delta function.
\begin{equation}
S_{\alpha\beta} = \delta_{\alpha\beta}
\end{equation}
where $\alpha$ and $\beta$ specify the state.  Since there are no degeneracies, they uniquely determine the energy of the in and out states, $E_\alpha$ and $E_\beta$.  That is to say, the S matrix is trivial for us.  The reason is that we have truncated the Hilbert space so severely that we have removed the possibility of non-trivial scattering.  In a future work, when we have included basis states with greater numbers of free particles, the S matrix will become non trivial and scattering will be possible.

\section{\label{sec:dependence on parameters}Dependence on Parameters}
Now that we have a basic understanding of the results of the numerical diagonalization for one particular set of parameters, we would like to explore how the results depend on the parameters.  We describe the dependence on $\Delta p$ in Sec.~\ref{sec:Delta p}, $\lambda$ in Sec.~\ref{sec:lambda}, $m$ in Sec.~\ref{sec:m} and the cutoff momentum in Sec.~\ref{sec:E_cutoff}.

\subsection{\label{sec:Delta p}Dependence on $\mathbf{\Delta p}$}
The physical momentum is continuous and therefore, the spacing between momenta $\Delta p$ is not a physical parameter.  Therefore, the physical values should be independent of $\Delta p$ and the results of our calculation should converge to the physical values in the limit that $\Delta p\to 0$.  Achieving this will require a renormalization of the physical parameters $\lambda$ and $m$.  We will discuss renormalization further in Sec.~\ref{sec:renormalization}.  In order to clarify this, we begin by studying the dependence on $\Delta p$ by itself.  In this section, we continue to use the same numerical values for $m$ and $\lambda$ defined in Eq.~(\ref{eq:parameters}), but we now vary $\Delta p$.  We numerically diagonalized the Hamiltonian for four values, namely $\Delta p=$0.05GeV, 0.025GeV, 0.01GeV and 0.005GeV, covering one order of magnitude.  

We begin with a discussion of the eigenvalues.  We found that, with the exception of the vacuum, all the eigenvalues were 
essentially unaffected by the change in $\Delta p$.  In fact, we found that the change in the eigenvalues was on the order of 0.01\% or smaller for all the eigenstates other than the vacuum.  This was true both for the eigenvalues obtained by numerical diagonalization of the Hamiltonian as well as for the single-particle energy obtained directly from Eq.~(\ref{eq:1 part energy}).  Because of this, all the eigenvalues other than the vacuum can be seen in Fig.~\ref{fig:EnSpectrum2} for all values of $\Delta p$.  

\begin{figure}[!]
\begin{center}
\includegraphics[scale=0.85]{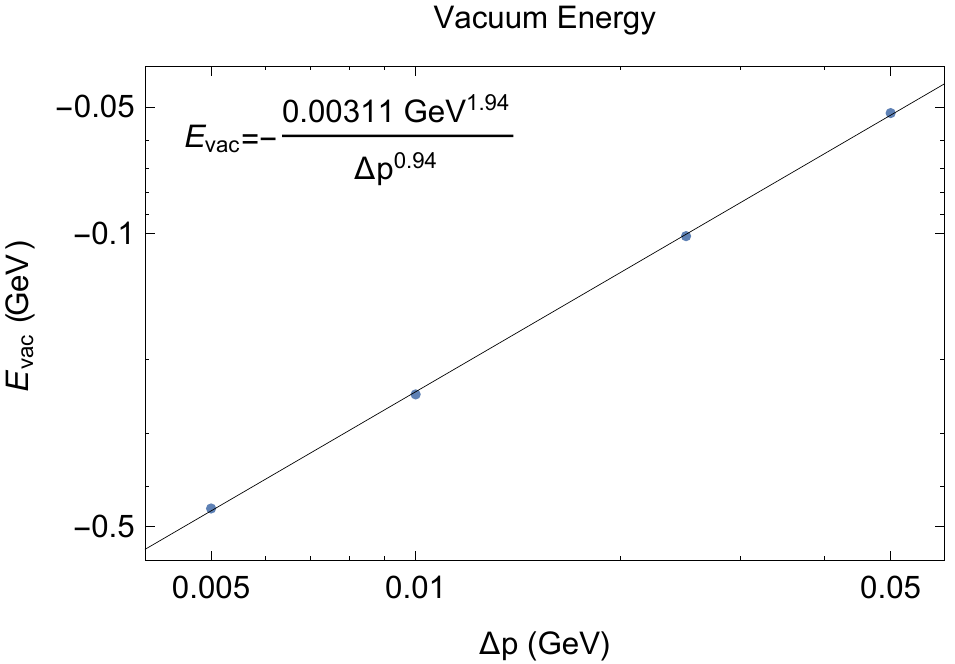}
\end{center}
\caption{\label{fig:EnVacuumDp}A plot of the energy of the vacuum as a function of $\Delta p$.  The other parameters are given in Eq.~(\ref{eq:parameters}).  The eigenvalues were obtained by a numerical diagonalization of the Hamiltonian in Eq.~(\ref{eq:Discrete Hamiltonian}) and are plotted as blue dots.  The black line is the best fit line on a log-log plot.  The inset formula is the expression for the best fit line.}
\end{figure}

The vacuum, on the other hand, was significantly affected by the change in $\Delta p$.  We plot the vacuum energies in Fig.~\ref{fig:EnVacuumDp} on a log-log scale.  
We can see that the vacuum energy satisfies a simple power law scaling with the size of $\Delta p$ given by
\begin{equation}
E_{vac}(\Delta p) = - \frac{0.00311\mbox{GeV}^{1.94}}{\Delta p^{0.94}}\ .
\label{eq:E_vac(Dp)}
\end{equation}
Although we can not derive this result fully analytically, we give a brief argument for this power-law dependence in App.~\ref{Evac 1/Delta p}.
From this result, we can see that the vacuum energy diverges towards negative infinity as $\Delta p\to0$.  This should not surprise us as quantum field theory is well known to have infinities that are required to be absorbed into other parameters by the process of renormalization.   In the present approach to calculating observables in quantum field theory, these infinities come precisely from $\Delta p\to0$, as well as from the cutoff on energy discussed in Sec.~\ref{sec:E_cutoff}.  We will discuss renormalization further in Sec.~\ref{sec:renormalization}.

The mass gap is the difference between the single-particle state energy and the vacuum energy.  Since the single-particle energy does not change, we see that the physical mass increases as $\Delta p$ decreases.  It also becomes infinite and renormalization must be performed to achieve the physically measured value.  Such an adjustment will not only affect the single-particle mass but all the other energies as well and also the wavefunctions.  We will discuss this further in Sec.~\ref{sec:renormalization}.  For now, we simply point out that the height of the energies of the excited states above the vacuum grows as $\Delta p\to0$ due to the negative growth of the vacuum energy as shown in Fig.~\ref{fig:EnVacuumDp}.

\begin{figure}[!]
\begin{center}
\includegraphics[scale=0.85]{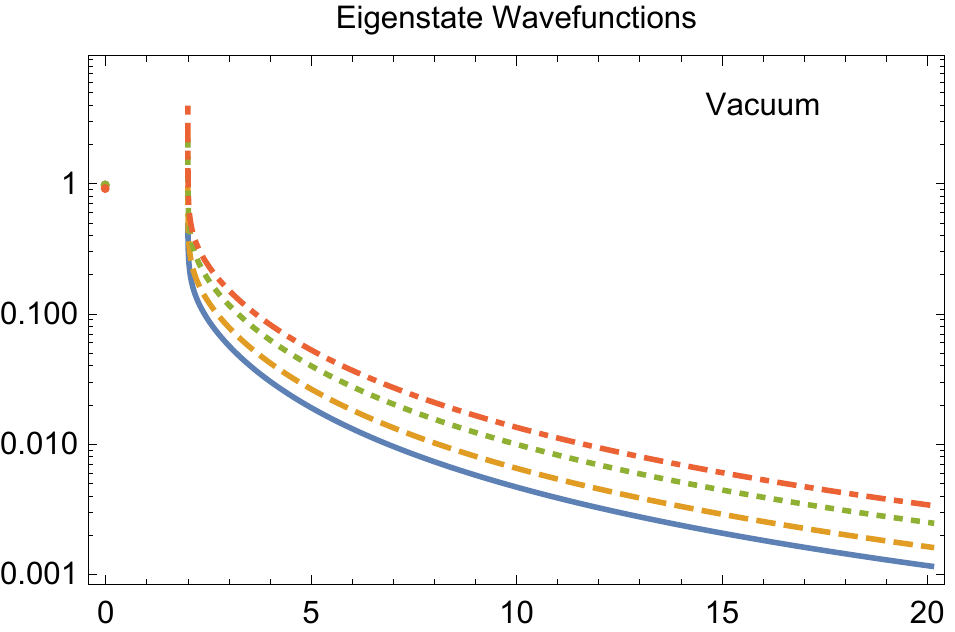}\\
\includegraphics[scale=0.85]{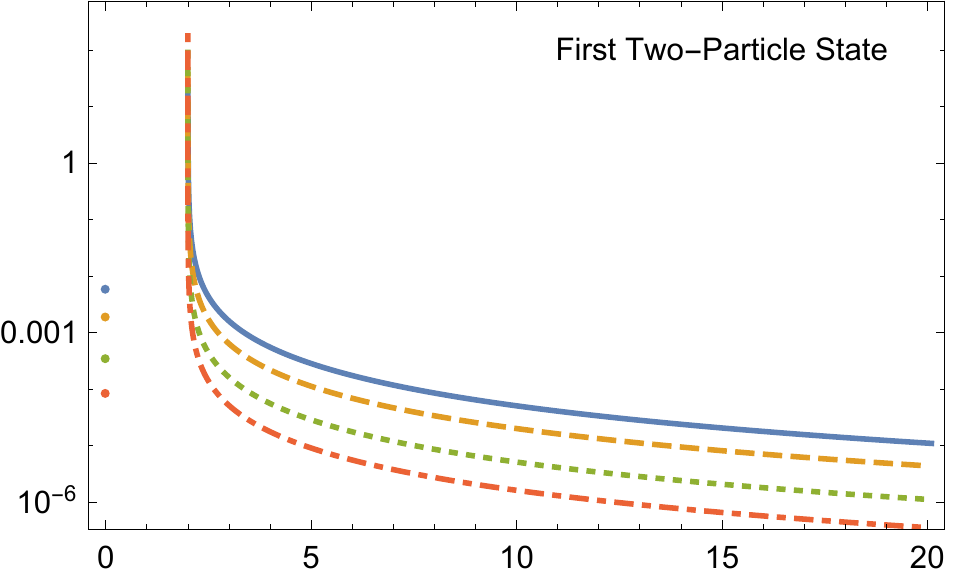}\\
\includegraphics[scale=0.85]{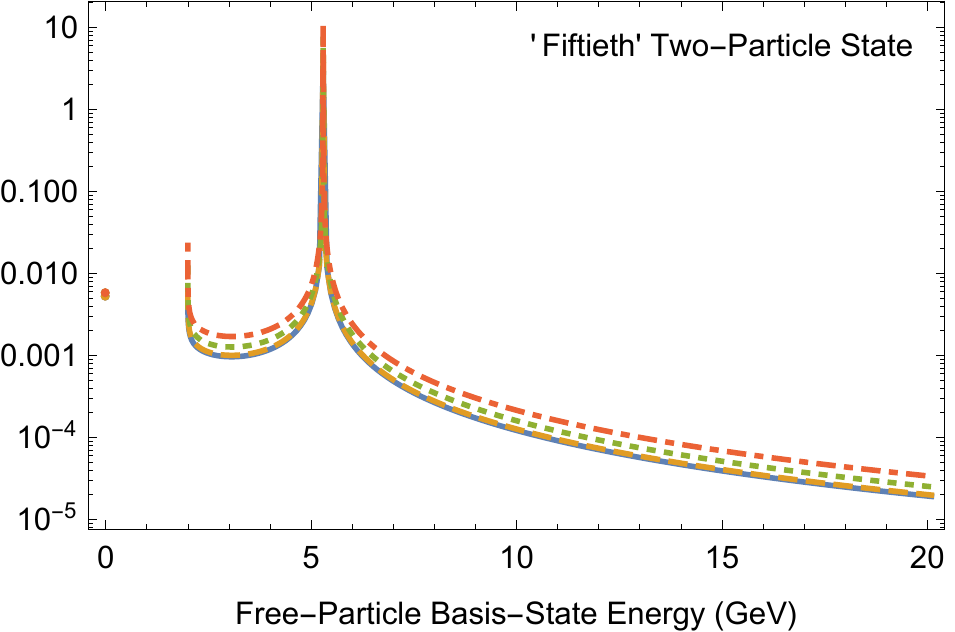}
\end{center}
\caption{\label{fig:Dp VacuumWaveFunction2}A plot of the absolute values of the coefficients of the wavefunction divided by the square root of the basis-state spacing for a few eigenstates.  These wavefunctions were obtained by a numerical diagonalization of the Hamiltonian in Eq.~(\ref{eq:Discrete Hamiltonian}) with the parameters given in Eq.~(\ref{eq:parameters}) except for $\Delta p$.  The value of $\Delta p$ for each series are as follows: (solid blue) $\Delta p=0.05$GeV, (dashed orange) $\Delta p=0.025$GeV, (dotted green) $\Delta p=0.01$GeV, and (dot-dashed red) $\Delta p=0.005$GeV.  The top plot is for the vacuum state, the middle plot is for the first excited state and the bottom plot is for the `fiftieth' excited state.  Further details can be found in Subsection~\ref{sec:Delta p}.}
\end{figure}
We have also investigated the dependence of the wavefunctions on the momentum grid spacing.  In Fig.~\ref{fig:Dp VacuumWaveFunction2}, we have plotted the same wavefunctions that we plotted in Fig.~\ref{fig:VacuumWaveFunction2}, but with different values of $\Delta p$.  We have also changed how we normalize the plot.  In Fig.~\ref{fig:VacuumWaveFunction2}, we plotted the raw coefficients of the basis states so that the discrete sum of the squares of the points adds to one.  However, in this plot, by contrast, we have normalized the continuum by dividing by the square root of the spacing between the points.  With this normalization, the integral of the square of the continuum curve plus the square of the free-vacuum coefficient equals one.  Because of this, although the continuum curves in this plot look qualitatively different than those in Fig.~\ref{fig:VacuumWaveFunction2}, the solid blue curves are actually exactly the same.  The reason  the change in normalization was necessary is that when we change the momentum grid spacing, the number of basis states changes with it.  As the number of basis states increases, each coefficient is reduced to keep the discrete normalization the same.  This multiplicity of basis states reduction is not the crucial point, however.  What we want to compare is the coefficient after removing this multiplicity of states reduction.  By dividing by the basis-state spacing, we have removed this reduction and can directly compare the coefficients for different values of $\Delta p$, as given in Fig.~\ref{fig:Dp VacuumWaveFunction2}.

A comment is in order about the bottom plot of Fig.~\ref{fig:Dp VacuumWaveFunction2}, labeled as the `fiftieth' two-particle state.  This is the wavefunction for the fiftieth two-particle state when $\Delta p=0.05$GeV.  However, it is actually the ninety-ninth two-particle state when $\Delta p=0.025$GeV, the two-hundred-fourty-sixth two-particle state when $\Delta p=0.01$GeV and the four-hundred-ninety-first two-particle state when $\Delta p=0.005$GeV.  The property that all of these eigenstates have in common is that they are the states that are dominated by the basis state $|(-2.45\mbox{GeV},1),(2.45\mbox{GeV},1)\rangle$.  On the other hand, the top plot is the vacuum for all $\Delta p$ and the middle plot is the first two-particle state for all $\Delta p$.  

\begin{figure}[!]
\begin{center}
\includegraphics[scale=0.85]{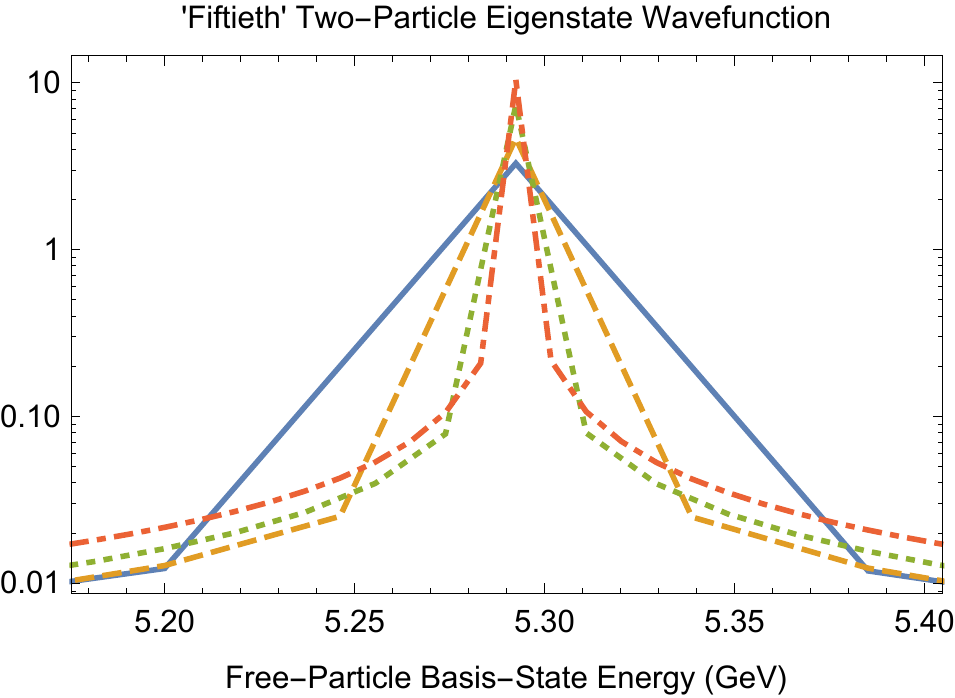}
\end{center}
\caption{\label{fig:Dp VacuumWaveFunction2Zoom}A magnification of the bottom plot of Fig.~\ref{fig:Dp VacuumWaveFunction2}.}
\end{figure}

The dots and curves correspond with the following values of $\Delta p$:  (solid blue) $\Delta p=0.05$GeV, (dashed orange) $\Delta p=0.025$GeV, (dotted green) $\Delta p=0.01$GeV and (dot-dashed red) $\Delta p=0.005$GeV.  In general, all the eigenstates have the feature that the peak in the continuum becomes narrower and taller with decreasing $\Delta p$ (see Fig.~\ref{fig:Dp VacuumWaveFunction2Zoom} for example).  In all three cases, the dot-dashed red peak (for $\Delta p=0.005$GeV) is the narrowest and tallest peak.  For the bottom plot, that is true for both the peak at $2$GeV and at just above $5$GeV and is a general feature of the eigenstates.  

There are two properties that transition as we move from lower eigenstates to higher ones.  The first is the contribution from the continuum outside of the peak region.  Beginning with the first two-particle state and continuing through the eigenstate with an energy of approximately 2.5GeV, the contribution outside the peak decreases as $\Delta p$ decreases.  However, as we move towards higher eigenstates, this difference is reduced until it vanishes and is reversed.  By the time the `fiftieth' eigenstate is reached, for example, we can see that the contribution from the continuum increases outside the peak as $\Delta p$ decreases.  This pattern is also followed by the continuum in the case of the vacuum.  The second property to transition is the discrete contribution of the free vacuum.  For the lowest eigenstates, including the vacuum and first two-particle eigenstate shown as the top two plots, the contribution of the free vacuum decreases with decreasing $\Delta p$.  The decrease in the free vacuum is very slight, however, and it still dominates the vacuum eigenstate for all values of $\Delta p$ as can be seen in the top plot where all the dots are practically on top of each other.  For concreteness, we list the free-vacuum contribution for the vacuum.  Its coefficient is 0.99, 0.98, 0.96 and 0.94, respectively for decreasing values of $\Delta p$.  Again, a transition occurs around the eigenstate with a peak at approximately 2.5GeV.  For lower eigenstates, the trend is for the the free-vacuum to make a smaller contribution for smaller $\Delta p$.  However, for higher eigenstates, the trend is for the free-vacuum to make larger contributions for smaller $\Delta p$.  

This transition may at first seem strange since it appears at the surface that the entire wavefunction is increasing for the higher eigenstates which would seem to violate normalization.  However, we note again that the peak gets taller and narrower as $\Delta p$ decreases.  In Fig.~\ref{fig:Dp VacuumWaveFunction2Zoom}, we have plotted a magnified view of the peak from the bottom plot of Fig.~\ref{fig:Dp VacuumWaveFunction2}, where we can see this behavior for the `fiftieth' two-particle state.  As this occurs, there is a competition between the area under the peak decreasing due to a narrower peak and the area increasing due to a taller peak, and this affects the rest of the wavefunction.  For the vacuum, the peak is discrete.  Since it decreases slightly, the continuum must increase to make up for it.  For the excited states on the other hand, it depends on how close the peak is to the edge of the continuum.  If it is very close, if the peak is to the left of approximately 2.5GeV, then the area under the peak is larger and the continuum and free vacuum go down.  If, on the other hand, it is far from the edge, if the peak is to the right of approximately 2.5GeV, then the area under the peak is smaller and the continuum and free-vacuum go up.  For concreteness, the free-vacuum contributions to the `fiftieth' two-particle state are 0.00541, 0.00546, 0.00561 and 0.00589, respectively for decreasing values of $\Delta p$.

\subsection{\label{sec:lambda}Dependence on $\mathbf{\lambda}$}
Both the eigenstates and their energies depend on the coupling $\lambda$.  If $\lambda\to0$, the theory becomes absolutely free and the eigenvalues match the free energies exactly and the wavefunctions become delta functions.  We have checked this limit with our numerical code and get agreement.  On the other hand, we would also like to know what happens as we vary $\lambda$ over a broad range of values.  Although, we usually prefer to keep $\lambda$ positive to maintain the stability of the vacuum, we will consider both positive and negative values of $\lambda$ in this subsection.  Although negative values of $\lambda$ may indeed destabilize the vacuum in the full theory, in our severely truncated Hilbert space, we find that the vacuum remains stable for $\lambda$ between the $-2$GeV$^2$ and $2$GeV$^2$ that we tested.

\begin{figure}[!]
\begin{center}
\includegraphics[scale=0.85]{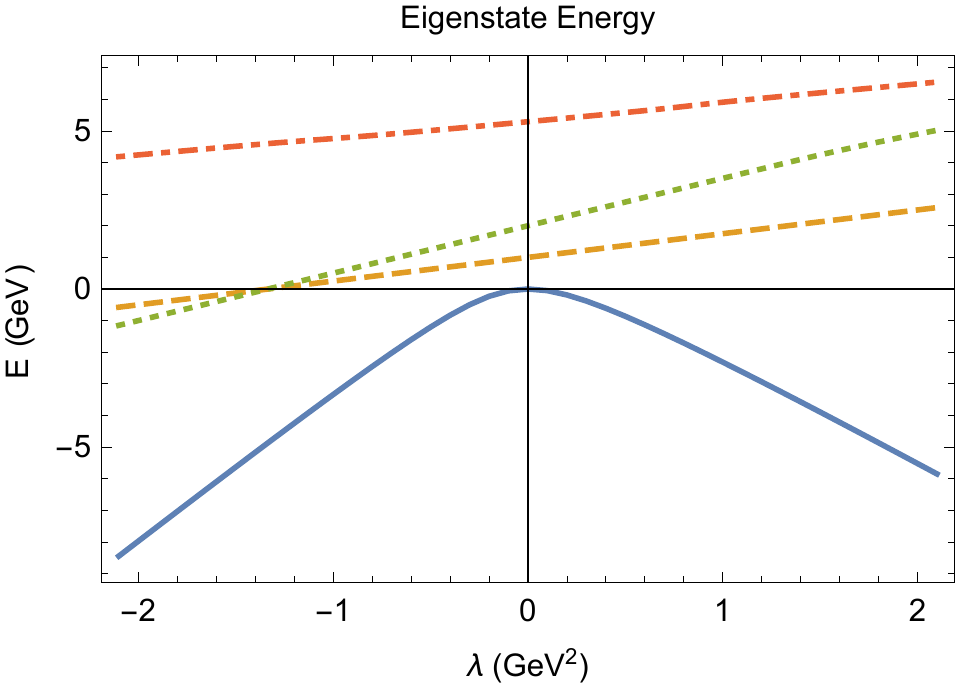}\\
\includegraphics[scale=0.85]{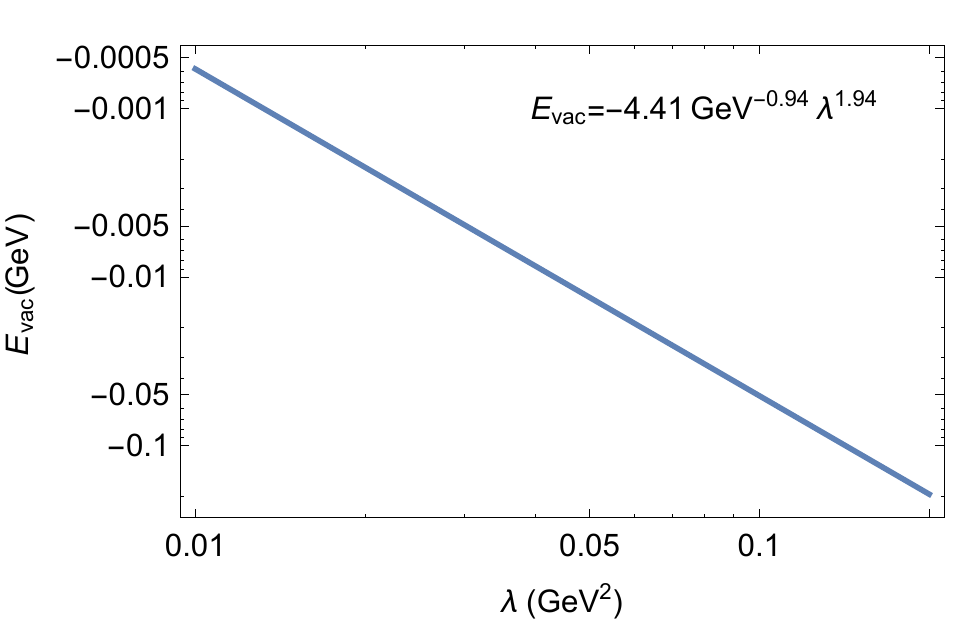}
\end{center}
\caption{\label{fig:EnFewDlambda}A plot of the energy eigenvalues as a function of $\lambda$.  In the top plot, the vacuum is in solid blue, the single-particle state is in dashed orange, the first two-particle state is in dotted green and the fiftieth two-particle state is in dot-dashed red.  The bottom plot is a magnification of the vacuum energy on a log-log plot.  The inset formula is the best fit curve for the vacuum energy.  The other parameters are given in Eq.~(\ref{eq:parameters}).  The eigenvalues were obtained by a numerical diagonalization of the Hamiltonian in Eq.~(\ref{eq:Discrete Hamiltonian}).}
\end{figure}
We begin by plotting the dependence of the energies over a broad range of $\lambda$ for a few eigenstates in the top plot of Fig.~\ref{fig:EnFewDlambda}.  In solid blue, at the bottom, we see that the vacuum energy peaks at 0GeV when $\lambda=0$ and drops towards negative values for both positive and negative $\lambda$.  Its shape appears to be quadratic near $\lambda=0$ and nearly linear for larger absolute values of $\lambda$.  The other eigenvalues, on the other hand, all appear to have a linear dependence on $\lambda$ with a positive slope, increasing towards larger values of $\lambda$.  This is not surprising since the non-free part of the Hamiltonian depends linearly on $\lambda$ [see Eqs.~(\ref{eq:1 part energy}) through (\ref{eq:<-p,p|H|-q,q>})].  The single-particle state energy is plotted in dashed orange and is directly above the vacuum when $\lambda\gtrsim-1.3$GeV$^2$.  The lowest two-particle state is plotted in dotted green and is directly above the one-particle state for $\lambda\gtrsim-1.3$GeV$^2$.  It has a greater slope than the one-particle state eigenvalue and so crosses the one-particle eigenvalue at $\lambda\sim-1.3$GeV$^2$ at an energy of 0GeV (the energy of the lowest two-particle state is always twice that of the single-particle state).  This greater slope is also not surprising since the coefficient of $\lambda$ for $\langle(0$GeV$,2)|H|(0$GeV$,2)\rangle$ in Eq.~(\ref{eq:<0,2|H|0,2>}) is a little more than double the coefficient of $\lambda$ for $\langle(0$GeV$,1)|H|(0$GeV$,1)\rangle$ in Eq.~(\ref{eq:1 part energy}).   This comes from the fact that the basis state in $\langle(0$GeV$,2)|H|(0$GeV$,2)\rangle$ has twice as many particles and, therefore, twice the free energy.  To the left of $\sim-1.3$GeV$^2$, the eigenvalues of the one-particle state and the lowest two-particle state are inverted and the lowest two-particle state becomes lower in energy than the single-particle state.  Of all the eigenvalues dependence on $\lambda$, the lowest two-particle state energy has the greatest slope.  As we go to higher two-particle states, they begin with a greater energy at $\lambda=0$ and have slopes that are slightly smaller than the state below them.  Other two-particle state energies cross below the single-particle state energy for more negative values of $\lambda<-1.3$GeV$^2$.  The dot-dashed red line is the energy of the fiftieth two-particle state.

We do not know for certain whether having two-particle states pass below the single-particle state in energy is a sickness of the theory.  However, it certainly seems strange.  Moreover, since it occurs for large negative values of $\lambda$, where we do not usually consider the theory stable, we will henceforth consider only $\lambda>-1.3$GeV$^2$.  Furthermore, since the slopes of the two-particle states become increasingly shallower as the energy increases, the two-particle state energies will begin crossing each other at large positive values of $\lambda$.  For this reason, we will focus on $|\lambda|\lesssim$1GeV$^2$ for the remainder of this subsection.  

\begin{figure*}[!]
\begin{center}
\textbf{Eigenstate Wavefunctions}\\\vspace{0.1in}
\includegraphics[scale=0.85]{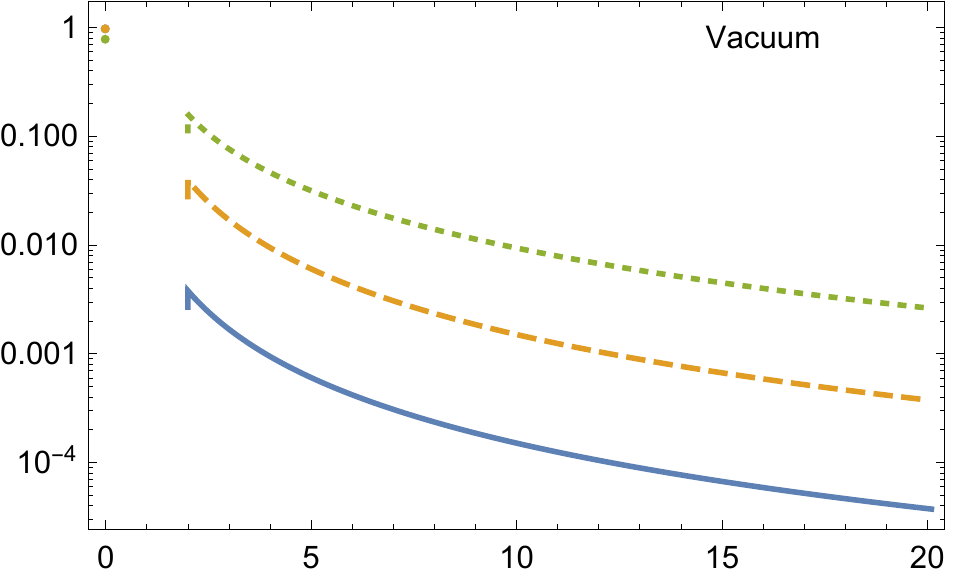}\
\includegraphics[scale=0.85]{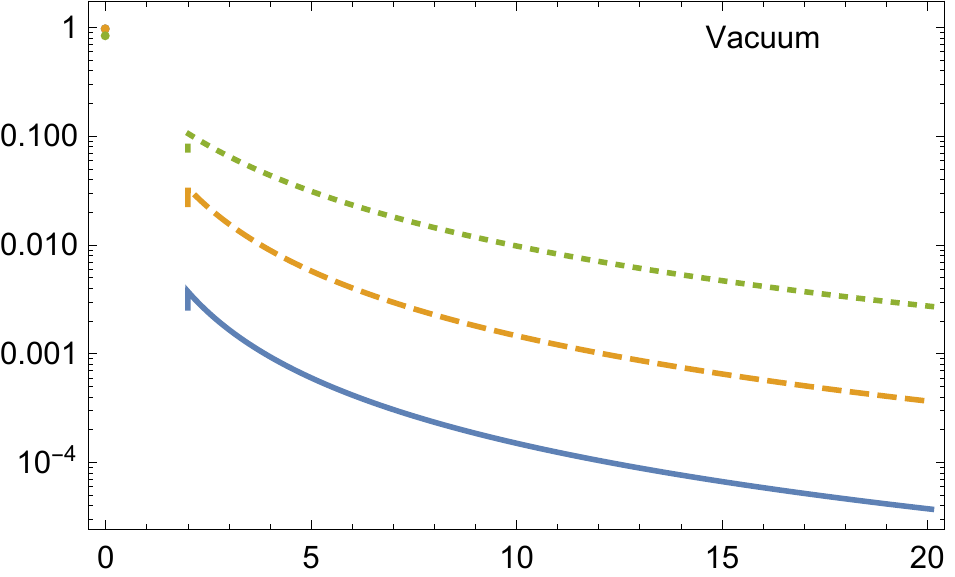}\\
\includegraphics[scale=0.85]{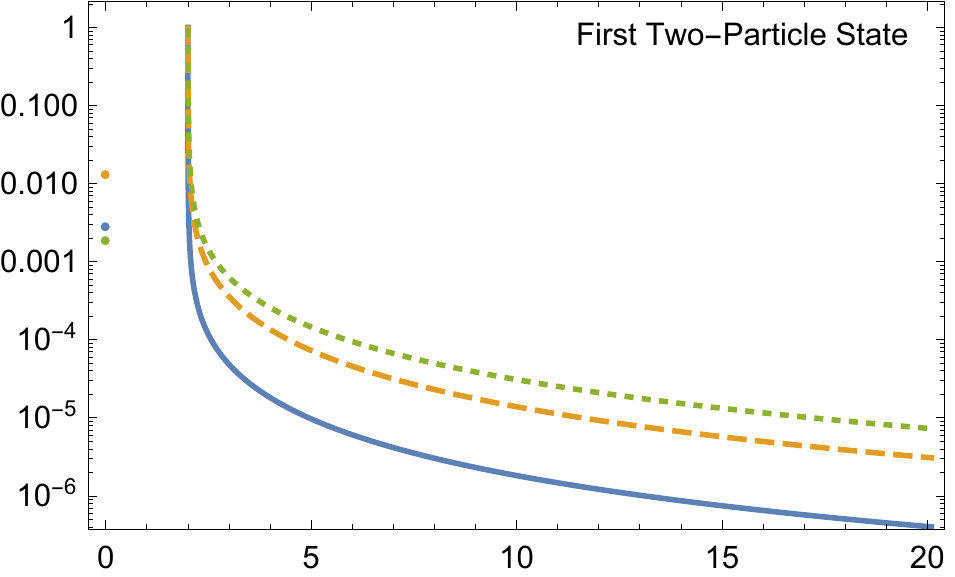}\
\includegraphics[scale=0.85]{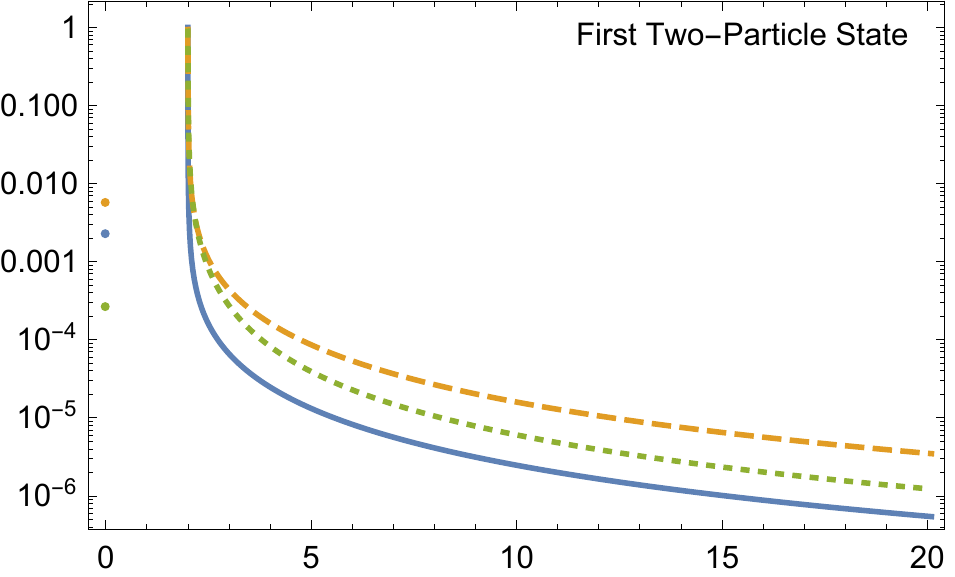}\\
\includegraphics[scale=0.85]{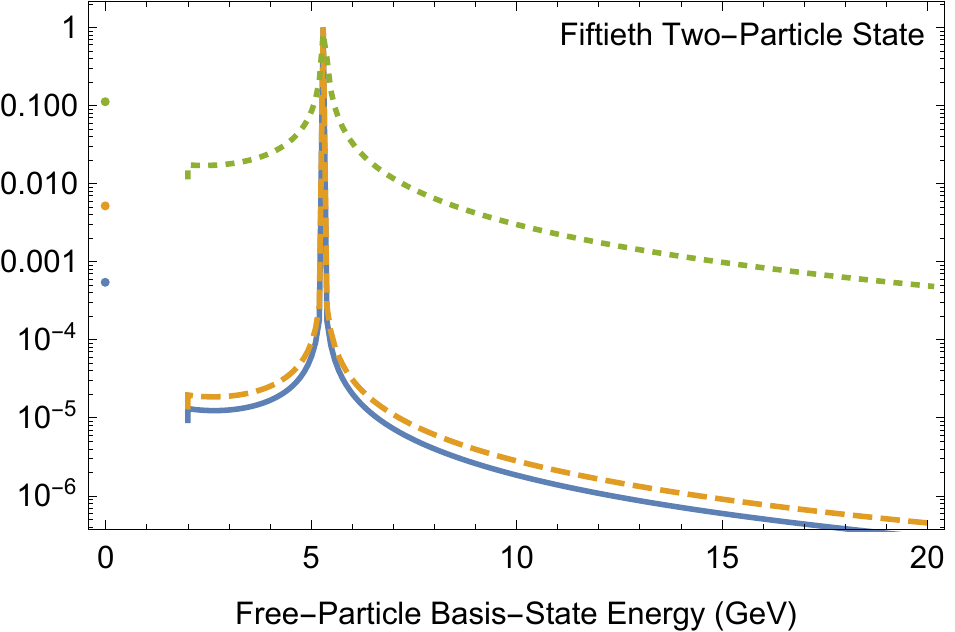}\
\includegraphics[scale=0.85]{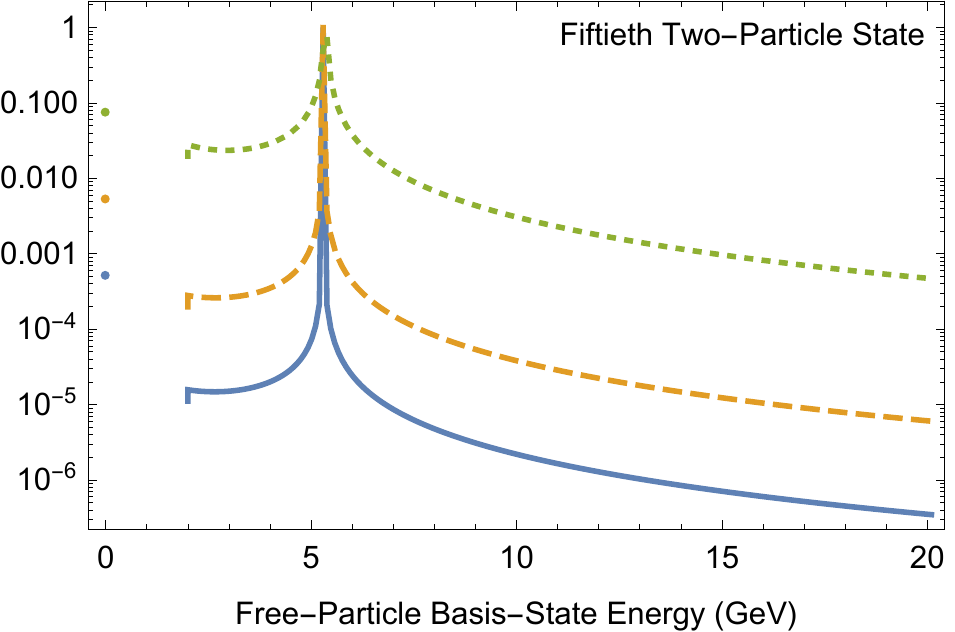}
\end{center}
\caption{\label{fig:Dlambda VacuumWaveFunction2}A plot of the absolute values of the coefficients of the wavefunction for a few eigenstates.  These wavefunctions were obtained by a numerical diagonalization of the Hamiltonian in Eq.~(\ref{eq:Discrete Hamiltonian}) with the parameters given in Eq.~(\ref{eq:parameters}) except for $\lambda$.  The value of $\lambda$ for each series are as follows: (solid blue) $\lambda=\pm0.01$GeV$^2$, (dashed orange) $\lambda=\pm0.1$GeV$^2$ and (dotted green) $\lambda=\pm1$GeV$^2$.  The left column is for negative $\lambda$ while the right column is for positive $\lambda$.  The top row is for the vacuum state, the middle row is for the first two-particle state and the bottom row is for the fiftieth two-particle state.  Further details can be found in Subsection~\ref{sec:lambda}.}
\end{figure*}

In the bottom plot of Fig.~\ref{fig:EnFewDlambda}, we show a magnification of the vacuum energy on a log-log plot where we see that it follows a simple power-law scaling.  We have found the best fit line for the vacuum energy.  It is
\begin{equation}
E_{vac}(\lambda)=-4.41\mbox{GeV}^{-0.94}\lambda^{1.94}\ .
\label{eq:E_vac(lambda)}
\end{equation}
App.~\ref{Evac 1/Delta p} gives a brief argument for this dependence on $\lambda$.
As expected, we find that the vacuum energy goes to zero as $\lambda$ does.  We also find that the dependence on $\lambda$ is quadratic as we observed from the top plot.  We find it interesting that the power dependence on $\lambda$ is slightly less than 2 in exactly the same amount (0.06) as the power dependence on $\Delta p$ is different than -1 [see Eq.~(\ref{eq:E_vac(Dp)})].

We next turn to the wavefunctions of the eigenstates.  We plot the absolute values of the coefficients of the basis states for each wavefunction in Fig.~\ref{fig:Dlambda VacuumWaveFunction2}.  We have plotted the wavefunctions for both negative (left column) and positive (right column) values of $\lambda$.    The curves are as follows: (solid blue) $\lambda=\pm0.01$GeV$^2$, (dashed orange) $\lambda=\pm0.1$GeV$^2$, and (dotted green) $\lambda=\pm1$GeV$^2$.  We have plotted the coefficients of the basis states directly and have not divided by the square root of the energy spacing between states.  So, for example, the dashed orange curves on the right correspond directly with those of Fig.~\ref{fig:VacuumWaveFunction2}.  

The first thing that we notice is that the wavefunction is most like a delta function for the smallest absolute value of $\lambda$, namely the solid blue curve with $\lambda=\pm0.01$GeV$^2$.  For larger absolute values of $\lambda$, the peak broadens and other basis states become more important.  This makes sense as mentioned previously.  When the coupling is $\lambda=0$, the theory is absolutely free and the wavefunction is an exact delta function.  When we turn the coupling on, the basis states surrounding the peak of the delta function immediately begin to contribute.  This contribution goes to zero in the limit of zero coupling and increases with increasing coupling.  If we focus on the vacuum (the top row of Fig.~\ref{fig:Dlambda VacuumWaveFunction2}), we see this process continuing without much subtlety for all the scanned values of $\lambda$.  As $\lambda$ becomes larger in absolute value, the peak's height diminishes and the contribution from the other basis states increases.

If we next focus on the first two-particle state in the middle row of Fig.~\ref{fig:Dlambda VacuumWaveFunction2}, we see this same trend on the left for negative values of $\lambda$.  However, for positive values of $\lambda$, we see this trend occur at first, but then as the value of $\lambda$ grows large, the trend begins to reverse somewhat.  Although the wavefunction with the smallest $\lambda$ (the solid blue curve) is still the most delta-function like, the dotted green curve of $\lambda=0.1$GeV$^2$ is more delta-function like than the dashed orange curve of $\lambda=1$GeV$^2$.  This is because a transition is in the process of occurring where the first two-particle state wavefunction becomes peaked at a basis state higher than $|(0$GeV$,2)\rangle$.  We will see this happen for the fiftieth two-particle state within our range of $\lambda$ next.

\begin{figure}[!]
\begin{center}
\includegraphics[scale=0.89]{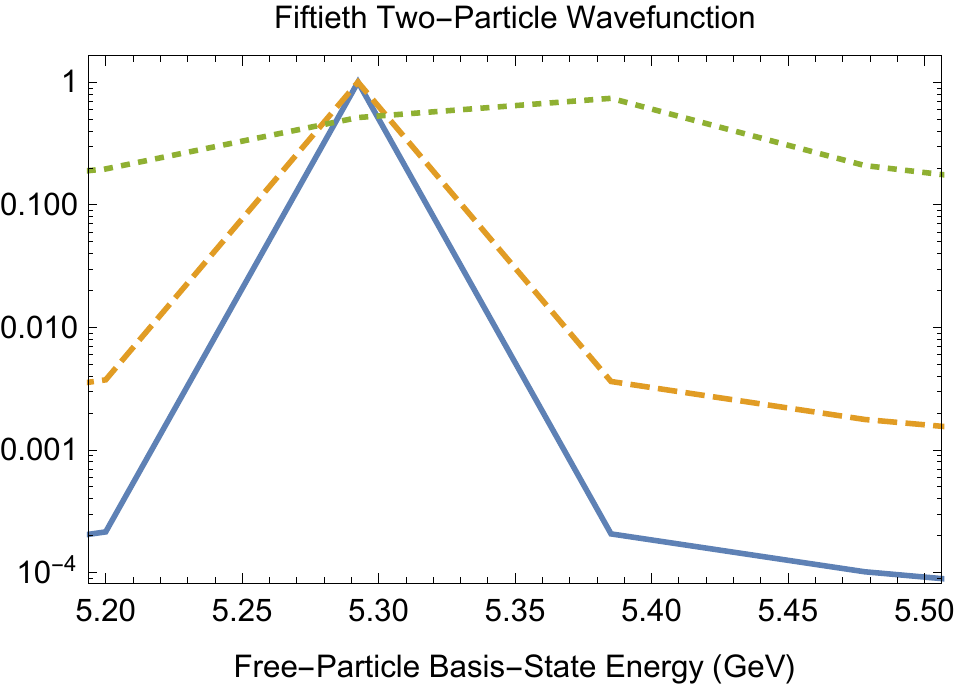}\\
\vspace{0.1in}
\includegraphics[scale=0.85]{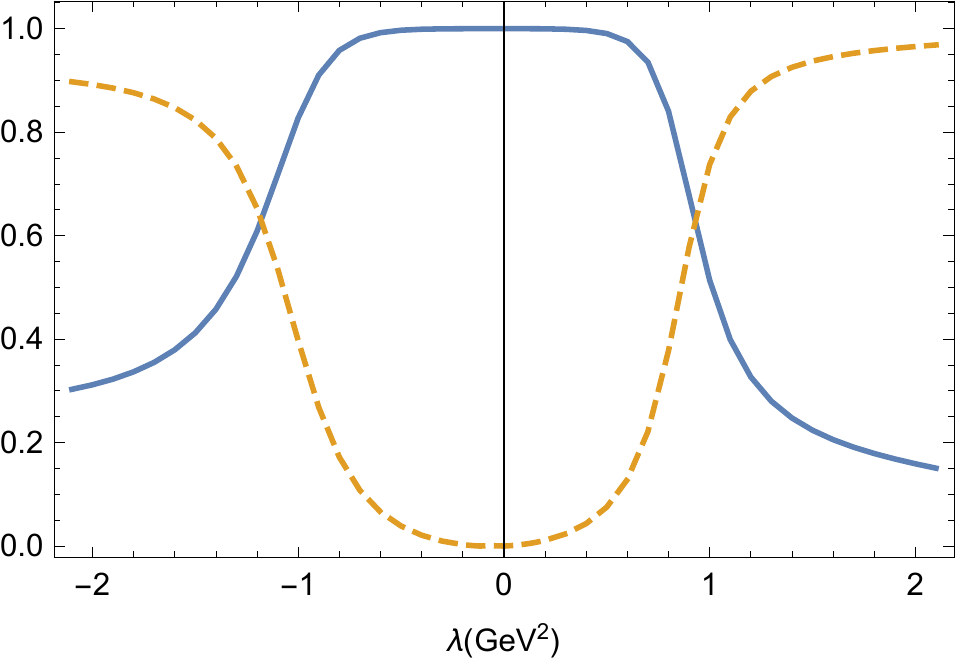}
\end{center}
\caption{\label{fig:Dlambda VacuumWaveFunction2 Cross Over}On the top is a magnification of the peak region of the bottom right plot of Fig.~\ref{fig:Dlambda VacuumWaveFunction2}.  The bottom plot is of the coefficient of the fiftieth (solid blue) and fifty-first (dashed orange) basis states as a function of $\lambda$.  These wavefunctions were obtained by a numerical diagonalization of the Hamiltonian in Eq.~(\ref{eq:Discrete Hamiltonian}) with the parameters given in Eq.~(\ref{eq:parameters}) except for $\lambda$.  Further details can be found in Subsection~\ref{sec:lambda}.}
\end{figure}

Focusing now on the bottom row of Fig.~\ref{fig:Dlambda VacuumWaveFunction2} with the fiftieth two-particle state wave function, we see that the same general trend occurs, but with a new twist.  On the right side, for positive $\lambda$, we see that the dotted green curve ($\lambda=1$GeV$^2$) has a slightly different peak position.  The new peak is at the fifty-first two-particle basis state rather than the fiftieth, although the eigenvalue is still the fiftieth two-particle state energy.  To make this transition more clear, we have plotted a magnification of the peaks at the top of Fig.~\ref{fig:Dlambda VacuumWaveFunction2 Cross Over}.
We can see that for small values of $\lambda$, the fiftieth two-particle eigenstate has a peak at the fiftieth two-particle basis-state, namely $|(-2.45$GeV$,1),(2.45$GeV$,1)\rangle$.  However, as $\lambda$ becomes large, the peak shifts over to the fifty-first two-particle basis-state, namely $|(-2.50$GeV$,1),(2.50$GeV$,1)\rangle$.  This transition is gradual as we can see in the bottom plot of Fig.~\ref{fig:Dlambda VacuumWaveFunction2 Cross Over}, where we have plotted the coefficient of $|(-2.45$GeV$,1),(2.45$GeV$,1)\rangle$ in solid blue and the coefficient of $|(-2.50$GeV$,1),(2.50$GeV$,1)\rangle$ in dashed orange.  For small values of $\lambda$ the solid blue curve is large and the dashed orange curve is small.  However, around $\lambda\sim\pm1$GeV$^2$, a transition occurs where the solid blue curve and dashed orange curves cross.  For larger absolute values of $\lambda$, the curves are inverted with the dashed orange curve large and the solid blue curve small.

\subsection{\label{sec:m}Dependence on $\mathbf{m}$}
\begin{figure}[!]
\begin{center}
\includegraphics[scale=0.85]{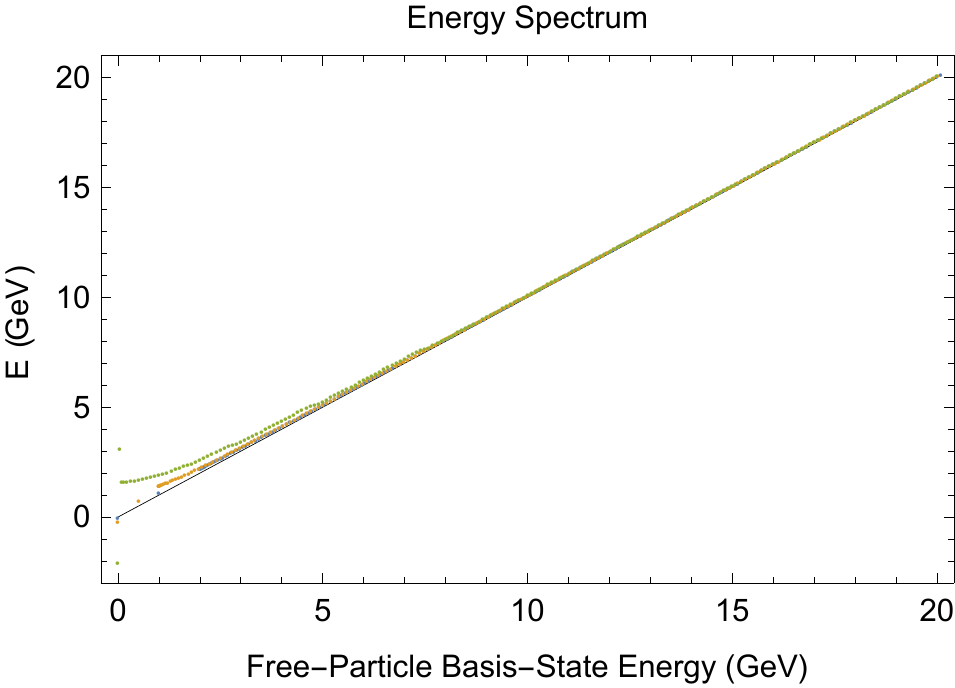}\\
\vspace{0.1in}
\includegraphics[scale=0.85]{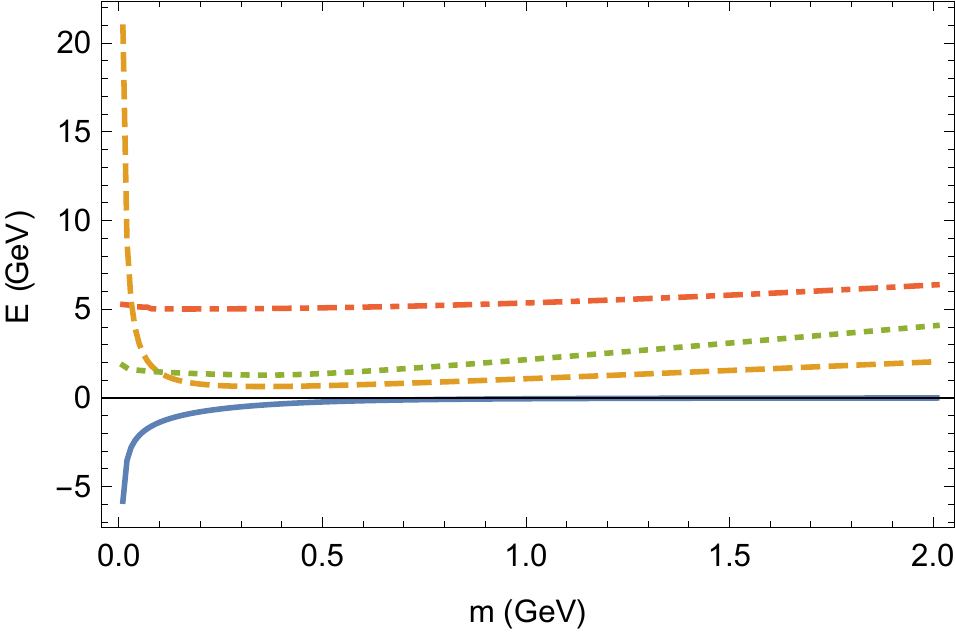}
\end{center}
\caption{\label{fig:Dmass energy spectrum}The top frame is a plot of the eigenstate energies for three values of $m$.  The black line gives the diagonal line of the free energies of the basis states while the colored dots give the eigenstate energies.   The value of $m$ for each series are as follows.  The green dots are for $m=0.05$GeV, the orange dots are for $m=0.5$GeV and the blue dots are for $m=1$GeV.  The other parameters are given in Eq.~(\ref{eq:parameters}).  The eigenstates were obtained by a numerical diagonalization of the Hamiltonian in Eq.~(\ref{eq:Discrete Hamiltonian}).  The bottom plot is the energy of a few eigenstates as a function of $m$.  The solid blue line is the vacuum energy, the dashed orange line is the single-particle state, the dotted green line is the first two-particle state and the dot-dashed red line is the fiftieth two-particle state.}
\end{figure}

We now turn to the dependence of the eigenstates on the mass parameter.  Although an analysis of $m^2<0$ would be interesting for its role in spontaneous symmetry breaking, we will here focus on $m^2>0$.  We have scanned over a range of masses from $m=0.02$GeV up to $m=2.00$GeV in steps of $0.01$GeV.  At the lower end, the mass is below both the other dimensionful parameters, the momentum spacing of $\Delta p=0.05$GeV and the coupling of $\lambda=0.1$GeV$^2$, while at the higher end, the mass is greater than both the momentum spacing and the coupling.  We begin by plotting the energy eigenvalues for a few representative masses in the top plot of Fig.~\ref{fig:Dmass energy spectrum}.   The solid black line represents the free-particle energies of the basis states for comparison.  The dots correspond with the eigenvalues.  The blue eigenvalues for $m=1$GeV are completely covered by the orange eigenvalues for $m=0.5$GeV, which are partially covered by the green eigenvalues for $m=0.05$GeV.  
%In general, we can see that the trend is for the lowest eigenvalues to move down and to the left.  This makes sense because the mass parameter is decreasing resulting in a lower basis-state energy and a lower eigenvalue.  We can also see that the lower eigenvalues become a little more separated from the free-particle energies of the basis state as $m$ is lowered.  

\begin{figure}[!]
\begin{center}
\includegraphics[scale=0.85]{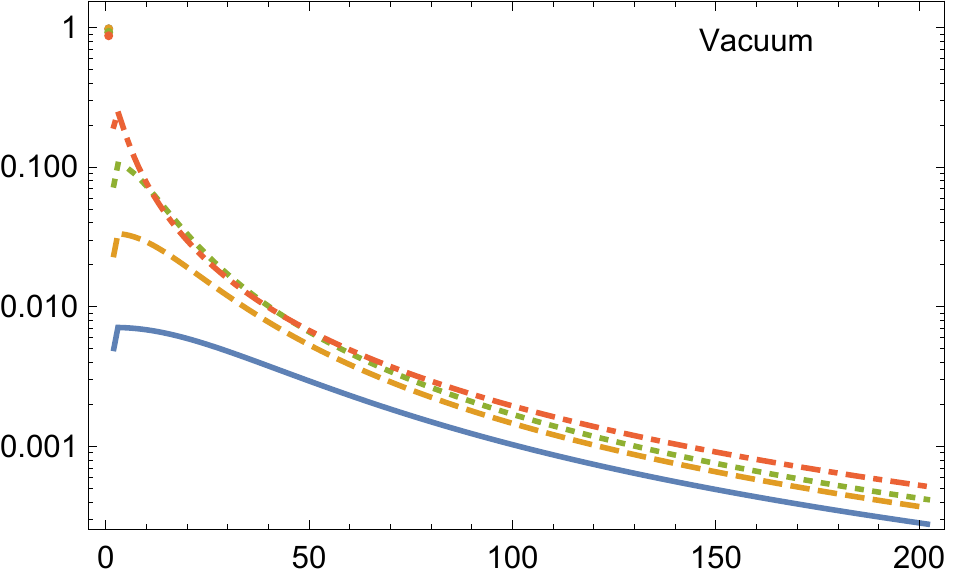}\\
\includegraphics[scale=0.85]{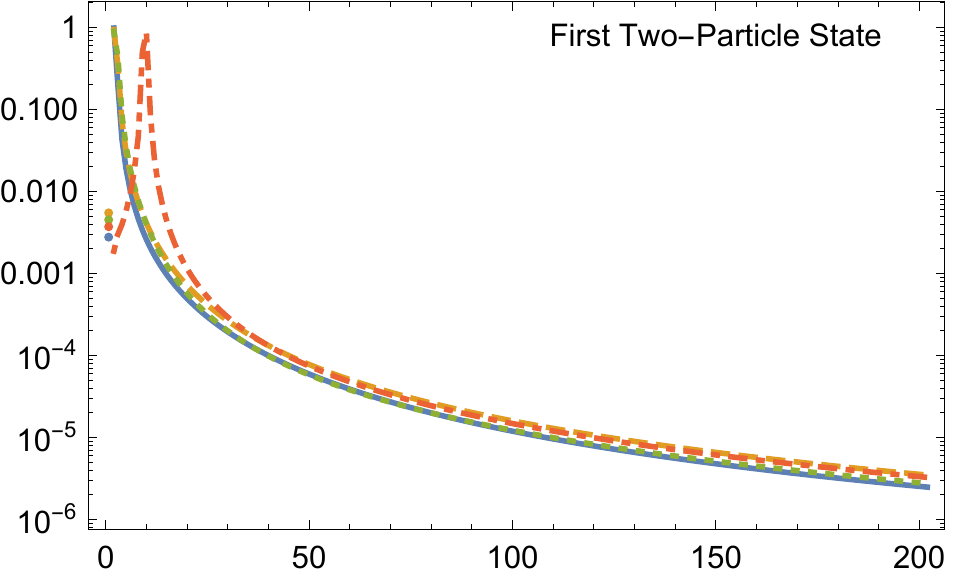}\\
\includegraphics[scale=0.85]{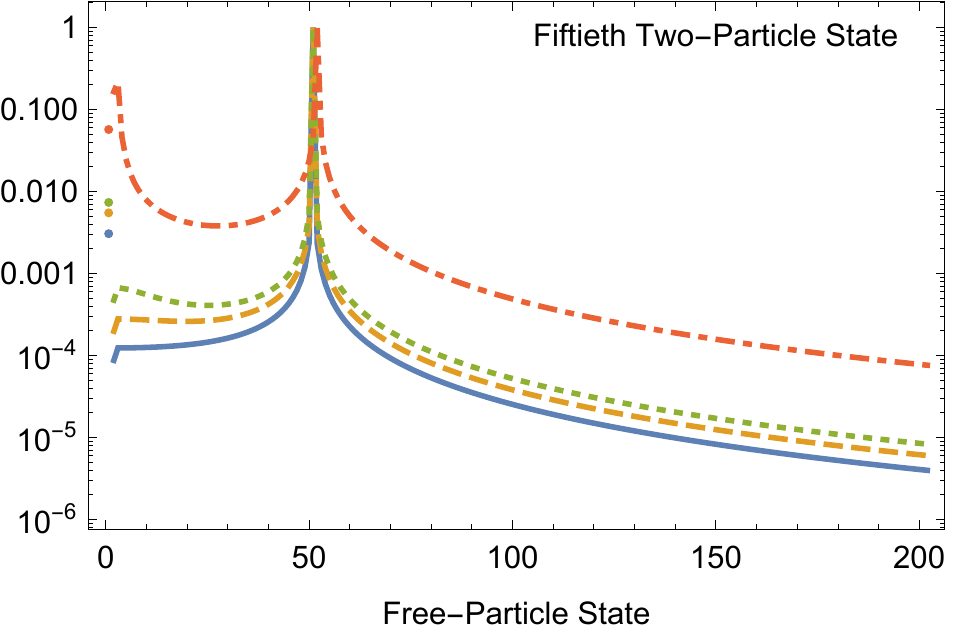}
\end{center}
\caption{\label{fig:Dmass VacuumWaveFunction2}A plot of the absolute values of the coefficients of the wavefunction for a few eigenstates.  These wavefunctions were obtained by a numerical diagonalization of the Hamiltonian in Eq.~(\ref{eq:Discrete Hamiltonian}) with the parameters given in Eq.~(\ref{eq:parameters}) except for $m$.  The value of $m$ for each series are as follows from bottom to top: (solid blue) $m=2$GeV, (dashed orange) $m=1$GeV, (dotted green) $m=0.5$GeV, and (dot-dashed red) $m=0.05$GeV.  The top plot is for the vacuum state, the middle plot is for the first two-particle state and the bottom plot is for the fiftieth two-particle state.  The horizontal axis is different than in the other wavefunction plots and corresponds with the left-most column of Table~\ref{tab:basis states} for $m=1$GeV.  Further details can be found in Subsection~\ref{sec:m}.}
\end{figure}

We find that when $m\gg\sqrt{\lambda}$, the energy spectrum asymptotically approaches the free-particle energy of the basis states.  That is, it asymptotically approaches the diagonal black line.  This is because, in this limit, the coupling is asymptotically vanishing relative to the mass.  Furthermore, in this limit, we find that the only change to the spectrum, as the mass is increased, is a shift up and to the right by $2\Delta m$.  This makes sense since the eigenvalues are nearly identical with the free-energies of the basis states and $E_{free}=2\sqrt{p^2+m^2}$.  On the other hand, when the mass is small $m\lesssim\sqrt{\lambda}$, the eigenvalues begin to diverge from the free energies of the basis states, especially at the lower end.  Furthermore, we see that when the mass parameter drops below the momentum spacing $m\lesssim2\Delta p$, the energy of the single-particle state and the vacuum begin to diverge.  In order to clarify this behavior, we have plotted these energy eigenvalues as a function of $m$ in the bottom plot of Fig.~\ref{fig:Dmass VacuumWaveFunction2}.  This behavior is an artificial feature of the discretization of the momentum space.  Strictly speaking, in Nature, the momentum spacing is infinitesimal, if not strictly zero.  Therefore, we see that it is important to set the momentum spacing to be well below the free-particle mass in order to achieve reliable results.

We next turn to the mass dependence of the wavefunctions.  As in previous subsections, we plot the vacuum, first two-particle state and fiftieth two-particle state wavefunctions in Fig.~\ref{fig:Dmass VacuumWaveFunction2}.  However, we now use a different horizontal axis.  Instead of the free-particle energies of the basis states, we simply use the ordinal number of the basis states.  That is, $|\rangle$ is $0$, $|(0,2)\rangle$ is $1$, $|(-0.05$GeV$,1),(0.05$GeV$,1)\rangle$ is 2, and so on.  These ordinal numbers are listed in Table~\ref{tab:basis states} for $m=1$GeV.  The reason we have done this is to emphasize the essential aspects of their similarities and differences as we vary the mass rather than the shift in the free-particle energies of the basis states on the horizontal axis.  The masses for the lines are as follows, starting with the bottom curve.  The solid blue lines are for $m=2$GeV, the dashed orange lines are for $m=1$GeV, the dotted green lines are for $m=0.5$GeV, and the dot-dashed red lines are for $m=0.05$GeV.  The dashed orange lines ($m=1.00$GeV) correspond with the curves in Fig.~\ref{fig:VacuumWaveFunction2}, the solid blue curves in Fig.~\ref{fig:Dp VacuumWaveFunction2}, and the dashed orange lines on the right in Fig.~\ref{fig:Dlambda VacuumWaveFunction2}.

We see that as the mass increases, the height of the peak increases and the contribution from the other states decreases.  This is because, as the mass increases, the relative size of the coupling diminishes leading the theory closer to the free theory.  As this occurs, the wavefunction is approaching the delta function of the free theory.  On the other hand, as the mass decreases, this trend is reversed as long as the mass is still larger than the square root of the coupling and the momentum spacing.  However, when $m\lesssim2\Delta p$, the curves begin to change qualitatively.  The peak begins to shift for all but the vacuum.  In the case of the first two-particle state (the middle plot of Fig.~\ref{fig:Dmass VacuumWaveFunction2}), the peak has already shifted to the tenth two-particle basis state by the time $m=0.05$GeV.  In the case of the fiftieth two-particle state, the peak has shifted up one basis state higher.  Moreover, the contribution of the first two-particle basis state has significantly increased when $m\sim0.05$GeV.  As in the discussion of the energy eigenvalues, the theory is no longer sensible when the mass drops near or below the momentum spacing.  We see again that for sensible results, we must keep the mass well above $\Delta p$.

\subsection{\label{sec:E_cutoff}Dependence on the Momentum Cutoff}
\begin{figure}[!]
\begin{center}
\includegraphics[scale=0.85]{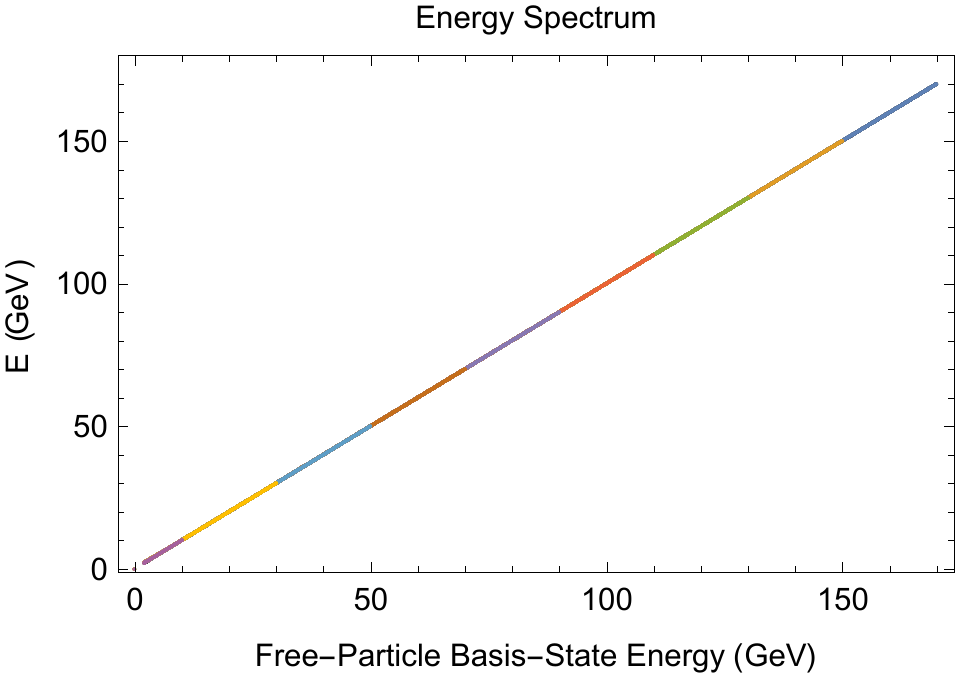}\\
\vspace{0.1in}
\includegraphics[scale=0.85]{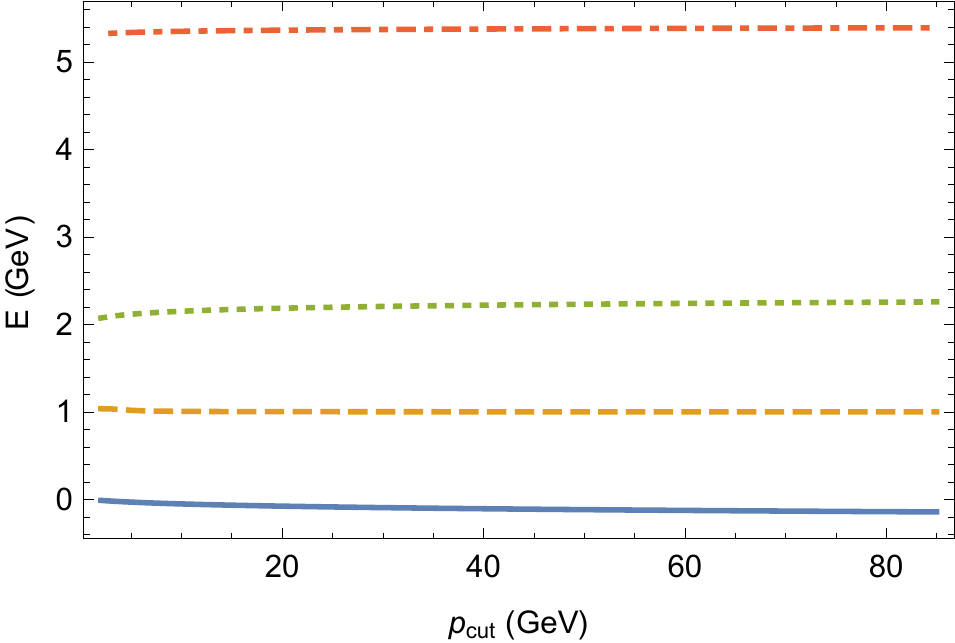}
\end{center}
\caption{\label{fig:Ecutoff energy spectrum}Plots of the eigenstate energies for different values of the cutoff momentum.  In the top plot, the black dots give the free energies of the basis states while the colored dots give the eigenstate energies.   The value of the cutoff momentum for each series are as follows from top right to bottom left: (blue) 85GeV, (orange) 75GeV, (green) 65GeV, (red) 55GeV, (purple) 45GeV, (brown) 35GeV, (blue) 25GeV, (yellow) 15GeV and (purple) 5GeV.  The other parameters are given in Eq.~(\ref{eq:parameters}).  The eigenstates were obtained by a numerical diagonalization of the Hamiltonian in Eq.~(\ref{eq:Discrete Hamiltonian}).  The bottom plot is of individual eigenvalues as a function of the cutoff energy.  The solid blue line is for the vacuum, the dashed orange line is for the single-particle state, the dotted green line is for the first two-particle state and the dot-dashed red line is for the fiftieth two-particle state.}
\end{figure}
The final parameter we consider is the momentum cutoff.  As in the case of the momentum spacing, this parameter is unphysical.  Nature corresponds with the limit $p_{cut}\to\infty$.  In our numerical calculations, we would like to take it as large as possible while still achieving efficient diagonalization.  In this section, we analyze the dependence of the solutions on this cutoff.  

As we extend the momentum cutoff towards higher energies, a greater number of basis states are allowed in our truncated Hilbert space.  Therefore, we expect our energy spectrum to also extend towards larger energies.  This is indeed, what we find.  In the top plot of Fig.~\ref{fig:Ecutoff energy spectrum}, we have plotted the energy spectrum for a series of cutoff momenta.  The spectrum of the smallest momentum cutoff (5GeV) is plotted in purple and only extends to just above 10GeV.  The spectrum of the next higher momentum cutoff (15GeV) is plotted in yellow and extends a bit further to just above 30GeV.  This trend continues with the spectrum of each momentum cutoff extending a bit further towards the top right.  The momentum cutoff in this plot is taken in increments of 10GeV with the final momentum cutoff being 85GeV and plotted in blue and extending to just above 170GeV.  

\begin{figure}[!]
\begin{center}
\includegraphics[scale=0.85]{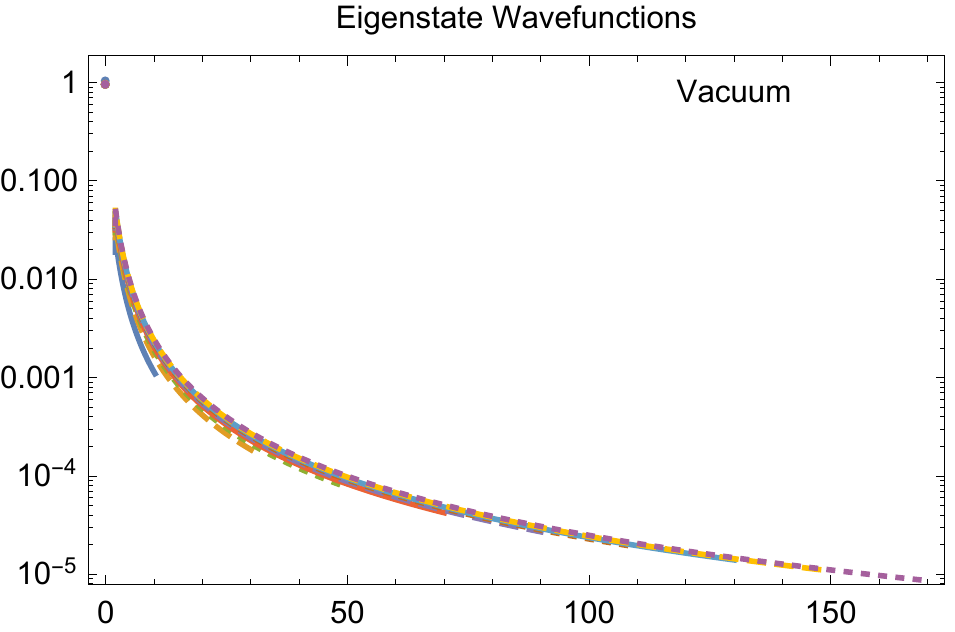}\\
\includegraphics[scale=0.85]{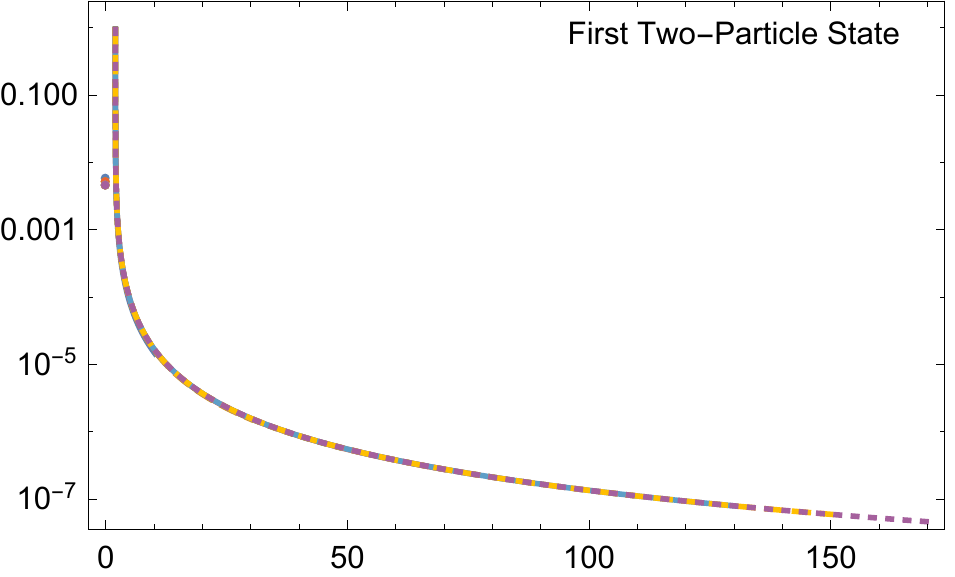}\\
\includegraphics[scale=0.85]{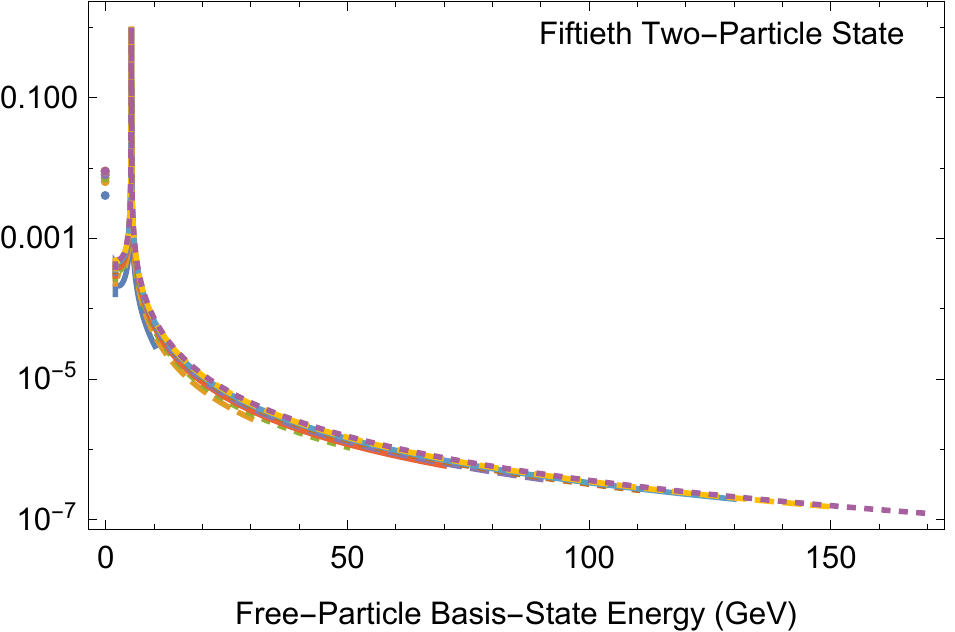}
\end{center}
\caption{\label{fig:Ecutoff VacuumWaveFunction2}A plot of the absolute values of the coefficients of the wavefunction for a few eigenstates.  These wavefunctions were obtained by a numerical diagonalization of the Hamiltonian in Eq.~(\ref{eq:Discrete Hamiltonian}) with the parameters given in Eq.~(\ref{eq:parameters}) except for the cutoff momentum.  The value of the cutoff momentum for each series are as follows from bottom to top of the continuum: (solid blue) 5GeV, (dashed orange) 15GeV and so on in increments of 10GeV up to a cutoff of 85GeV (in dashed purple).  The top plot is for the vacuum state, the middle plot is for the first two-particle state and the bottom plot is for the fiftieth two-particle state.  Further details can be found in Subsection~\ref{sec:E_cutoff}.}
\end{figure}

As we can see, on this scale, the energy eigenvalues essentially lie directly on top of each other where the basis states overlap at the bottom left.  The main difference between the spectra is that, for larger momentum cutoffs, the spectrum extends to larger energies.  This is a good sign and appears to signal that the energies are largely insenstive to the cutoff energy.  To clarify this, we plot in the bottom of Fig.~\ref{fig:Ecutoff energy spectrum} a few eigenvalues as functions of the cutoff momentum.  The solid blue line is for the vacuum, the dashed orange line is for the single-particle state, the dotted green line is for the first two-particle state and the dot-dashed red line is for the fiftieth two-particle state.  We can see that all of these eigenvalues are practically constant except at the left edge where a small deviation can be seen.  This is reassuring and suggests that our cutoff of 10GeV is probably sufficient for our purposes.  We would not expect this behavior in full four-dimensional $\phi^4$ theory as the mass is quadratically sensitive to the cutoff.  However, here in two dimensions, the mass is only logarithmically sensitive to the cutoff so, in the context of the current severely truncated Hilbert space, this behavior seems reasonable.  It will be interesting to see what happens when greater numbers of free particles are included in the basis states.

We now turn to the wavefunctions, which we plot for different values of the cutoff momentum in Fig.~\ref{fig:Ecutoff VacuumWaveFunction2}.  The first thing we see is that, as the cutoff momentum grows, so does the extension of the wavefunction towards higher basis states.  Because of this, we can clearly distinguish the curves for different cutoff energies.  The bottom solid blue curve is for a cutoff of 5GeV, the dashed orange line directly above this is for a cutoff of 15GeV and so on in increments of 10GeV.  The top dashed purple curve is for a cutoff of 85GeV.  

Another feature of the cutoff dependence of the wavefunctions is that they broaden around the peak as the cutoff grows.  The height of the peak lowers a tiny amount and the states to the sides of the peak increase slightly in importance.  The change is greatest from a cutoff of 5GeV to 15GeV, and decreases as the cutoff grows larger.  That is, the difference between the wavefunctions for cutoffs of 75GeV and 85GeV is much smaller than the difference between the wavefunctions for cutoffs of 5GeV and 15GeV.  The reason for this is that as the wavefunction extends to higher basis states, the contribution to the normalization from these states suddenly turns on.  Since the higher basis states are contributing, the lower basis states must adjust to keep the normalization fixed.  On the other hand, the higher basis states contribute a very small, exponentially decreasing, amount to the lower eigenstate wavefunctions, so the importance of this diminishes as the cutoff momentum grows.  

We see that, although the wavefunctions are affected by the momentum cutoff, and therefore we see that higher momentum cutoffs are better, the effect is not enormous.  Moreover, since the number of basis states in the Hilbert space grows with the cutoff momentum and reduces the effectiveness of the numerical diagonalization, a balance is necessary.  It appears to us that a cutoff of 10GeV in the present calculation is sufficient to understand the behavior of the energies and the wavefunctions and that the numerical power might be better used analyzing the effect of a reduction of $\Delta p$ when we consider renormalization in the next section.  We do not expect the dependence on the cutoff momentum to add much to the discussion of renormalization.

\section{\label{sec:renormalization}Renormalization}
\begin{figure}[!]
\begin{center}
\includegraphics[scale=0.85]{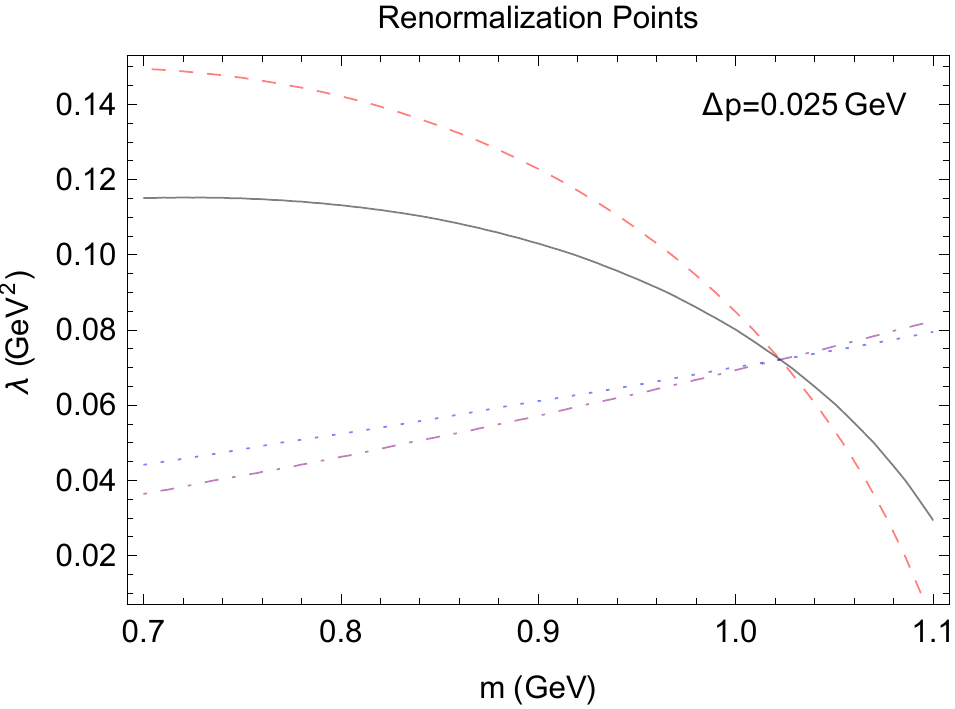}\\
\includegraphics[scale=0.85]{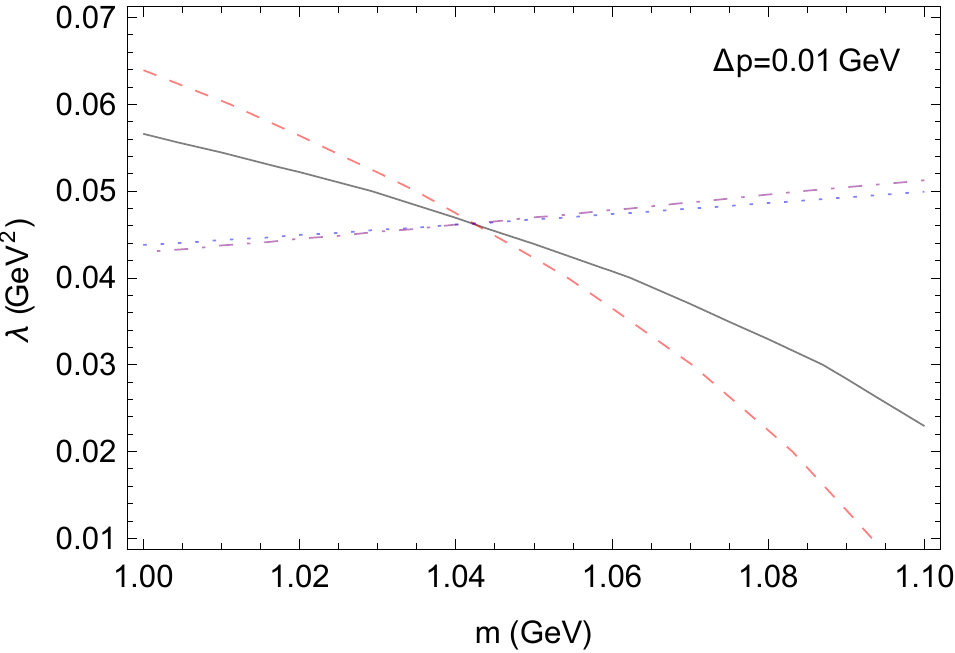}\\
\includegraphics[scale=0.85]{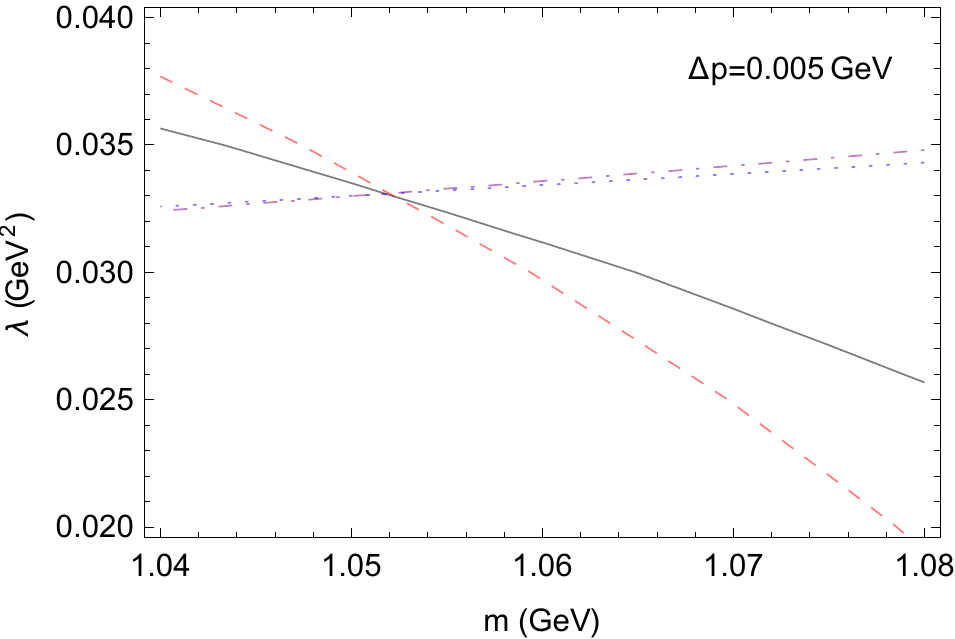}
\end{center}
\caption{\label{fig:Renormalization Point}Plots of the curves used to determine the renormalization points for $\Delta p=0.025$GeV (top), $\Delta p=0.01$GeV (middle) and $\Delta p=0.005$GeV (bottom).  The reference point for the parameters is given in Eq.~(\ref{eq:parameters}).  The solid black curve represents when the single-particle energy is the same height above the vacuum as for the reference point.  The red dashed curve represents when the lowest two-particle energy is the same height above the vacuum as for the reference point.  The blue dotted curve represents when the vacuum has the same value as for the reference point.  The purple dot-dashed curve represents when the vacuum has the same contribution from the free vacuum as for the reference point.}
\end{figure}
In Subsection~\ref{sec:Delta p}, we saw that as $\Delta p$ becomes smaller, the energy gap grows ever larger due to the growth of the vacuum energy towards negative infinity while the other eigenvalues remain essentially unchanged.  In fact, we found that the vacuum energy was inversely proportional to $\Delta p$ [see Eq.~(\ref{eq:E_vac(Dp)})], suggesting that it would diverge as $\Delta p\to0$.  Of course, the physical energies are finite in this limit and, in any case, we must renormalize the parameters in order to keep the observables at their physical values.  In this section, we will renormalize $\lambda$ and $m$.  In Subsection~\ref{sec:lambda}, we saw the dependence of the eigenvalues on $\lambda$.  The non-vacuum eigenvalues grew linearly with $\lambda$ while the magnitude of the vacuum energy grew proportional to $\lambda^2$ [see Eq.~(\ref{eq:E_vac(lambda)})].  Furthermore, we know that, whatever the value of $\Delta p$, as we take the limit $\lambda\to0$, the theory becomes completely decoupled, the eigenvalues merge with the free energies and the wavefunctions become delta functions.  So, it is clear that we can reverse the trend resulting from a decrease in $\Delta p$ with a suitable reduction of $\lambda$.  As we do this, we also expect to be required to adjust the mass as well to precisely fit the observable values.  This is indeed what we find.

In order to achieve a renormalization, we must choose a set of observables to fix as we decrease $\Delta p$ and adjust the parameters.  Two natural choices are the energy gap of the single-particle state which is the height of the single-particle state above the vacuum and the energy gap of the lowest two-particle state, again the height of its eigenvalue above the vacuum.  We set these two observables equal to the value we obtained with the reference parameters given in Eq.~\ref{eq:parameters}, respectively 1.08GeV+0.05GeV=1.13GeV and 2.15GeV+0.05GeV=2.20GeV, which we take to be observable.  We then scan the values we obtain for these observables for a range of values of $\lambda$ and $m$ for a series of decreasing $\Delta p$ in Fig.~\ref{fig:Renormalization Point}.  The observable mass gap of the single-particle state is plotted in solid black while the observable energy gap of the lowest two-particle state is plotted in dashed red.  We then take the intersection of these curves as the renormalized values of the parameters $\lambda$ and $m$.  Additionally, we plot in dotted blue the parameter values where the vacuum has the same energy as for the reference parameters, namely -0.05GeV.  Interestingly, it intersects the solid black and dashed red curves at the same parameter point where they cross each other.  The reason for this is that the eigenvalue of the lowest two-particle state is twice the eigenvalue of the single-particle state as described in Subsec.~\ref{sec:lambda}.  Therefore, the point at which their energy gaps from the vacuum agree would naturally coincide with the point where the vacuum had the same energy as well.  In other words, these two energy gaps are correlated and may not serve as independent observables for the purpose of renormalization.  

Normally in high energy physics, we can use the scattering amplitude as an independent observable for renormalization.  As discussed in Sec.~\ref{sec:scattering}, our Hilbert space is too severely truncated to allow non-trivial scattering, so this won't work here.  But, we consider as the closest alternative the magnitude of the coefficient of one of the basis states for one of the eigenstates.  We choose the coefficient of the free vacuum in the vacuum state, the left-most red dot in the top plot of Fig.~\ref{fig:VacuumWaveFunction2}, which has a value of 0.991 for the reference parameters in Eq.~(\ref{eq:parameters}).  We fix that as one of our observables and plot in Fig.~\ref{fig:Renormalization Point}, in dot-dashed purple, the parameter values where this coefficient is equal to 0.991.  We observe that this curve also crosses the other three curves at exactly the same parameter point so that all four curves cross at the same point.  Therefore, we find that several reasonable choices of observables give the same renormalization point and, in fact, we will see soon that all the energy eigenvalues coincide with the reference values for this choice of renormalized parameters.  We take this as a positive sign for our calculation and settle on the intersection points in Fig.~\ref{fig:Renormalization Point} as our renormalized parameters.  Their values are $\lambda=0.0723$GeV$^2$ and $m=1.02$GeV for $\Delta p=0.025$GeV, $\lambda=0.0463$GeV$^2$ and $m=1.04$GeV for $\Delta p=0.01$GeV, and $\lambda=0.0331$GeV$^2$ and $m=1.05$GeV for $\Delta p=0.005$GeV.

\begin{figure}[!]
\begin{center}
\includegraphics[scale=0.85]{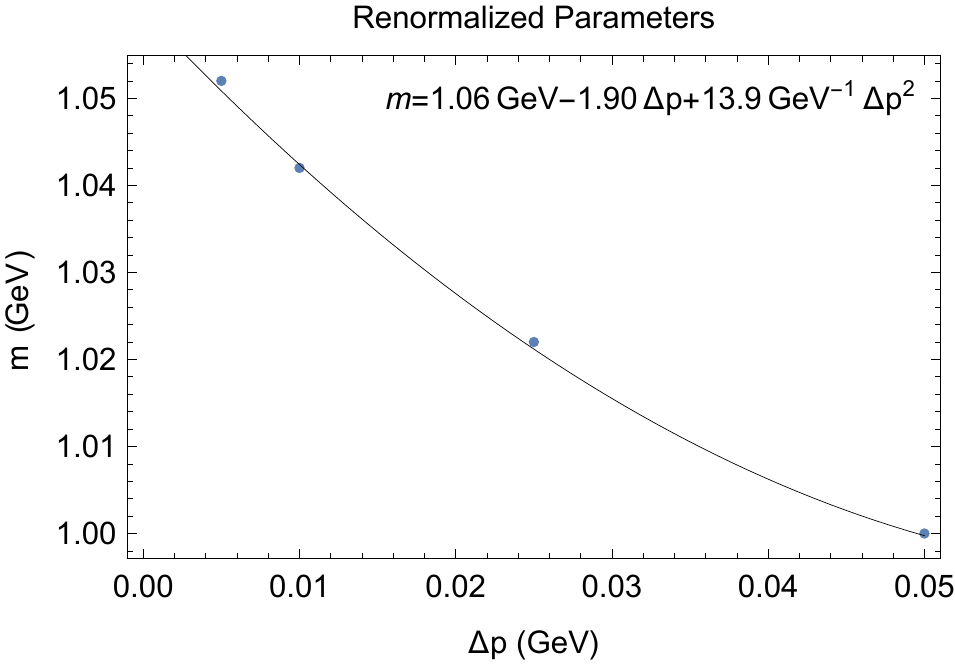}\\
\includegraphics[scale=0.85]{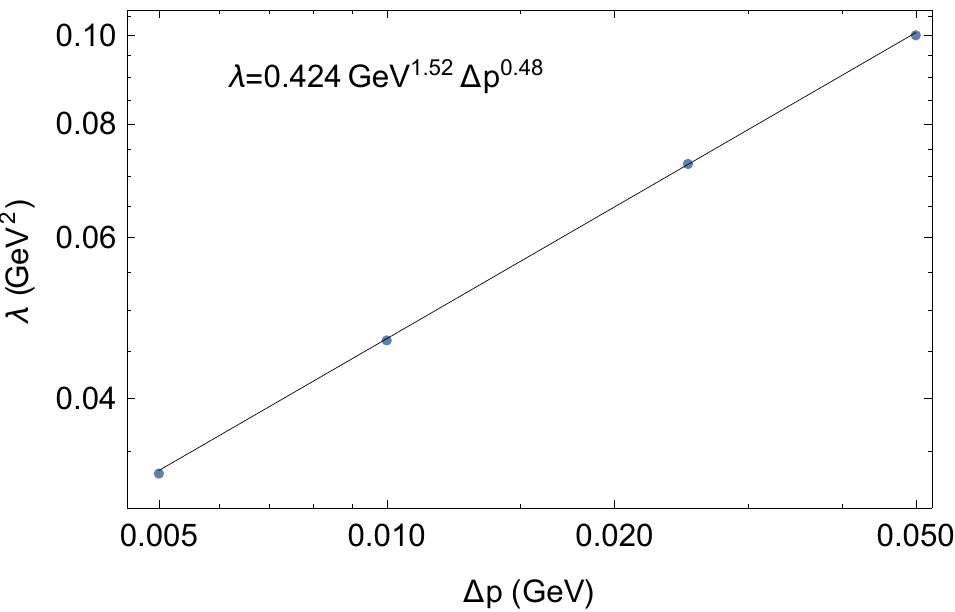}
\end{center}
\caption{\label{fig:Renormalized Parameters}Plots of the renormalized parameters as a function of $\Delta p$.  The top plot is for the mass parameter $m$ while the bottom plot is for the coupling parameter $\lambda$.  Additionally, the formulas for the best-fit curves are displayed.  The renormalized parameters come from the intersections of the curves in Fig.~\ref{fig:Renormalization Point}.}
\end{figure}
We have plotted the renormalized values of $\lambda$ and $m$ as functions of $\Delta p$ in Fig.~\ref{fig:Renormalized Parameters} along with their best fit curves.  The top plot is for the mass and the best fit curve of lowest order was quadratic and has the form
\begin{equation}
m(\Delta p) = 1.06\mbox{GeV}-1.90\Delta p+13.9\mbox{GeV}^{-1}\Delta p^2\ .
\end{equation}
The mass does not appear to require an infinite renormalization as $\Delta p\to0$.  In fact, it appears to approach the finite value of approximately 1.06GeV in the limit $\Delta p\to0$ which is only approximately six percent above the value of the parameter $m$ when $\Delta p=0.05$GeV.  Furthermore, it is only approximately six percent below the observable mass gap of 1.13GeV and only a couple percent below the energy eigenvalue for the single-particle state of 1.08GeV.  Perhaps more data extending to lower $\Delta p$ would alter this extrapolation and bring it closer to one of these.  We are unable to do this at this point but hope to do so in a future work.

The renormalized coupling $\lambda$ is plotted in the bottom of Fig.~\ref{fig:Renormalized Parameters} on a log-log plot.  We can see that the renormalized points are fit very well by a straight line on a log-log plot giving us the expression
\begin{equation}
\lambda(\Delta p) = 0.424\mbox{GeV}^{1.52}\Delta p^{0.48}\ .
\label{eq:lambda(Delta p)}
\end{equation}
We can clearly see that $\lambda\to0$ as $\Delta p\to0$ as we expected.  We also see that the power of $\Delta p$ is essentially $1/2$ and we understand that it comes from the combined power-law dependence of the vacuum energy on $\Delta p$ and $\lambda$ seen in Eqs.~(\ref{eq:E_vac(Dp)}) and (\ref{eq:E_vac(lambda)}). If we simply take the product of these at the renormalization point, then we have for the renormalized value of the vacuum energy
\begin{equation}
E^2_{vac} = 0.0137\mbox{GeV} \Delta p^{-0.94} \lambda^{1.94}
\end{equation}
Taking $E_{vac}=-0.05$GeV and solving for $\lambda$, we have
\begin{equation}
\lambda = 0.416\mbox{GeV}^{1.52} \Delta p^{0.48}
\end{equation}
in very good agreement with the dependence we measured in Eq.~(\ref{eq:lambda(Delta p)}).

\begin{figure}[!]
\begin{center}
\includegraphics[scale=0.85]{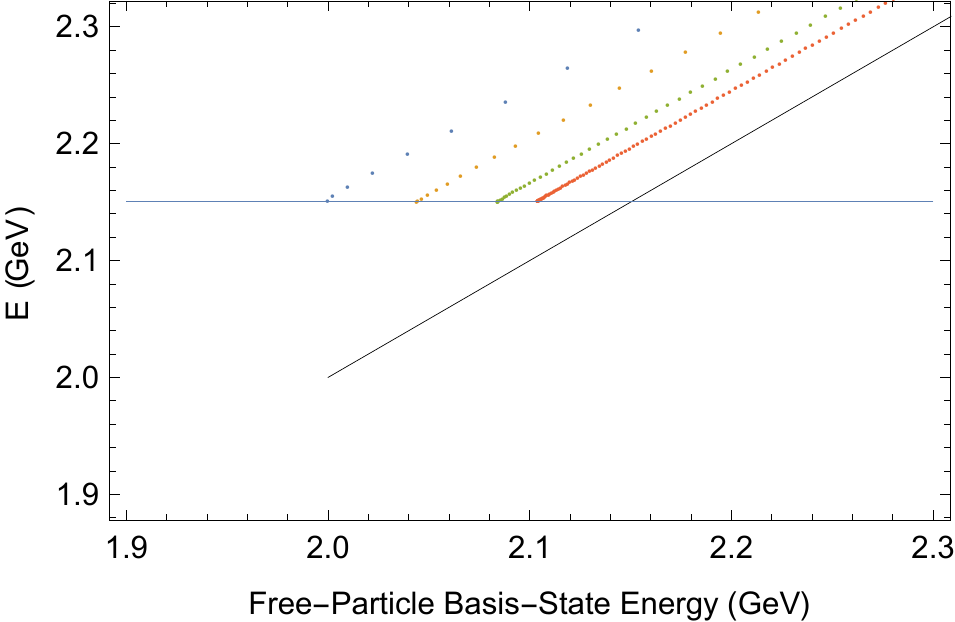}
\end{center}
\caption{\label{fig:EnSpectrumDpDlambdaDmassZoom}A magnification of the beginning of the two-particle continuum of energies.  The black line is the free energies for reference.  The other eigenvalues are as follows: (blue) $\Delta p=$0.05GeV, (yellow) $\Delta p=$0.025GeV, (green) $\Delta p=$0.01GeV, and (red) $\Delta p=$0.005GeV.  The values of $\lambda$ and $m$ are given in Sec.~\ref{sec:renormalization}.}
\end{figure}

Now that we have the renormalized values of the parameters for different values of $\Delta p$, we would like to see how the other observables are affected by this change.  First of all, the eigenvalues are all exactly the same as for the reference point in Eq.~(\ref{eq:parameters}).  This is extremely encouraging as it suggests that our numerical procedure does a very good job of obtaining these eigenvalues and they don't differ significantly from the limit $\Delta p\to0$.  In order to see what is happening with the eigenvalues during renormalization, we plot in Fig.~\ref{fig:EnSpectrumDpDlambdaDmassZoom} a magnification of the two-particle eigenvalues at the beginning of the continuum.  $\Delta p$ gets smaller as we move from the blue points at the left towards the red points at the right.  The first thing we see is that the spacing becomes smaller as $\Delta p$ becomes smaller as we expect.  Secondly, we see that the eigenvalue of the first two-particle state is constant.  This is as we expect since this is precisely one of our renormalization conditions.  In particular, we have noted that as we decrease $\Delta p$, we are forced to decrease $\lambda$ to compensate.  However, this alone would cause the energy spectrum to shift vertically down from the blue dots.  In order to compensate this effect, we also increase $m$ slightly, which causes the lowest eigenvalue to increase up and to the right to end at the position of the lowest red point (for $\Delta p=0.005$GeV).  With this combination, the renormalized energies have exactly the same energy as before, when considering the equivalent state as we will see when we discuss the energy of the `fiftieth' eigenstate.

\begin{figure}[!]
\begin{center}
\includegraphics[scale=0.85]{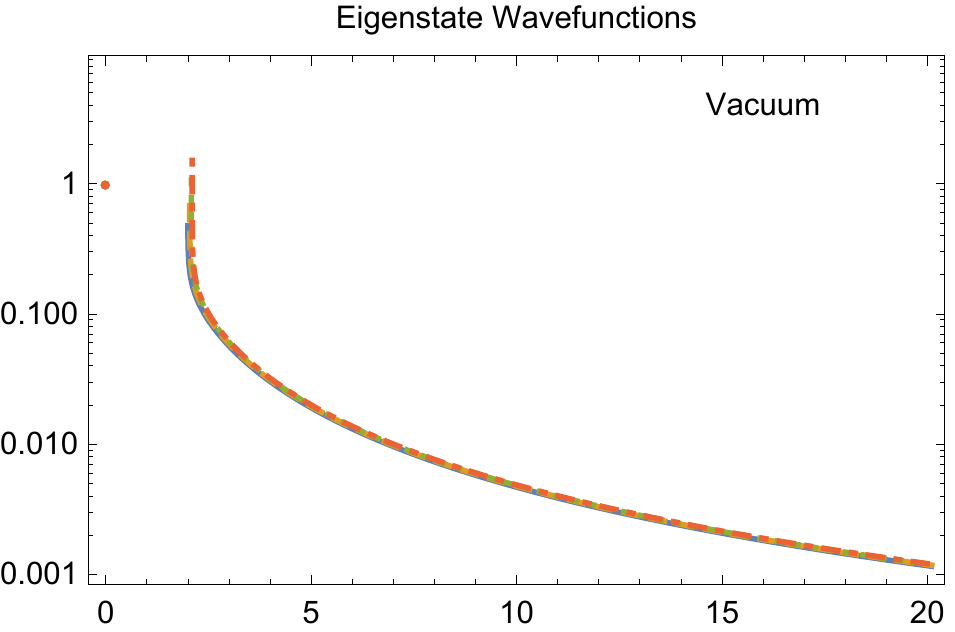}\\
\includegraphics[scale=0.85]{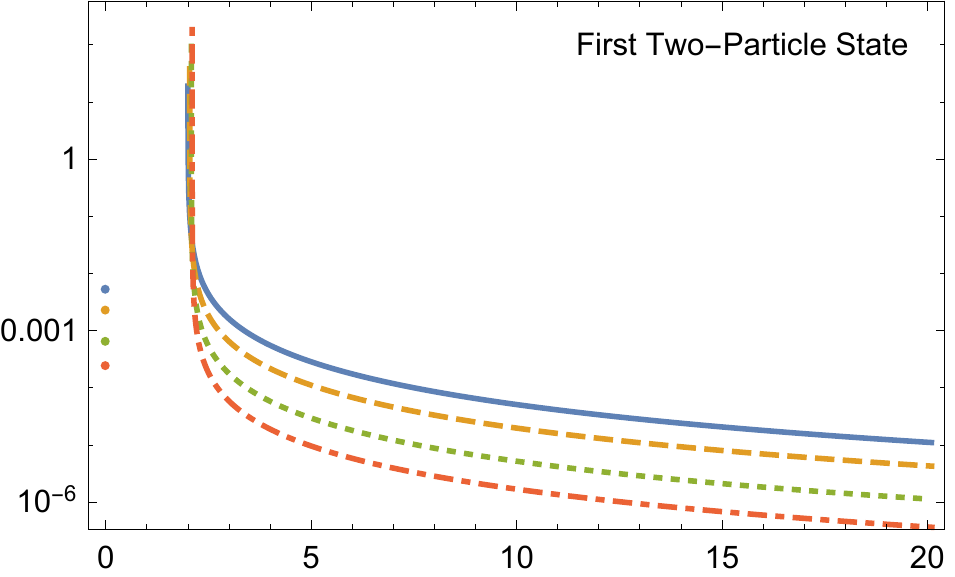}\\
\includegraphics[scale=0.85]{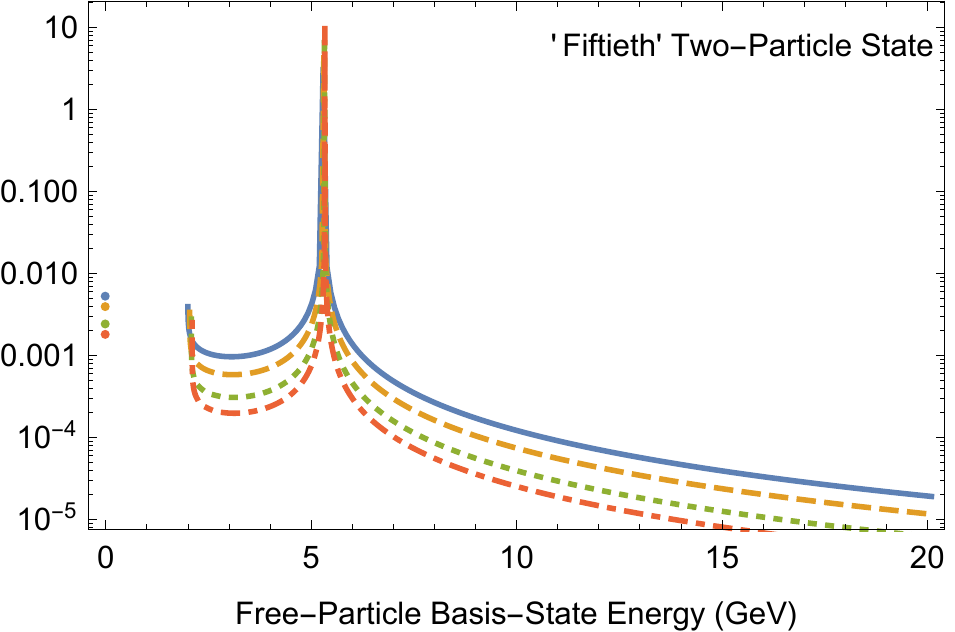}
\end{center}
\caption{\label{fig:Renormalized Wave Functions}A plot of the absolute values of the coefficients of the wavefunction for a few eigenstates.  These wavefunctions were obtained by a numerical diagonalization of the Hamiltonian in Eq.~(\ref{eq:Discrete Hamiltonian}) with the renormalized parameters described in Sec.~\ref{sec:renormalization}.  The lines are as follows: (solid blue) $\Delta p=0.05$GeV, (dashed orange) $\Delta p=0.025$GeV, (dotted green) $\Delta p=0.01$GeV, and (dot-dashed red) $\Delta p=0.005$GeV.  The top plot is for the vacuum state, the middle plot is for the first two-particle state and the bottom plot is for the `fiftieth' two-particle state.}
\end{figure}

We next plot the renormalized wavefunctions in Fig.~\ref{fig:Renormalized Wave Functions}.  This plot presents a challenge for comparison because we have changed both $\Delta p$ and $m$ between the curves, each of which makes direct comparison of wavefunctions more difficult.  Because the change in $\Delta p$ is much more significant than the change in $m$, we have chosen to normalize the wavefunctions in the same way as in Sec.~\ref{sec:Delta p}; we have divided the coefficients by the square root of the spacing between basis states.  This gives a better comparison between the wavefunctions for different $\Delta p$.  On the other hand, as we vary $m$, we shift the free-particle energies of the basis states.  In the context of the wavefunctions, this shifts the continuum to the right by roughly $2\Delta m$.  However, this shift is not physical since the free-particle energies of the basis state is unmeasureable.  All we can measure are the eigenstate energies.  In principle, we could compensate by shifting all the wavefunctions to the left by an equal amount in order to have the peaks line up again, although each coefficient would no longer be above the free-particle energy of its basis state.  This appears unnecessary to us, especially since the change in mass is so small.  Moreover, we feel clarity is better served to simply leave the wavefunction coefficients above the free-particle energies of the basis states and simply note that the slight shift to the right of the peaks is unphysical.  

\begin{figure}[!]
\begin{center}
\includegraphics[scale=0.85]{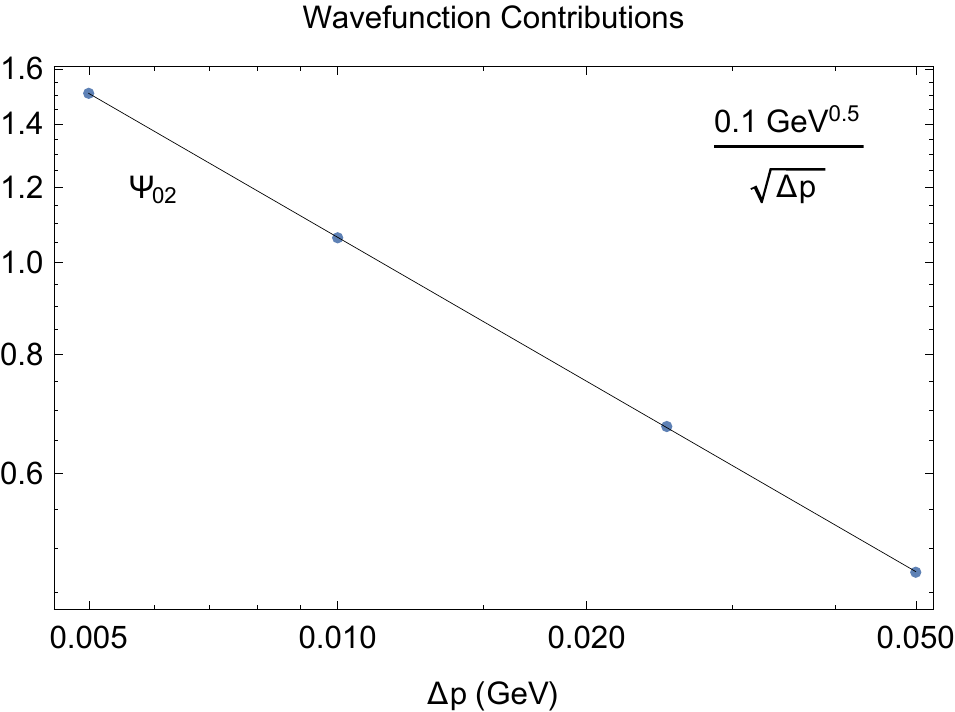}\\
\includegraphics[scale=0.85]{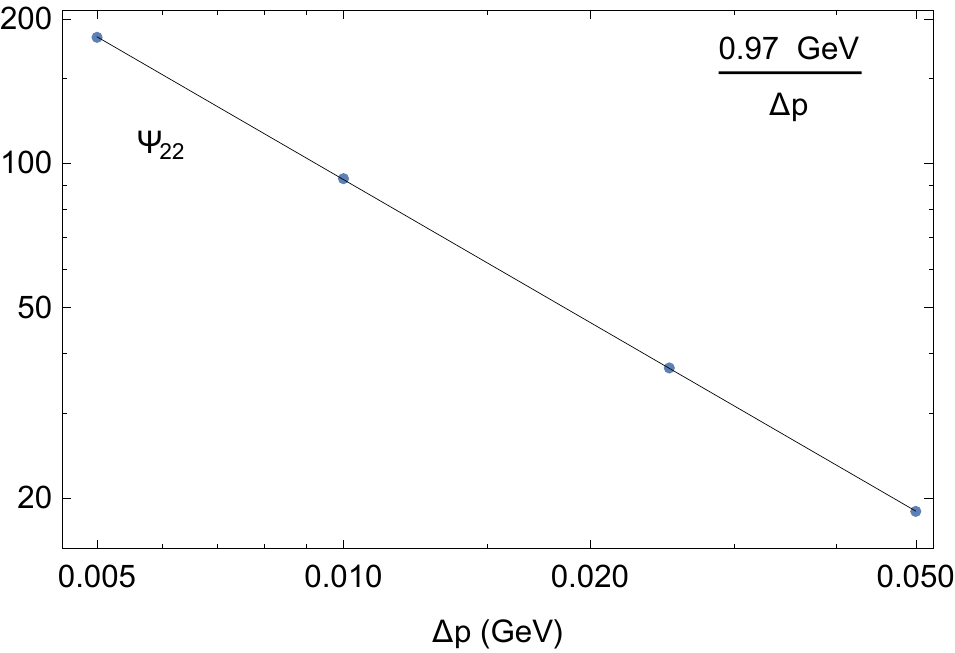}\\
\includegraphics[scale=0.85]{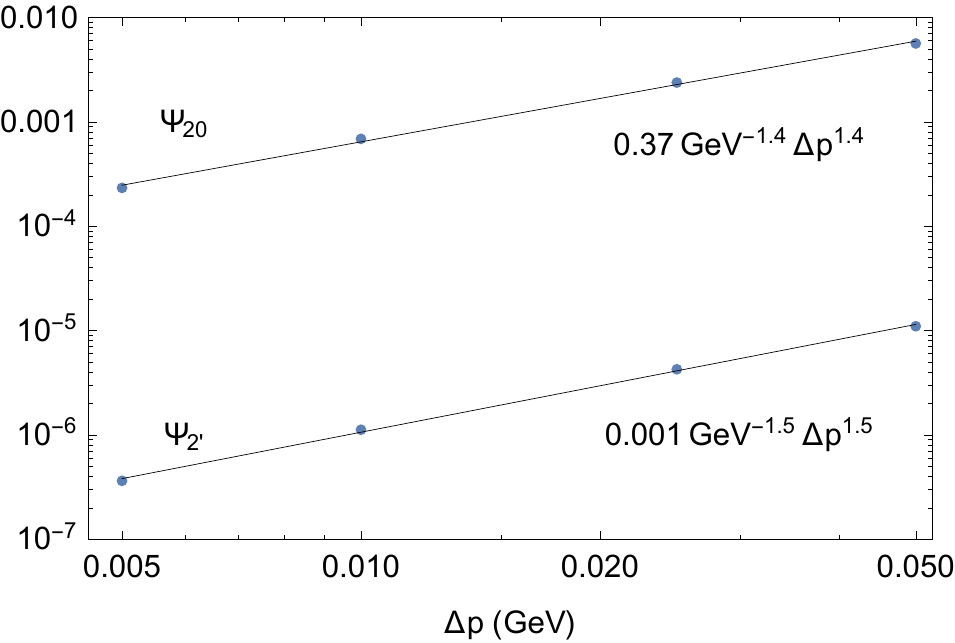}
\end{center}
\caption{\label{fig:VacuumHeight2DpDlambdaDmass}A plot of: the height of the peak at the beginning of the continuum of the vacuum (top), the top of the peak in the first two-particle state (middle), the free vacuum contribution in the first two-particle state (top line of bottom), and the contribution of the right-most point of the continuum in the first two-particle state (bottom line of bottom).  All the data come from Fig.~\ref{fig:Renormalized Wave Functions}.  The lines and formulas are the best-fit curves.}
\end{figure}

The lines in Fig.~\ref{fig:Renormalized Wave Functions} are as follows.  The solid blue lines are for $\Delta p=0.05$GeV, the dashed orange lines are for $\Delta p=0.025$GeV, the dotted green lines are for $\Delta p=0.01$GeV, and the dot-dashed red lines are for $\Delta p=0.005$GeV.  The top plot is for the vacuum state.  Interestingly, the wavefunctions nearly exactly coincide.  The free-vacuum contribution is exactly the same for all $\Delta p$, as expected from our normalization condition as described above, and dominates the wavefunction with a value of 0.991.  If we consider the continuum, the only major difference is that the peak at the position of the first two-particle state is shifted slightly to the right, as described in the preceding paragraph, and also that the peak becomes slightly higher as $\Delta p$ decreases.  To understand this better, we have plotted the height of this peak as a function of $\Delta p$ in the top plot of Fig.~\ref{fig:VacuumHeight2DpDlambdaDmass}.
We find that the height of this peak is given by $0.1$GeV$^{0.5}/\sqrt{\Delta p}$ and, therefore, diverges as $\Delta p\to0$.  What is happening seems clear.  The total area under the wavefunction squared is constant.  The contribution of the free vacuum dominates and does not change.  The contribution from most of the continuum grows larger as $\Delta p$ diminishes [see the top plot of Fig.~\ref{fig:Dp VacuumWaveFunction2}] while it grows smaller as $\lambda$ diminishes [see the top right plot of Fig.~\ref{fig:Dlambda VacuumWaveFunction2}].  The combined effect of reduced $\Delta p$ and $\lambda$ is that the two effects largely cancel each other and the contribution from most of the continuum remains the same.  The peak is slightly different.  Since it gets shifted slightly to the right, it must make up for the loss of area under the previously extended curve and, therefore, it must grow.  It also becomes narrower with the result that the area under this small section of the wavefunction (squared) remains constant.  In the limit of $\Delta p\to0$, this peak becomes initesimal in width and infinite in height in such a way as to preserve the area under the squared wavefunction.  Although similar to a delta function, it is not exactly the same due to its normalization and shape.  Remembering that the total area under the squared wavefunction is one and that the free vacuum dominates it with a contribution of 0.991$^2$=0.982, we see that the area under this squared peak approaches a finite constant that is less than 0.02, and therefore, much less than one.

In the middle plot of Fig.~\ref{fig:Renormalized Wave Functions}, we find the wavefunctions for the first two-particle state.  In this case, we see from the middle plot of Fig.~\ref{fig:Dp VacuumWaveFunction2} that the continuum surrounding the peak decreases as $\Delta p$ decreases and we explained this behavior in Sec.~\ref{sec:Delta p}.  In the present context, we see that when combining this with the fact that the continuum surrounding the peak also decreases with decreasing $\lambda$ (due to greater decoupling), the continuum surrounding the peak has nothing to do but decrease when renormalized here and that is what we find.  The same statement applies to the contribution from the free-vacuum.  The peak, on the other hand, grows larger.  In part this is because the position of the peak shifts slightly to the right and therefore must make up for the loss of area under it by becoming higher as discussed for the vacuum in the previous paragraph.  In part, this is also due to the reduction of importance of the other states to the wavefunction.  We have plotted the height of the peak at $|(0$GeV$,2)\rangle$ in the middle of Fig.~\ref{fig:VacuumHeight2DpDlambdaDmass}.  We see that it grows as $1/\Delta p$ and becomes infinite as $\Delta p\to0$. This is similar to our remarks in the previous paragraph except that here this peak dominates the area under the squared wavefunction.  In fact, in this case, the rest of the wavefunction diminishes as $\Delta p\to0$ so that this peak becomes more and more dominant the smaller $\Delta p$ becomes.  In fact, we find that the contribution from the free vacuum (the left-most dot in the middle plot of Fig.~\ref{fig:Renormalized Wave Functions}) falls off as $1/\Delta p^{1.4}$ and the contribution from the highest basis state (the right-most point of the middle plot of Fig.~\ref{fig:Renormalized Wave Functions}) falls off as $1/\Delta p^{1.5}$.  We have plotted both of these in the bottom frame of Fig.~\ref{fig:VacuumHeight2DpDlambdaDmass}.  It appears that the wavefunction of the first two-particle state approaches very close to a true delta function. 

\begin{figure}[!]
\begin{center}
\includegraphics[scale=0.85]{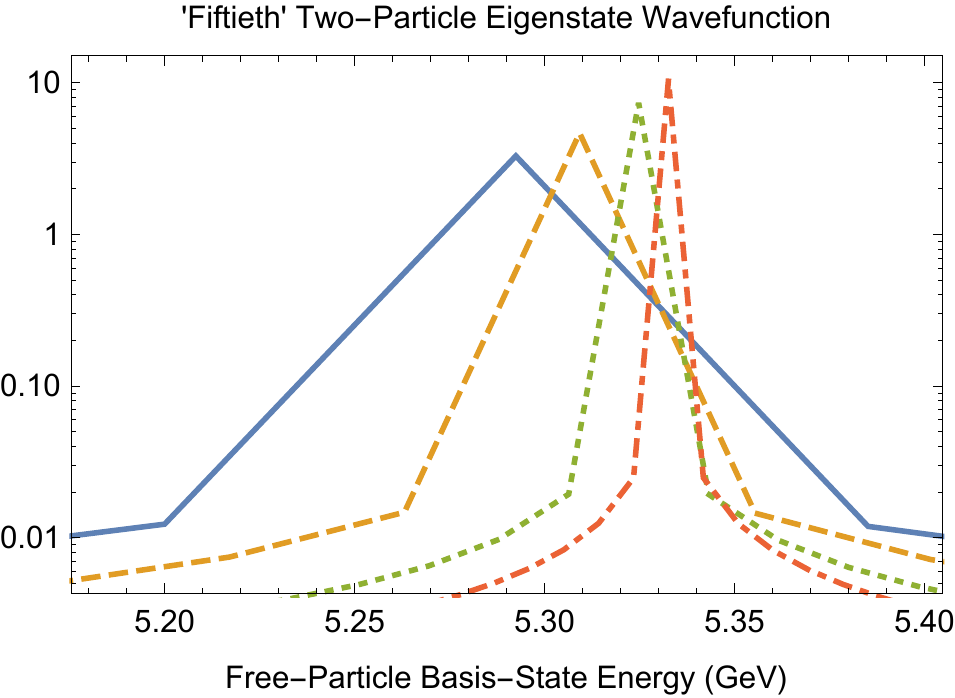}\\
\vspace{0.05in}
\includegraphics[scale=0.85]{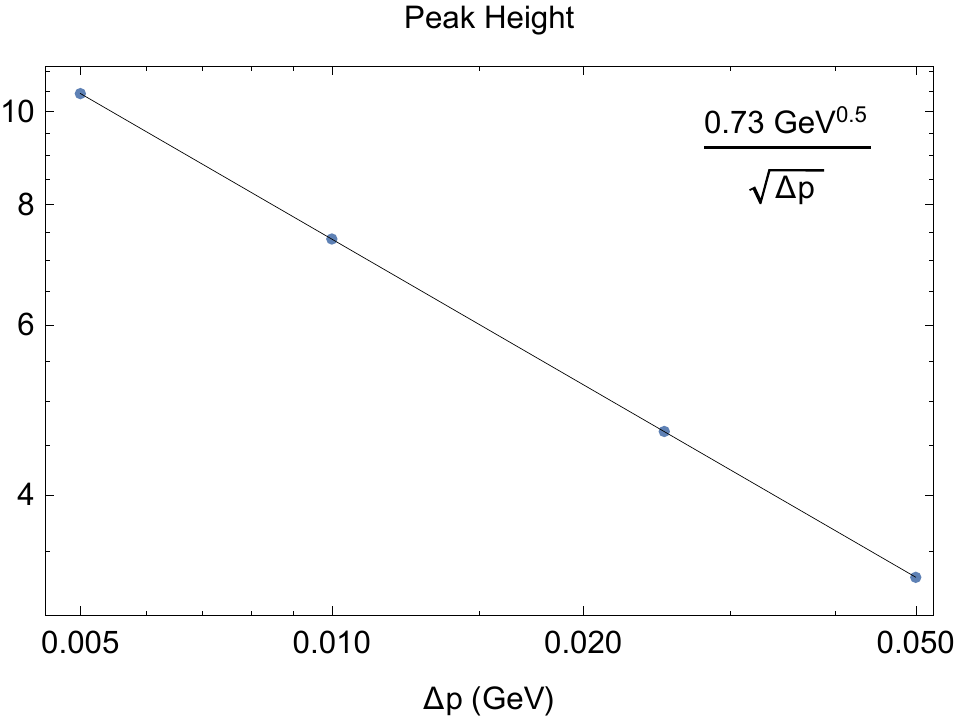}\\
\vspace{0.05in}
\includegraphics[scale=0.85]{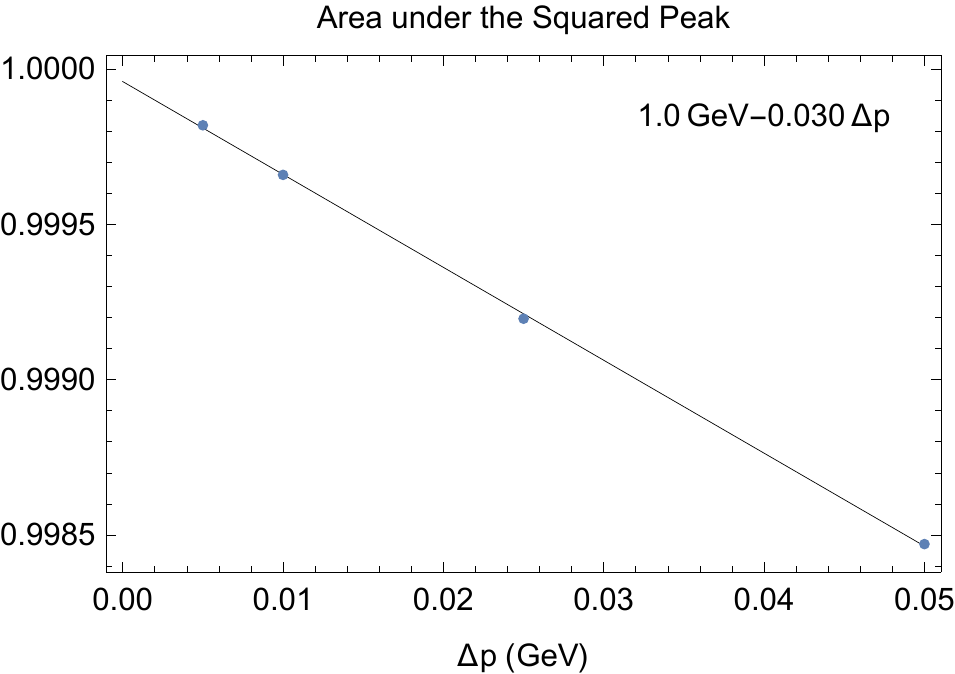}
\end{center}
\caption{\label{fig:FiftiethExcitedWaveFunction2DpDlambdaDmassZoom}The top plot is a magnification of the bottom plot of Fig.~\ref{fig:Renormalized Wave Functions} around the peak.  The middle plot gives the height of the peak as a function of $\Delta p$.  The bottom plot gives the area under the squared peak as a function of $\Delta p$.}
\end{figure}

Finally, the bottom plot of Fig.~\ref{fig:Renormalized Wave Functions} contains the `fiftieth' two-particle state, where in order to determine the `fiftieth' two-particle state we use the same procedure as in Sec.~\ref{sec:Delta p}.  It is defined as the fiftieth two-particle state when $\Delta p=0.05$GeV.  It is defined to be the state that has a peak at the same free-particle energy for the other values of $\Delta p$ when $m$ is held constant.  That is, we have chosen the same states here as we did in Sec.~\ref{sec:Delta p}.  It is the ninety-ninth two-particle state when $\Delta p=0.025$GeV, the two-hundred-fourty-sixth two-particle state when $\Delta p=0.01$GeV and the four-hundred-ninety-first two-particle state when $\Delta p=0.005$GeV.  Because we chose it exactly as we did in Sec.~\ref{sec:Delta p}, the peak shifts slightly to the right here in Fig.~\ref{fig:Renormalized Wave Functions}.  A magnified view of the peak can be seen in the top plot of Fig.~\ref{fig:FiftiethExcitedWaveFunction2DpDlambdaDmassZoom}.  However, we reiterate, this shift is unphysical and only due to the change in the mass parameter $m$.  It is only because the free-particle energy of the basis states is slightly increased as a result.  To further make the point that this shift is unphysical, we note that the eigenstate energy for all of these states is 5.35GeV, exactly the same and independent of $\Delta p$ (after renormalization).  We again see that the renormalization point is unique.  We could also have used this energy for the renormalization, but it would have given us the same renormalized parameter points.  Thus, we see that, although shifted on this plot, these do correspond to the same physical state.

In the case of the `fiftieth' two-particle state, we see that the continuum surrounding the peak decreases with decreasing $\Delta p$.  As in the previous two cases, this is the result of the combined effect of a reduced $\Delta p$ and a reduced $\lambda$.  Similarly to the vacuum, we see in the bottom plot of Fig.~\ref{fig:Dp VacuumWaveFunction2} that the continuum surrounding the peak increases with decreasing $\Delta p$.  However, the increase is much smaller than in the case of the vacuum seen in the top plot of Fig.~\ref{fig:Dp VacuumWaveFunction2}.  So, in this case, the reduction due to a decreasing $\lambda$, as seen in the bottom plot of Fig.~\ref{fig:Dlambda VacuumWaveFunction2}, dominates and the renormalized wavefunction decreases around the peak as $\Delta p$ decreases.  Focusing on the top plot of Fig.~\ref{fig:FiftiethExcitedWaveFunction2DpDlambdaDmassZoom}, we see that, additionally, the peak narrows and increases in height.  This is the behavior resulting from both a decrease in $\Delta p$ as we can see in Fig.~\ref{fig:Dp VacuumWaveFunction2Zoom} and a decrease in $\lambda$ as we can see in the top plot of Fig.~\ref{fig:Dlambda VacuumWaveFunction2 Cross Over} (for small $\lambda$).  So, this behavior is not surprising here.  In fact, we have plotted the height of the peak as a function of $\Delta p$ in the middle plot of Fig.~\ref{fig:FiftiethExcitedWaveFunction2DpDlambdaDmassZoom} and can see that the height of the peak grows as $1/\sqrt{\Delta p}$ and so becomes infinite as $\Delta p\to0$.  Once again, we are seeing the formation of a delta function as we approach the continuum limit where $\Delta p\to0$.  We suspect that the area under the squared curve is approaching nearly one, and indeed it is as we can see in the bottom plot of Fig.~\ref{fig:FiftiethExcitedWaveFunction2DpDlambdaDmassZoom}.  In this plot, we have shown the area under the squared peak where the width is taken from the points directly to the sides of the peak.  For example, for $\Delta p=0.05$GeV, we have taken the area from 5.2GeV to 5.39GeV whereas for $\Delta p=0.005$GeV, we have taken the area from 5.32GeV to 5.34GeV.  In greater precision, our $\Delta p=0$ intercept is 0.99996GeV suggesting that there is a very small amount of the area (0.00004GeV) given by the regions of the wavefunction outside the peak.

The renormalization appears to be very satisfying to our eyes.  We are able to fix several observables and, in this severely truncated Hilbert space at least, they all give the same renormalized values for the parameters as $\Delta p\to0$.  Second, after renormalizing the parameters, the eigenstate energies, and indeed the energy gap above the vacuum, appear stable and, in fact, retain the same values as $\Delta p\to0$.  This gives us hope that this method of numerically diagonalizing the truncated Hamiltonian of field theory may be able to produce sensible and useful results as this method is further developed.  We also find that the vacuum wavefunction is largely unaffected by $\Delta p\to0$ after renormalization and that the only major change is that the minor peak at the beginning of the continuum becomes narrower and taller while preserving its small contribution to the overall area under the squared wavefunction.  For the other eigenstates, the wavefunction is more greatly altered by the reduction of $\Delta p$, even after the renormalization.  However, it appears that we understand its behavior as $\Delta p\to0$.  The wavefunction outside the peaks drops with decreasing $\Delta p$ reducing its contribution to the area under the squared wavefunction while the peaks become narrower and taller preserving their contribution to the area under the squared wavefunction.  In fact, we explicitly showed this for the main peak of the `fiftieth' two-particle state, where we plotted the area under the squared peak and showed that it approached one as $\Delta p\to0$.

\section{\label{sec:conclusions}Summary and Conclusions}
In this paper, we have suggested a new non-perturbative approach to calculating the S matrix that complements existing perturbative efforts.  It involves the direct diagonalization of the Hamiltonian in a suitably truncated Hilbert space that is finite and uses the inner product of the eigenstates to determine the S matrix.  In Sec.~\ref{sec:theory}, we have introduced a simple test theory to develop this technique, a $\lambda\phi^4$ theory in two space-time dimensions.  We have shown how to discretize the Hamiltonian and cut it off in order to make it finite and amenable to numerical diagonalization.  We have also worked out the Hamiltonian matrix elements for a Hilbert space that only includes up to two free particles in its basis states.

We wrote a code that calculated this Hamiltonian matrix numerically and diagonalized it and we discussed our results in the remainder of this paper.  In Sec.~\ref{sec:results}, we introduced our illustrative reference set of parameter values and described our results for this reference set of parameters, including their dependence on the time.  In Subsec.~\ref{sec:results:energy}, we showed the energy spectrum resulting from our diagonalization in Fig.~\ref{fig:EnSpectrum2}.  We found an energy gap between the vacuum, the one-particle state and the beginning of the two-particle state continuum.  We found that the energies of the two-particle eigenstates grew nearly linearly with their free-particle eneriges and we found that the eigenvalues were independent of time.

In Subsec.~\ref{sec:Results:absolute wave function}, we displayed the absolute value of the coefficients of the wavefunctions in Fig.~\ref{fig:VacuumWaveFunction2}.  We found that these wavefunctions were strongly peaked at one particular basis state.  We found that the vacuum is peaked at the free vacuum, the first two-particle state is peaked at the lowest two-particle basis state and so on.  Each state is peaked at its related basis state with a small contribution from other basis states.  We also found that the moduli of the wavefunction coefficients is independent of time.  In Subsec.~\ref{sec:Results:wave function phases}, we plotted the phases of the coefficients of the wavefunctions in Fig.~\ref{fig:wave function phases} and found that these phases grew linearly with time at the same rate independent of the eigenstate.  We plotted these rates of change of the phases in Fig.~\ref{fig:phase plot} and showed that these phase rates agreed exactly with the free energies of the basis states.  We also noted that there was a universal unitary operator that advanced the eigenstate wavefunction in time and that the Hamiltonian at a non-zero time could be obtained as this unitary transformation acting on the Hamiltonian at $t=0$.

In Subsec.~\ref{sec:scattering}, we discussed the S matrix in this simple theory and noted that it was trivial due to the severe truncation of the Hilbert space to only include basis states with a product of two or fewer free particles.  In particular, each eigenvalue was unique and each eigenstate was nondegenerate.

In Sec.~\ref{sec:dependence on parameters}, we began a discussion of how the eigenvalues and eigenstate wavefunctions depended on each of the parameters of the theory.  We began with their dependence on the discretized momentum spacing in Subsec.~\ref{sec:Delta p}.  We noted that this parameter $\Delta p$ is unphysical and that it should, in principle, be taken to zero.  Although this is impossible in practice, we are able to study its dependence on $\Delta p$ over a wide range of values of $\Delta p$ and extrapolate.  In this subsection, we found that all the eigenvalues other than the vacuum energy were unsensitive to the value of $\Delta p$.  The vacuum energy, on the other hand, depended inversely on $\Delta p$, as we showed in Fig.~\ref{fig:EnVacuumDp}.  It grew towards larger negative values as $\Delta p$ decreased and became infinite in the limit of $\Delta p\to0$.  We also displayed the wavefunction's dependence on $\Delta p$ in Fig.~\ref{fig:Dp VacuumWaveFunction2}.  The peak became taller and narrower and approached a constant times a delta function.  We showed a magnified view of the `fiftieth' wavefunction in Fig.~\ref{fig:Dp VacuumWaveFunction2Zoom} where this change to the shape was clearly visible.

In Subsec.~\ref{sec:lambda}, we turned to the behavior of the solution as we varied the coupling constant $\lambda$.   We noted that as $\lambda\to0$, the theory becomes absolutely decoupled and that our numerical results agreed with this result.  In Fig.~\ref{fig:EnFewDlambda}, we showed the dependence of the eigenstate energies on $\lambda$.  All the eigenvalues other than the vacuum energy depended linearly on $\lambda$ while the vacuum energy fell off quadratically with $\lambda$ for small $\lambda$ and nearly linearly for large $\lambda$.  We also plotted the wavefunctions in Fig.~\ref{fig:Dlambda VacuumWaveFunction2} and showed that they approached delta functions as $\lambda\to0$.

In Subsec.~\ref{sec:m}, we described the dependence on the mass parameter $m$.  We showed in Fig.~\ref{fig:Dmass energy spectrum} a plot of the energy spectrum and how it depended on $m$.  We found that for sensible results, $m$ must be much greater than the momentum spacing $\Delta p$.  In this limit, we also showed that the major effect of increasing $m$ was to shift the entire energy spectrum up and to the right, when the energies are plotted as a function of the free-particle energies of the basis states.  In Fig.~\ref{fig:Dmass VacuumWaveFunction2}, we showed how the wavefunctions depend on $m$.  Once again, we noted that the results are only sensible when $m\gg\Delta p$.  Within that region, we noted that the larger $m$ became, the smaller the relative value of $\lambda$ and therefore, the closer to decoupling.  This explained the fact that the wavefunctions approached (slowly) delta functions as $m$ was increased.

In Subsec.~\ref{sec:E_cutoff}, we described the dependence of our solutions on the cutoff momentum $p_{cut}$.  In Fig.~\ref{fig:Ecutoff energy spectrum}, we showed the dependence of the eigenvalues on this parameter.   In particular, other than the availability of more states in the Hilbert space at higher energy, there was very little change until the cutoff momentum approached the physical parameters of the theory.  In Fig.~\ref{fig:Ecutoff VacuumWaveFunction2}, we plotted the wavefunctions.  We saw that, as for the energy eigenvalues, the wavefunctions were extended to higher basis states as a result of increasing the cutoff.  Since these new basis states contributed to the normalization of the wavefunction, the wavefunction was mildly affected at the lower basis states as well.  However, we found that the significance of this effect was small if the cutoff momentum was taken large compared to dimensionful parameters of the theory.

In Sec.~\ref{sec:renormalization}, we finally turned our attention to the renormalization of our parameters $\lambda$ and $m$ as we took the physical limit $\Delta p\to0$.  We found that several reasonable observables gave the same renormalization points as we decreased $\Delta p$.  We showed the intersection of these observables in Fig.~\ref{fig:Renormalization Point}, where we can see that all of the lines cross at the same $\lambda, m$ point.  We further found that with this set of renormalized values of $\lambda$ and $m$, all the eigenvalues remained constant as $\Delta p\to0$, which we took as a very positive sign for our renormalization procedure.  In Fig.~\ref{fig:Renormalized Parameters}, we plotted the dependence of the renormalized $\lambda$ and $m$ on $\Delta p$.  We found that the renormalized $m$ varied slowly and roughly quadratically, but did not blow up or go to zero.  We extrapolated its value in the limit $\Delta p\to0$ to be $1.06$GeV and noted how close that was to both the original parameter value and the observable value of the physical mass.  We also found that $\lambda\propto\sqrt{\Delta p}$ and described how this came from the dependence of the vacuum energy on both $\lambda$ and $\Delta p$.  With this, we found that in order to renormalize the observable values, we must take $\lambda\to0$ as we take $\Delta p\to0$.  We noted that this makes sense since $\Delta p\to0$ causes the eigenvalues to blow up whereas $\lambda\to0$ causes the eigenvalues to fall down to the free-particle energies of the basis states.  They have opposite effects, so it makes sense that by taking them both to zero in an appropriate way, their effects can largely cancel.  Very interestingly, we see that this is the procedure that replaces absorption of infinities into bare parameters in perturbation theory.  In Fig.~\ref{fig:EnSpectrumDpDlambdaDmassZoom}, we showed how the energy eigenvalues change as $\Delta p$ becomes smaller and $\lambda$ and $m$ are renormalized.  In effect, the eigenvalues for the same physical state remain constant.  

In Fig.~\ref{fig:Renormalized Wave Functions}, we show how the wavefunctions are affected by renormalization as $\Delta p\to0$.  We described how its behavior was a combination of the behavior when $\Delta p$ becomes smaller and $\lambda$ becomes smaller along with a slight increase in $m$.  We also noted that each of the peaks approaches a delta function.  In fact, in Fig.~\ref{fig:VacuumHeight2DpDlambdaDmass}, we showed how the height of several different basis states changed with $\Delta p$ for the vacuum and the first two-particle state.  The peaks grow taller and narrower while the other basis states diminish in size.  In Fig.~\ref{fig:FiftiethExcitedWaveFunction2DpDlambdaDmassZoom}, we focused on the main peak of the `fiftieth' two-particle state and showed that its height increased towards infinity while its width decreased towards zero.  However, we showed that the area inside this squared peak was stable and slowly approached nearly one giving very close to a true delta function.  

Finally, in this section, we discussed that the renormalization procedure appears very robust.  The renormalization points appear to be unique in the present case removing the potential ambiguity in how this should be done.  The energies of the eigenstates are extremely stable as $\Delta p\to0$ after renormalization and therefore can be taken, in the limit, as the physical values.  The wavefunctions approach closer to delta functions at the dominant basis state with a small amount of contribution from the other basis states.

So far, we have only scratched the very tip of the iceberg.  Our long-term hope is to use this technique to non-perturbatively study the SM and beyond.  We will tackle projects incrementally as we develop the numerical technology as well as our theoretical understanding.  For the remainder of this section, we will discuss a few of our future plans.

Now that we have warmed up with our $\lambda\phi^4$ theory in two space-time dimensions with a Hilbert space that included only up to two free particles in the basis states, there are two important immediate projects to complete.  One is to increase the Hilbert space to include up to four free particles in the basis states.  This will already involve a planned improvement in the efficiency of our diagonalization code.  With this enlarged Hilbert space, we will see how greatly truncating the number of free particles in the basis states has effected our results.  We will also have the possibility of non-trivial S-matrix elements for the first time since the eigenstates will no longer all be nondegenerate.  Once we have non-trivial S-matrix elements, we will also be able to analyze the dependence of scattering amplitudes on the collision energy.  

In another related direction, we would like to diagonalize the Hamiltonian of $\lambda\phi^4$ theory in three space-time dimensions and better determine the benefits and drawbacks of only including two space-time dimensions.  Furthermore, although we will only include up to two-particle basis states in the Hilbert space, because the particles can have momentum in two dimensions, degeneracies will be possible and non-trivial scattering will be allowed.  

Finally, as we increase our understanding of $\lambda\phi^4$ theory in the context of this new technique, we would like to extend the methods described here to other particles and interactions of the SM, including QED, QCD and beyond.

\begin{acknowledgments}
The author would like to thank R.~Grobe, Q.~Su and Q.Z.~Lv for bringing the diagonalization of the Hamiltonian to the author's knowledge and for helpful comments in the early stages of this project.  

\end{acknowledgments}

\appendix

\section{\label{app:a and adagger}Creation and Annihilation Operators}
In this appendix, we will briefly review the derivation of the properties of the creation and annihilation operators.  As described in Section~\ref{sec:theory}, we define our Hilbert-space basis as the direct product of single-particle states and label these states as
\begin{equation}
|(p_1,n_1),(p_2,n_2),\cdots\rangle\ ,
\end{equation}
where $p_i$ represents the one-dimensional momentum and $n_i$ represents the multiplicity for that momentum.  This momentum multiplicity notation is not really ideal for continuous momenta.  In contrast, it is very convenient for discrete momenta, which is the main subect of this paper.  Therefore, in this appendix, we will derive the properties of the creation and annihilation operators in discrete momentum space.  Since the single-particle states are orthogonal (and by construction they are normalized), we have the inner product rule
\begin{eqnarray}
\langle(p'_1,n'_1),(p'_2,n'_2)\cdots|(p_1,n_1),(p_2,n_2)\cdots\rangle =\hspace{0.5in}\\
\sum \prod \delta_{p'_i,p_j}\delta_{n'_i,n_j}\ ,\nonumber
\label{eq:<...> def}
\end{eqnarray}
where the product is over all pairs of momenta and the sum is over all orderings of the products.  Any inner products where momenta with nonzero multiplicities do not match up are, of course, zero.  For example,
\begin{eqnarray}
\langle(p',n')|(p,n)\rangle = \delta_{p',p}\delta_{n',n}\hspace{1in}&&\\
\langle(p'_1,n'_1),(p'_2,n'_2)|(p_1,n_1),(p_2,n_2)\rangle =\hspace{1in}&&\\
 \delta_{p'_1,p_1}\delta_{n'_1,n_1}\delta_{p'_2,p_2}\delta_{n'_2,n_2}&&\nonumber\\
 +\delta_{p'_1,p_2}\delta_{n'_1,n_2}\delta_{p'_2,p_1}\delta_{n'_2,n_1}\ .\nonumber
\end{eqnarray}

With this set of basis states and inner-product rules, we now \textit{define} the creation operator to have the property that it adds one particle to the basis state
\begin{equation}
a^\dagger_{p_i}|\cdots(p_i,n_i)\cdots\rangle = \sqrt{n_i+1}|\cdots(p_i,n_i+1)\cdots\rangle\ .
\label{eq:adagger definition}
\end{equation}
Other normalizations would not give the correct physical values for the masses.
As a \textit{consequence} of this definition, we get that the hermitian conjugate of the creation operator annihilates a particle from the basis state.  We can see this by inserting the hermitian conjugate of the creation operator between two general basis states as in
\begin{eqnarray}
\langle(p_1,n_1)\cdots(q,n_q)\cdots|a_{q}|(p'_1,n'_1)\cdots(q,n'_q)\cdots\rangle = \\
\langle(p'_1,n'_1)\cdots(q,n'_q)\cdots|a^\dagger_q|(p_1,n_1)\cdots(q,n_q)\cdots\rangle^* = \nonumber\\
\sqrt{n_q+1}\langle\cdots(q,n'_q)\cdots|\cdots(q,n_q+1)\cdots\rangle^* = \nonumber\\
\sqrt{n_q+1}\left(\sum\prod \delta_{p'_i,p_j}\delta_{n'_i,n_j}\right) \delta_{n'_q,n_q+1}\ ,\nonumber
\end{eqnarray}
where we have separated the $q$ term from the rest of the product.  This last line is equal to
\begin{equation}
\sqrt{n'_q}\langle(p_1,n_1)\cdots(q,n_q)\cdots|(p'_1,n'_1)\cdots(q,n'_q-1)\cdots\rangle
\end{equation}
(since $n'_q=n_q+1$).
As a result, we obtain from this the rule
\begin{equation}
a_{q}|(p_1,n_1)\cdots(q,n_q)\cdots\rangle = \sqrt{n_q}|(p_1,n_1)\cdots(q,n_q-1)\cdots\rangle\ ,
\label{eq:a definition}
\end{equation}
which is exactly the definition of the annihilation operator.  

Now that we have the creation and annihilation operators, we can also determine their commutator by acting on a given basis state in both orders
\begin{eqnarray}
a^\dagger_pa_p|\cdots(p,n_p)\cdots\rangle &=& n_p|\cdots(p,n_p)\cdots\rangle\ ,\\
a_pa^\dagger_p|\cdots(p,n_p)\cdots\rangle &=& (n_p+1)|\cdots(p,n_p)\cdots\rangle\ ,\nonumber
\end{eqnarray}
and then taking the difference
\begin{equation}
\left(a_pa^\dagger_p-a^\dagger_pa_p\right)|\cdots(p,n_p)\cdots\rangle = |\cdots(p,n_p)\cdots\rangle\ ,
\end{equation}
which gives us
\begin{equation}
\left[a_p,a^\dagger_p\right] = 1\ .
\end{equation}
On the other hand, if the momenta are not the same, we have
\begin{eqnarray}
a^\dagger_pa_{p'}|(p,n_p),(p',n_{p'})\cdots\rangle =\hspace{1.25in}&&\\
\hspace{0.5in}\sqrt{(n_p+1)n_{p'}}|(p,n_p+1),(p',n_{p'}-1)\cdots\rangle&&\ ,\nonumber\\
a_{p'}a^\dagger_p|(p,n_p),(p',n_{p'})\cdots\rangle =\hspace{1.25in}&&\\
\hspace{0.5in}\sqrt{n_{p'}(n_p+1)}|(p,n_p+1),(p',n_{p'}-1)\cdots\rangle&&\ ,\nonumber
\end{eqnarray}
and then taking the difference
\begin{equation}
\left(a_{p'}a^\dagger_p-a^\dagger_pa_{p'}\right)|(p,n_p),(p',n_{p'})\cdots\rangle = 0\ ,
\end{equation}
where $p\neq p'$.  Putting these together, we find
\begin{equation}
\left[a_p,a^\dagger_{p'}\right] = \delta_{p',p}\ .
\end{equation}

\section{\label{app:numerical techniques}Numerical Techniques}
In order to evaluate the complex hermitian Hamiltonian matrix numerically and diagonalize it, we wrote a C++ code.  We implemented a complex number, a matrix based on that number and a diagonalization routine that works in two steps.  The first step uses the Householder method to tridiagonalize the matrix.  This is followed by a series of Jacobi transformations that drive the final off-diagonal elements to zero.  Once this was done, the Hamiltonian in Eq.~(\ref{eq:Discrete Hamiltonian}) was implemented and diagonalized.  We will very briefly describe the diagonalization steps below.

As we mentioned, the first step of the diagonalization uses the Householder method to transform the matrix to tridiagonal form where all entries are zero except the diagonal and the elements directly above and below the diagonal.  This step occurs as a series of $n$ unitary transformations, where the Hamiltonian matrix is $n\times n$ and each transformation zeros all the elements of a row and column except the diagonal and the element directly above and below the diagonal in that row and column.  Furthermore, this routine keeps track of the product of unitary transformations in order to obtain the eigenvectors at the end.  Our code is based on the description in Numerical Recipes \cite{Press:1992C}, but modifications to allow for complex numbers and parallelization using OpenMP \cite{OpenMP} were added.  Several checks of this code were performed including testing whether the final unitary transformation (the product of the individual unitary transformations) directly triadiagonalized the original matrix and gave the same result as that obtained by the routine in the first place.

The second step uses the so-called QR diagonalization method to bring the matrix into a final diagonal form.  Each cycle of the QR diagonalization is implemented as a series of $n$ Jacobi transformations and brings the final off-diagonal entries closer to zero.  The cycle must then be run multiple times until the off-diagonal entries are below an acceptable level.  As before, our routine keeps track of the product of the transformations in order to obtain the eigenvectors as well as the eigenvalues.  Our code is again based on the description in Numerical Recipes \cite{Press:1992C}, and again has modifications to allow for complex numbers and parallelization using OpenMP.  We performed several checks of this code as well, including testing whether the final unitary transformation obtained from both routines actually diagonalizes the original matrix and gives the eigenvalues obtained by the code.

\section{\label{Evac 1/Delta p}Dependence of Vacuum Energy on $\mathbf{\lambda}$ and  $\mathbf{\Delta p}$}
In this appendix, we would like to give a very rough understanding of the vacuum energy dependence on $\Delta p$ and $\lambda$ in Eqs.~(\ref{eq:E_vac(Dp)}) and (\ref{eq:E_vac(lambda)}).
The Hamiltonian from Eq.~(\ref{eq:Discrete Hamiltonian}) has the following matrix form when the basis states (other than the single-particle basis state) are given in Table~\ref{tab:basis states}
\begin{equation}
H = \left(\begin{array}{ccc}
0 & \frac{\sqrt{2}\lambda\Delta p}{16m}\sum\frac{1}{\omega} &  \cdots\\
\frac{\sqrt{2}\lambda\Delta p}{16m}\sum\frac{1}{\omega} & 2m + \frac{\lambda\Delta p}{4m}\sum\frac{1}{\omega}+\frac{\lambda\Delta p}{8m^2} & \cdots\\
\vdots & \vdots & \ddots
\end{array}\right)\ .
\end{equation}
The dependence on $\lambda$ is all explicitly given in the Hamiltonian.  The dependence on $\Delta p$, on the other hand, comes in three forms.  The first is in the explicit dependence in the matrix, the second is in $\sum 1/\omega$ which grows larger with decreasing $\Delta p$, and the third is in the size of the matrix which also grows larger with decreasing $\Delta p$.  In order to understand the vacuum energy's dependence on $\Delta p$, all three of these sources are important.  Although we can not derive a comprehensive formula showing the dependence of the vacuum energy on both $\Delta p$ and $\lambda$, we can give an argument showing where some of that dependence comes from.

We begin by studying the effect of $\Delta p$ on $\sum 1/\omega$.  In addition to $\Delta p$, this sum also depends on $m$ and the cutoff momentum $p_{cut}$.  For a given value of $m$ and $p_{cut}$, the value of $\Delta p$ determines the number of terms in the sum.  For example, if $\Delta p=p_{cut}$, there are three terms in the sum, but if $\Delta p=p_{cut}/2$, there are five terms, and so on.  In general, when $\Delta p\ll p_{cut}$, we find that the number of terms in the sum grows as $1/\Delta p$.  If $\Delta p$ is halved, then the number of terms doubles.  Furthermore, when $\Delta p\ll p_{cut}$, each term is similar in size to its neighbors, so that $\sum 1/\omega$ grows like $1/\Delta p$.  Taken together, we find
\begin{equation}
\sum \frac{1}{\omega} = \frac{f(m,p_{cut})}{\Delta p}\ ,
\label{eq:app:1/omega}
\end{equation}
where $f(1GeV,10GeV)=6$ for the reference parameters given in Eq.~(\ref{eq:parameters}).  With this relation, we can see that the $1/\Delta p$ in $\sum 1/\omega$ cancels the $\Delta p$ in the numerator of many Hamiltonian matrix elements.  Plugging this in, we have
\begin{equation}
H = \left(\begin{array}{cccc}
0 & \frac{\sqrt{2}\lambda f}{16m} & \frac{\lambda f}{8\sqrt{m^2+\Delta p^2}}  \cdots\\
\frac{\sqrt{2}\lambda f}{16m} & 2m + \frac{\lambda f}{4m}+\frac{\lambda\Delta p}{8m^2} & \frac{\sqrt{2}\Delta p}{8m\sqrt{m^2+\Delta p^2}} \cdots\\
\frac{\lambda f}{8\sqrt{m^2+\Delta p^2}} & \frac{\sqrt{2}\lambda\Delta p}{8m\sqrt{m^2+\Delta p^2}} & \cdots \\
\vdots & \vdots & \ddots
\end{array}\right)\ ,
\end{equation}
where $f=f(m,p_{cut})$ is indepdent of $\Delta p$ and $\lambda$.  

Although this integral is still impossible to solve analytically, we can solve it analytically if we temporarily simplify by taking the remaining explicit dependence on $\Delta p\to0$.  Then, we have the matrix
\begin{equation}
H \sim \left(\begin{array}{cccccc}
0 & \frac{\sqrt{2}\lambda f}{16m} & \frac{\lambda f}{8m} & \frac{\lambda f}{8m} &  \cdots\\
\frac{\sqrt{2}\lambda f}{16m} & 2m + \frac{\lambda f}{4m} & 0 & 0 & \cdots\\
\frac{\lambda f}{8m} & 0 & 2m + \frac{\lambda f}{4m} & 0 & \cdots \\
\frac{\lambda f}{8m} & 0 & 0 & 2m + \frac{\lambda f}{4m}  & \cdots \\
\vdots & \vdots & \vdots  & \vdots & \ddots
\end{array}\right)\ .
\label{eq:App:HDp->0}
\end{equation}
Of course, this matrix does not give the full behavior of the complete Hamiltonian, but it turns out that we can solve it for the vacuum energy.  This matrix is of the form
\begin{equation}
H\sim\left(\begin{array}{cccccc}
0&\frac{A}{\sqrt{2}}&A&A&\cdots\\
\frac{A}{\sqrt{2}}&B&0&0&\cdots\\
A&0&B&0&\cdots\\
A&0&0&B&\cdots\\
\vdots&\vdots&\vdots&\vdots&\ddots
\end{array}\right)\ ,
\end{equation}
which has the eigenvalues
\begin{equation}
\frac{1}{2}\left(B\pm\sqrt{(4N-6) A^2+B^2}\right),B,B,B,\cdots\ .
\end{equation}
where $N$ is the size of the matrix (the number of rows).
Going back to the matrix in Eq.~(\ref{eq:App:HDp->0}), we find
\begin{equation}
E_{vac} \sim \frac{16m^2+2f\lambda-\sqrt{2}D}{16m}
\end{equation}
where
\begin{equation}
D=\sqrt{\left(\frac{g}{\Delta p}-1\right)f^2\lambda^2+32f\lambda m^2+128m^4}
\end{equation}
and $g=g(p_{cut})$ is a function of the cutoff momentum, but is independent of both $\Delta p$ and $\lambda$.  If we expand this in a Taylor series around $\lambda=0$ and $\Delta p=0$, we obtain for the leading term
\begin{equation}
E_{vac} \sim -\frac{g f^2\lambda^2}{256m \Delta p}\ .
\end{equation}
Although this form does not take into account the full explicit dependence on $\Delta p$ in the Hamiltonian(and some dependence on $\lambda$ is dropped when $\Delta p$ is taken to zero as well), it does include the dependence on $\Delta p$ in both $\sum 1/\omega$ and in the size of the matrix.  Notwithstanding the neglected dependence on $\Delta p$ and $\lambda$, this result does have the right form.  We find that $E_{vac}\sim1/\Delta p$ as in Eq.~(\ref{eq:E_vac(Dp)}) and $E_{vac}\sim\lambda^2$ as in Eq.~(\ref{eq:E_vac(lambda)}).

We also see from Eq.~(\ref{eq:1 part energy}) that the single particle energy, after using Eq.~(\ref{eq:app:1/omega}) is
\begin{equation}
E_{single} = m+\frac{\lambda f}{8m}\ ,
\end{equation}
which is independent of the momentum spacing, as described in Subsec.~\ref{sec:Delta p}, but depends linearly on $\lambda$ as described in Subsec.~\ref{sec:lambda}.

%Furthermore, if we accept the next higher eigenvalue from Eq.~(\ref{eq:App:HDp->0}), we have for the first two-particle state energy
%\begin{equation}
%E_{two} \sim 2m+\frac{\lambda f}{4m} = 2E_{single}\ ,
%\end{equation}
%independent of $\Delta p$ but linear in $\lambda$ as described in Subsecs.~\ref{sec:Delta p} and \ref{sec:lambda}, respectively.

\end{document}